# The competitive nature of STAT complex formation drives phenotype switching of T cells

Ildar I Sadreev[1,2,][*], Michael Z Q Chen[3], Yoshinori Umezawa[4], Vadim N Biktashev[1], Claudia Kemper[5], Diana V Salakhieva[6], Gavin I Welsh[7], Nikolay V Kotov[8].

[1] Centre for Systems, Dynamics and Control, College of Engineering, Mathematics and Physical Sciences, University of Exeter, Harrison Building, North Park Road, Exeter EX4 4QF, UK

[2] (current address) Section of Experimental Haematology, Leeds Institute of Cancer and Pathology, University of Leeds, Leeds, UK

[3] Department of Mechanical Engineering, The University of Hong Kong, Pokfulam Road, Hong Kong

[4] Department of Dermatology, The Jikei University School of Medicine, 3-25-8 Nishishimbashi, Minato-ku, Tokyo, 105-8461, Japan

[5] Division of Transplantation Immunology and Mucosal Biology, MRC Centre for Transplantation, King's College London, Guy's Hospital, London, United Kingdom, SE1 9RT

[6] Kazan (Volga Region) Federal University, 18 Kremlyovskaya St., 420008 Kazan, Russia

[7] Bristol Renal, School of Clinical Sciences, University of Bristol, Dorothy Hodgkin Building, Whitson Street, Bristol, BS1 3NY, UK

[8] Biophysics & Bionics Lab, Institute of Physics, Kazan Federal University, Kazan 420008, Russia

[*] Corresponding author. Ildar I Sadreev. Section of Experimental Haematology, Leeds Institute of Cancer and Pathology, University of Leeds, Leeds, UK. E-mail: ildar.sadreev@gmail.com

# Abstract

Signal transducers and activators of transcription (STATs) are key molecular determinants of T cell fate and effector function. A number of inflammatory diseases are characterized by an altered balance of T cell phenotypes and cytokine secretion. STATs, therefore, represent viable therapeutic targets in numerous pathologies. However, the underlying mechanisms of how the same STAT proteins regulate both the development of different T cell phenotypes and their plasticity during changes in extracellular conditions remain unclear. In this study, we investigated the STAT mediated regulation of T cell phenotype formation and plasticity using mathematical modeling and experimental data for intracellular STAT signaling proteins. The close fit of our model predictions to the experimental data allows us to propose a potential mechanism for T cell switching. According to this mechanism, T cell phenotype switching is due to the relative redistribution of STAT dimer complexes caused by the extracellular cytokine-dependent STAT competition effects. The developed model predicts that the balance between the intracellular STAT species defines the amount of the produced cytokines and thereby T cell phenotypes. The model predictions are consistent with the experimentally observed IFN-γ to IL-10 switching that regulates human Th1/Tr1 responses. The proposed model is applicable to a number of STAT signaling circuits.

# Author Summary

The immune system is a highly sophisticated and regulated complex of molecular interactions, the main function of which is to protect the host organism. Despite much recent progress in experimental studies, the underlying **molecular mechanisms of autoimmune diseases are still not clear and therefore new approaches to solve this problem are required. Systems biology can offer the necessary analytic tools for deeper elucidation of molecular mechanisms using available experimental data. In this work, we studied T cell plasticity and cytokine production focusing on STAT proteins, one of the key molecular elements of signal transduction in the immune system. We propose a new integrative systems approach to analyze STAT-STAT interactions in the immune response. The model developed in this study suggests that the T cell plasticity is due to the competition between the intracellular STAT pathways. We studied STAT-STAT interactions in the context of human immune-related diseases such as rheumatoid arthritis, inflammatory bowel disease and systemic lupus erythematosus and discuss the potential therapeutic implications of the proposed model in these diseases. The predictions of the model proposed in this study are supported by experimental data for IFN-γ and IL-10 production.**

## Introduction

Signal Transducers and Activators of Transcription (STATs) regulate cell differentiation, growth, apoptosis and proliferation by transducing signals from the cell membrane to the nucleus. There are seven members of the STAT family in mammalian cells: STAT1, STAT2, STAT3, STAT4, STAT5a, STAT5b and STAT6 [1, 2].

STAT proteins are activated by binding of a cytokine to its receptor followed by receptor dimerization and phosphorylation of their C-terminal transactivation domain (CTD) by Janus Kinases (JAKs). For example, phosphorylation of STAT1 occurs at Tyr701 in response to type II interferons [3] and phosphorylation of STAT3 occurs at Tyr705 in response to Interleukin 6 (IL-6) or Interleukin 10 (IL-10) [4, 5]. Phosphorylation at Tyr705 leads to dimerization [6] and regulates the activation of STAT3 [7-9]. STATs can form homo- or hetero- dimers only with their dimerization partners [10]. STAT dimers translocate to the nucleus and activate gene expression. Nuclear phosphatases can dephosphorylate STATs in the nucleus and initiate their return to the cytosol [11-14].

In the nucleus, activated STAT dimers induce cytokine production. Aberrations in the mechanism of cytokine production may give rise to various immune-related pathologies including autoimmune diseases such as rheumatoid arthritis (RA) [15, 16], systemic lupus erythematosus (SLE) [17-19], diabetes [20, 21] and cancer [22, 23]. These aberrations include inappropriate activation of Th1 cells, characterized by increased inflammatory IFN-γ production. It was reported in [24] that RA patients lack the so called IFN-γ to IL-10 switching – the transition of the inflammatory IFN-γ only state (Th1 cells) into a state characterized by a significant decrease of IFN-γ production and a gain of the regulatory IL-10 expression (Tr1 cells).

Several STATs, for example STAT3 and STAT5 [25, 26], lead to the production of IFN-γ and IL-10. The molecular mechanism of IFN-γ and IL-10 production via STAT3 and STAT5 is as follows. Extracellular IL-2, IL-6 and IL-21 bind to their complementary receptors. The receptors remain in complex with JAKs. Binding of the interleukins to their receptors induces the autophosphorylation of the receptors and the bound JAKs [27]. The phosphorylated receptor-JAK complexes can be dephosphorylated by SHP-1 [28]. Phosphorylation of STAT3 is performed by JAK in IL-2R:JAK and IL-6R:JAK complexes, while STAT5 is phosphorylated by IL-2R:JAK and IL-21R:JAK [29-31]. STAT3 and STAT5 are dephosphorylated by SHP-1 and SHP-2 phosphatases respectively [32, 33]. The phosphorylated STAT3 and STAT5 then can form either homo- or hetero- dimers [29, 30]. The dimers translocate into the cell nucleus and promote transcription of genes responsible for IFN-γ and IL-10 production. Both IFN-γ and IL-10 are degraded by some metalloproteases [34], which, in our model, we denote as Mp1 and Mp2, respectively.

There is an additional, STAT-independent, mechanism of the regulation of IL-10 production. It has been shown that pathogens can activate the c3-c3b complement system, which leads to the activation of CD46. For high IL-2 concentrations CD46 activates the SPAK/ERK pathway leading to the activation of SP1 [24, 35], which results in induction of genes responsible for IL-10 production [36].

Other STAT pathways, in addition to STAT3 and STAT5, can also lead to the induction of IFN-γ and IL-10 production. The production of IFN-γ is induced by the following interleukins: IL-12, IL-21, IL-2 and IL-35 [37-40]. Interleukins IL-12 and IL-35 activate STAT4 through the JAK-STAT pathway [30] while IL-21 and IL-2 activate STAT5 [41, 42]. The production of anti-inflammatory IL-10 is up-regulated by STAT1, STAT3 and STAT6 [26, 43, 44]. In particular, STAT1 is activated by IL-6 and IL-35; STAT3 by IL-2 and IL-6; STAT6 by IL-3 and IL-4 [30, 31, 45, 46].

Both STAT4 [37, 38, 46] and STAT5 [25, 47, 48] induce the production of IFN-γ. However, the data demonstrating a role of STAT5 in IFN-γ production are in contrast with results reported in [49], which shows that STAT5 can induce IL-10 production. At the same time, the fact that STAT4 activates IFN-γ production contradicts with the data in [50], where it was shown that STAT4 also induces the production of IL-10. This conflicting evidence about STAT signaling cannot be explained with the currently used approaches and therefore a new approach is required.

Several studies have offered insights into STAT signaling on a systems level. A mathematical model of JAK-STAT signaling pathway leading to the activation of STAT1 in liver cells was proposed in [51]. In that work, the dynamical properties of this system were investigated. The model showed that nuclear phosphatase is one of the most important regulators in this JAK-STAT pathway. Another attempt to model STAT1 activation was made in [52] studying JAK-STAT signaling in pancreatic stellate cells (PSC). By using Ordinary Differential Equations (ODEs) to describe the rates of the biochemical reactions in the JAK-STAT pathway, the model developed in that study could explain the temporal profiles of STAT1 activation.

In [53], a mathematical model for the JAK2-STAT5 pathway was proposed. The quantitative behavior of STAT5 phosphorylation was determined. In [54], the authors investigated IL-6 mediated JAK1-STAT3 pathway activation and the dynamics of JAK1 and STAT3 phosphorylation. They proposed a new approach to analyze JAK-STAT pathways by using Petri nets that describe the biomolecular mechanism of reactions and functional interactions with other components.

Thus there are several examples [51-54], where STAT signaling has been investigated using mathematical modeling. However, in these studies only one JAK-STAT pathway was investigated at a time. In phenotype development and plasticity in response to environmental changes in T-cells, more than one JAK-STAT pathways are involved [24, 55]. Due to this fact, these models of STAT signaling [51-54] cannot be used to explain the underlying molecular mechanisms of these processes. Therefore, to understand disease states such as autoimmune states and allergic reactions [24, 56], new approaches are needed to delineate the molecular mechanisms of T cell phenotype development and plasticity that are crucial for an efficient T-cell response.

In our previous paper [57] we studied phosphorylation of one STAT protein at a time. In this study, we investigated the role of multiple STAT proteins in T cell phenotype plasticity and in the IFN-γ to IL-10 switching [24]. The new developed model is based on previously published experimental results [29-31, 37-39, 41, 42, 45, 46, 58-64].

## Results

**A new integrative approach for STAT signaling**

In this study, we develop a novel approach to integrate multiple pieces of experimental data related to STAT signaling in T cells. We focus on integrating separate JAK-STAT signaling pathways into a unified cellular response. In most of the previous studies [51-54], only one signaling pathway was studied at a time, but this cannot fully predict the cellular behavior occurring in response to the interplay of STAT signaling pathways, which include cell fate determination, polarization and associated plasticity. Here we attempt to build a systems biology model that integrates STAT signaling pathways to analyze possible interplay.

Fig 1A summarizes previously published experimental results [29-31, 37-39, 41, 42, 45, 46, 58-63] and schematically represents the interdependent events in the cytokine/JAK-STAT signaling pathways. The diagram shows which cytokines activate the known STAT proteins and illustrates the fact that the STAT proteins form homodimers as well as heterodimers only with their dimerization partners [10]. In our model, STAT1 interacts with STAT2, STAT3 and STAT4 only, which is consistent with [29, 59, 60]. Heterodimerization partners for STAT3 are STAT4 and STAT5. STAT4 forms a heterodimer complex with STAT3 and STAT5 whereas the only partner for STAT6 is STAT2 [61-63]. In the scheme shown in Fig 1A, it is essential to study the mechanism of cross-talk as it influences the signaling of the participants and the outcome of the overall cellular response, which contributes to the T cell plasticity.

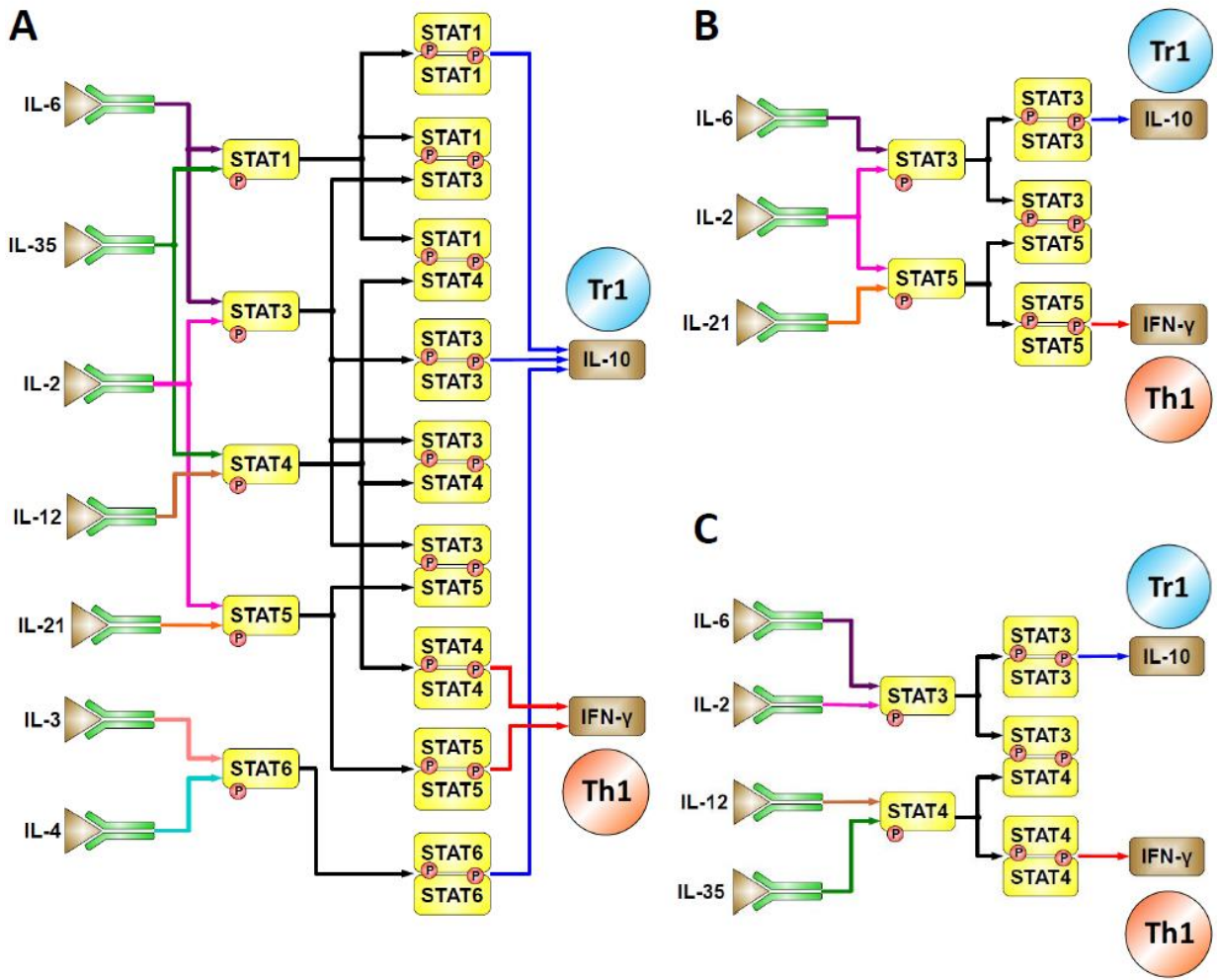

**Fig 1. The map of interleukins involved in induction of IFN-γ and IL-10 production via the STAT-activating mechanisms.** A. The map of interactions between the cytokines and the STATs based on experimental studies [29-31, 37-39, 41, 42, 45, 46, 58-63]. According to the map, STAT proteins are activated in response to extracellular cytokines. STATs can form dimer complexes only with certain dimerization partners. STAT dimers induce IFN-γ and IL-10 gene expression. B. STAT3-STAT5 subsystem extracted from the full map of interactions shown in (A). This subsystem activates IFN-γ and IL-10 production in response to IL-2, IL-6 and IL-21. In this model, IL-2 activates both STAT3 and STAT5 while the concentration of other cytokines is maintained constant. C. STAT3-STAT4 subsystem extracted from the full map (A). This subsystem induces the expression of IFN-γ and IL-10 in response to IL-2, IL-6, IL-12 and IL-35. In this case IL-2 activates STAT3 only.

In our model schematically shown in Fig 1A, we assume that the STAT homodimers are more effective in mediating gene induction and therefore have a greater contribution to cytokine production compared with the heterodimers, which is consistent with [65]. This assumption also means that the ratio between the STAT homodimers might define the type of produced cytokine and, thereby, the T cell phenotype. Our model suggests that there is a natural balance between STAT hetero- and homo- dimers. When there is no signal to switch (which corresponds to a certain concentration of input cytokines), the ratio between the homodimers is balanced. After the T cell receives the signal to switch, this balance is disrupted and then restored again, however there is now a new ratio between the competing STATs. This newly balanced ratio between the STAT homo- and hetero- dimers leads to the new type of produced cytokine and to the T cell phenotype switching.

Here for the first time to our knowledge we propose an underlying competition mechanism between the STAT homodimers modulated by the extracellular cytokines. Despite the fact that STAT interactions have been extensively studied in recent years, the necessity of heterodimer complex formation and the functional implications of these complexes still remain unclear [30]. According to our model, the role of the STAT heterodimers is to provide a "buffer", which can be defined as an intermediate state of species between two STAT homodimers that allows the STAT species to transfer from one state to another (for example, from STAT3- to STAT5- prevailing states).

The model proposed in this paper is applied to investigate previously reported plasticity effects between Th1 and Tr1 cell populations. One of the expressed cytokines, IFN-γ, is widely associated with the inflammatory Th1 phase while the other, IL-10, is dominant during the regulatory Tr1 phase [24]. The structure of the proposed model shown in Fig 1A allows for the assumption that the competition between the STAT proteins defines the expression levels of IFN-γ and IL-10. The direct result of the described experimental data integration followed by systems biology analysis is the proposition that cytokine-dependent STAT interactions lead to switching between two tightly regulated systems: the inflammatory IFN-γ only Th1 state into the regulatory Tr1 state characterized by increased IL-10 and decreased IFN-γ production levels [24].

**The model for coupled STAT3-STAT5 signal transduction**

Due to the relative complexity of Fig 1A, we have divided the full diagram into separate functional circuits, STAT3-STAT5 and STAT3-STAT4 based on the cytokine-cytokine interactions (Fig 1B and Fig 1C respectively). In our model, IFN-γ inducing STATs include STAT4 [37, 38, 46] and STAT5 [25, 47, 48] while IL-10 production is supported by STAT1, STAT3 and STAT6 [26, 43, 44]. To investigate the role of each STAT pair, the scheme is divided into subsystems in such a way that one STAT in the pairing induces the expression of IFN-γ while the other induces the expression of IL-10. Here we only focus on the STAT pairings that produce the opposite (inflammatory and regulatory) immune responses, therefore STAT1-STAT3 combination is not considered since both STAT1 and STAT3 lead to the expression of the regulatory IL-10.

Fig 1B shows that IL-2 is the only "input" cytokine in our model that can activate both STAT3 and STAT5, which are IL-10 and IFN-γ inducing STATs, respectively. At the same time, both IL-6 and IL-21 can activate only one STAT at a time, STAT3 and STAT5, respectively. The other IL-2-activated STAT combination, which induces the production of both IFN-γ and IL-10, is STAT3-STAT4 (Fig 1C). However, in contrast to the STAT3 and STAT5 pairing, in the STAT3-STAT4 system, IL-2 activates only STAT3 and not STAT4. Due to this role of IL-2 in STAT3, STAT4 and STAT5 activation, we studied the effects of varied IL-2 concentration assuming the continuous presence of other input cytokines (IL-6 and IL-21) by fixing their concentrations at constant levels. Therefore, we do not consider STAT1-STAT4 model since IL-2 does not activate any of the STATs involved in this pairing. As a result, the complicated scheme depicted in Fig 1A is divided into the two functionally similar but architecturally different submodules shown in Fig 1B and Fig 1C that can now be described mathematically.

We start our analysis with the STAT3-STAT5 system as illustrated in Fig 1B before moving on to analyze both the STAT3-STAT4 system (Fig 1C) and the combined STAT3-STAT4-STAT5 system. In order to build a model for the STAT3-STAT5 molecular system (Fig 1B), we include more biological details into the description of molecular mechanism: cytokine-receptor interactions, STAT phosphorylation/dimerization, CD46/SP1 signaling and cytokine production. The description of the governed reactions and equations can be found in the Materials and Methods and Supplementary Materials.

Next we obtained a set of parameters that fit our model predictions to the experimental data for IL-2 dependent IFN-γ and IL-10 production in Th1/Tr1 switching. The data were taken from [64] and is represented in Fig 2A as circles and crosses for the normalized IFN-γ and IL-10 concentrations as a function of IL-2 respectively. The details of parameter optimization, the set of optimized parameters and parameter sensitivity analysis are described in Supplementary Materials. Solid lines in Fig 2A illustrate the model predictions for the optimized set of parameters (set "O3" in Table S2).

Fig 2A visually demonstrates a good fit to the experimental data from [64], which is quantitatively supported by the small squared error (please see Table S2 in Supplementary Materials for more details) and also shows that with an increase of IL-2, the concentration of IFN-γ initially increases reaching a peak and then decreases while the concentration of IL-10 gradually increases. The model predictions (Fig 2A) clearly show the switching between the two populations of Th1 and Tr1 cells. At the same time there is an evidence that there is a population that produces both IFN-γ and IL-10, the origin of which has not yet been established [24].

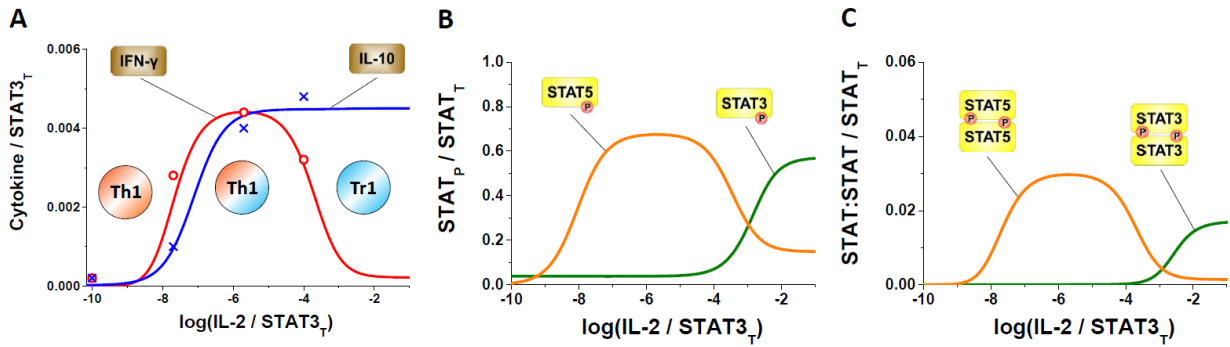

**Fig 2. The IL-2-dependent IFN-γ to IL-10 switching is due to the underlying STAT competition.** A. Model predictions (solid lines) compared with the experimental data for IFN-γ and IL-10 production as a function of IL-2 concentration (red circles and blue crosses respectively). Normalized experimental data from [64] show that the concentration of produced IFN-γ (circles) and IL-10 (crosses) depends on IL-2 concentration. With an increase of IL-2, the production of IFN-γ initially increases compared with the production of IL-10 for the same IL-2 concentration. Further increase of IL-2 leads to the decrease of IFN-γ and increase of IL-10 concentration. The low IL-10 and high IFN-γ correspond to the Th1 cell state, the medium IFN-γ and IL-10 concentrations correspond to the IL-10-producing Th1 cells and high IL-10 and low IFN-γ correspond to the Tr1 cell state, which is in line with the experimentally observed fact that the switching occurs for high amounts of IL-2 and that the activation of IFN-γ always precedes IL-10 [24]. B. Selective (bell-shaped) concentration-dependent STAT5 activation profile as a function of IL-2 concentration. C. STAT5:STAT5 homodimers also demonstrate selectivity to IL-2 due to the high dependence on STAT5p.

**IL-2-dependent selective regulation of the STAT competition**

STAT3 induces IL-10 production by forming STAT3:STAT3 homodimers [65, 66] and STAT5 induces IFN-γ production by forming STAT5:STAT5 complexes [25, 48]. We hypothesize here that the experimentally established IFN-γ to IL-10 switching (Fig 2A) is due to the STAT competition and caused by the STAT switching. In order to test our hypothesis, in this section we analyzed the model predictions for both STAT monomers and homodimers.

Fig 2B shows the STAT redistribution, which can be defined as the switching between the phosphorylation levels of the STATs as a function of IL-2 concentration. For low IL-2 concentrations STAT5 is more phosphorylated than STAT3, whereas for higher IL-2 concentrations phosphorylated STAT3 prevails over STAT5. Fig 2C illustrates the dependences of the STAT3:STAT3 and STAT5:STAT5 dimers, normalized by the total STAT3, on IL-2 concentration. The shapes of the curves for the dimers (Fig 2C) are similar to the shapes of their monomers (Fig 2B) due to the high dependence of the dimer concentrations on their monomers. Our model also predicts the bell-shaped dependence of phosphorylated STAT5 on IL-2 as well as STAT5:STAT5 homodimers on IL-2. This bell-shaped dependence means that STAT5 is selective to IL-2, or in other words, STAT5 has its maximum activity for a certain range of IL-2, where STAT5 phosphorylation level is high.

The predicted bell-shaped relationship may offer new insights into the dual role of STAT-mediated cytokine production. Experimental evidence suggests that STAT5 induces IFN-$\gamma$ production [25, 47, 48]. However, it was also shown in [49] that STAT5 can induce IL-10 production. Our model can explain this discrepancy by showing that the same amount of phosphorylated STAT5 may lead to low or high IL-10 production levels, depending on the extracellular concentration of IL-2 (Fig 2).

The model predictions presented in this section support our hypothesis that the switching between IFN-$\gamma$ producing Th1 cells and IL-10 producing Tr1 cells is due to the switching between the competing STAT proteins. Therefore, we suggest that the cells, that produce IFN-$\gamma$ and IL-10 simultaneously (IFN-$\gamma^+$IL-10$^+$ cells), develop from IFN-$\gamma$ producing Th1 cells that have received IL-2-dependent signal to switch, rather than represent a specific T cell population. This result is essential for understanding the mechanism of cytokine switching [24, 64].

## Investigation of possible mechanisms of JAK-STAT mediated inflammatory pathologies

We applied the developed model to investigate potential deviations in the immune system caused by changes in the tightly-regulated JAK-STAT signaling pathways. It has been reported that IFN-γ and IL-10 play a crucial role in autoimmune pathologies [67, 68]. Fig 3A illustrates the influence of changes in STAT pathways on IL-10 production. The model predicts that the production of IL-10 can be increased by attenuating the degradation of IL-10 by metalloproteases (increase of our model parameters $n_9$ and $M_{20}$) or, alternatively, by an increase of the total concentration of *IL-10* gene (modeled by parameter $g10_t$), as shown in Fig 3A and consistent with [69]. The changes in Fig 3A are illustrated for 15% parameter perturbation of $n_9$.

Our model suggests that the production of pro-inflammatory IFN-γ can be controlled by various intracellular mechanisms. For example, our model predicts (Fig 3B) that the magnitude of IFN-γ can be reduced by attenuation of the STAT5 pathway signaling (decrease of $s5_t$) [47] or by enhancement of metalloprotease-induced degradation of the produced IFN-γ (decrease of $M_{18}$ and $n_8$). This may appear to be an expected result as it follows from the structure of the model shown in (Fig 1B). However, another prediction of the model is that changes in the STAT3 pathway can also reduce the level of the STAT5-activated production of IFN-γ, which is not obvious from Fig 1B. This effect could be achieved by enhancing the formation of the STAT3:STAT5 heterodimer complex (decrease of $M_{14}$) or alternatively by attenuation of STAT3 dephosphorylation (increase of $M_9$). Fig 3B shows the effects of perturbations of these model parameters for a 1.5-fold change of $M_9$.

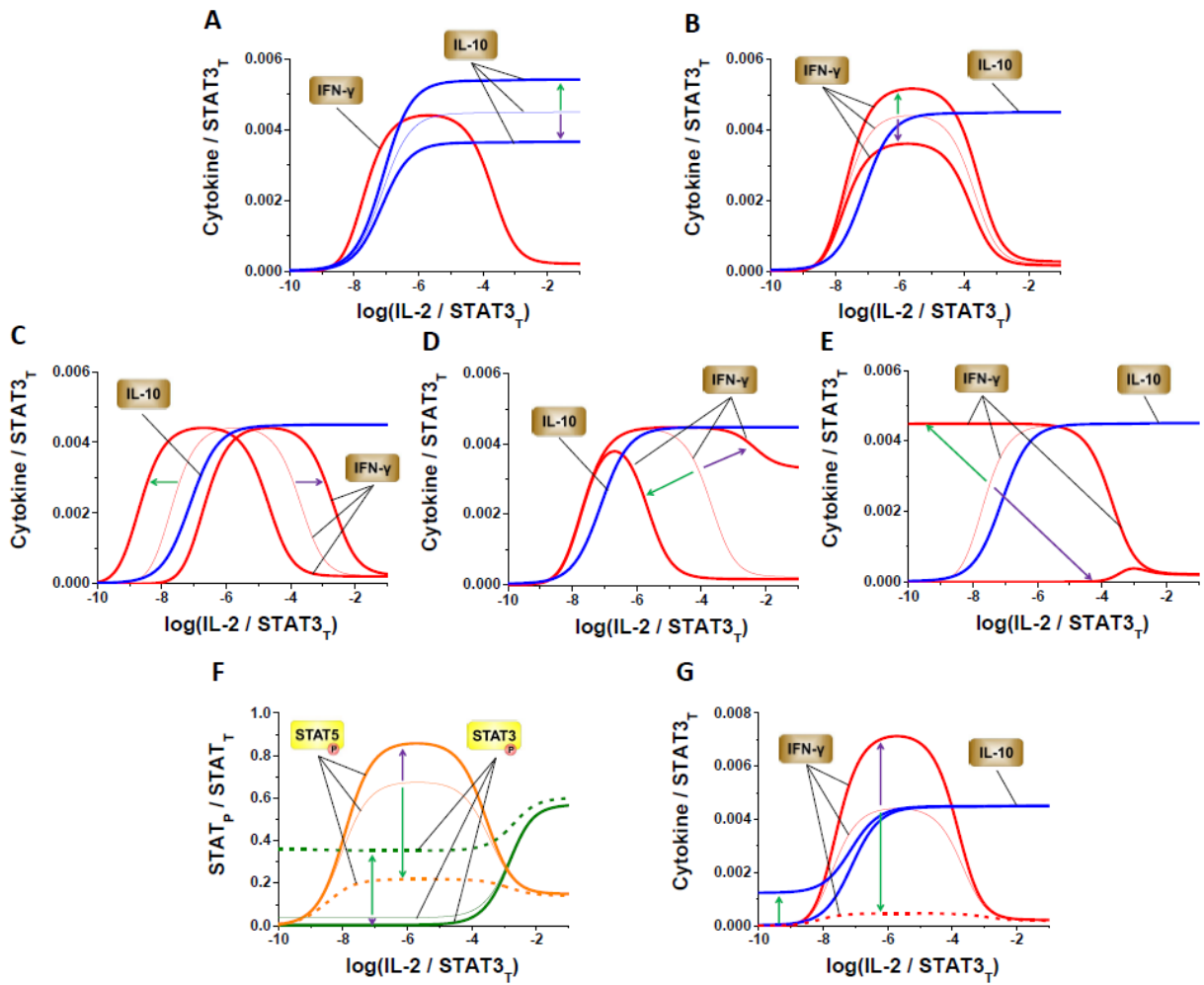

**Fig 3. The effects of changes in intracellular regulation on the developed model predictions.** The aberrations in both activating or competing JAK-STAT pathways regulate the amounts of produced IL-10 (A) and IFN-γ (B-E). The profile of IFN-γ production can be shifted along IL-2 axis by changes in IL-2R signaling (C). The aberrations in STAT3 and STAT5 phosphorylation may lead to IFN-γ profile shift along IL-2 axis with the reduction of its magnitude at the same time (D, E). The model predictions show that increased IL-6 leads to the lack of STAT redistribution (dotted lines in F) while reduced amount of IL-6 may lead to stronger STAT redistribution (solid lines in F). The lack of STAT redistribution observed for increased IL-6 leads to significantly lower level of IFN-γ production (G). Thus, our model predicts that the level of IFN-γ production can be reduced by the IL-6-dependent changes in the competing STAT3 pathway. Thin lines represent the model predictions for the optimized set of parameters shown in Supplementary Materials, while solid and dotted lines show the model predictions for the perturbed parameters shown in Table S3.

As a result of the parametric alterations, the shape of IFN-γ dependence shifts along the IL-2 axis (Fig 3C). According to our model, the alterations of the parameters that cause the IFN-γ dependence shown in Fig 3C to shift to higher IL-2 concentrations, represent the changes in IL-2 receptor activation. These changes include a decrease of the total amount of IL-2R (decrease of $r2_t$) or enhancement of the dephosphorylation of phosphorylated IL-2R by SHP-1 (decrease of $n_1$). It is notable that although the peak shifts along the IL-2 axis, the magnitude does not change during this transformation. These results might be important since it was shown that IL-2R signaling controls tolerance and immunity and that IL-2R deficiencies give rise to various pathologies including Inflammatory Bowel Disease (IBD) [24, 70]. Fig 3C shows the effects of $n_1$ perturbation of one order of its magnitude.

The proposed model suggests new potential strategies for the control of IFN-γ selectivity on IL-2 concentration. According to the model predictions, the IFN-γ peak shift along the IL-2 axis can be also achieved by alterations in the competing STAT3 signaling pathway [71], namely by IL-2 mediated STAT3 phosphorylation (parameter $M_7$) as shown in Fig 3D. This effect is a result of indirect interactions due to the redistribution of the STAT complexes as STAT3 does not directly regulate IFN-γ production. Our model predicts that due to the competition effects between STATs, STAT3 indirectly inhibits STAT5, which induces the production of IFN-γ.

The developed model also suggests that attenuation of IL-2-induced phosphorylation of STAT5 (increase of $M_{10}$) reduces IFN-γ magnitude and shifts the peak to the range of higher IL-2 concentrations [25, 72, 73] (Fig 3E). Our model predicts that the peak disappears when we apply the opposite changes (decrease of $M_{10}$).

Thus our model predictions demonstrate possible scenarios of alterations in JAK-STAT pathways that influence IFN-γ and IL-10 production. The scenarios include a change of IL-10 (Fig 3A) and IFN-γ (Fig 3B and Fig 3C) as well as the selective regulation of IFN-γ by IL-2 (Fig 3D and Fig 3E). The effects of the parametric changes on the concentrations of produced IFN-γ and IL-10 are summarized in Supplementary Table S3. The color of the arrows in Table

S3 corresponds to the color of the arrows that represent changes shown in Fig 3. The main message of Fig 3A-Fig 3E is that our model suggests alternative approaches for the regulation of IFN-γ and IL-10 production by employing their competing STAT pathways, STAT3 and STAT5 respectively.

In our model, in addition to IL-2, which controls the Th1/Tr1 switching [24], other cytokines also play an important role in T cell polarization. In the chosen example (Fig 1B) the production of anti-inflammatory IL-10 is up-regulated by IL-6 via the STAT3 signaling pathway [74]. Therefore, we employed our model to study the interplay between IL-2 and IL-6 through the STAT signaling pathways and their role in T cell polarization.

The developed model predicts how variations in the concentration of IL-6 may impact the switching (Fig 3F- Fig 3G). In our model, the concentration of IL-6 is described by parameter $Q_6$ as shown in Equation (6). Fig 3F illustrates that a decrease in IL-6 ($Q_6$ decrease) leads to stronger STAT5 to STAT3 switching, whereas an increase in IL-6 concentration ($Q_6$ increase) causes changes in both STAT3 and STAT5 phosphorylation levels as a function of IL-2 and thereby the lack of switching. It can be seen from Fig 3F that IL-6 activates STAT3 but at the same time inhibits STAT5 due to the STAT competition, which is consistent with observations in [75]. As a result of IL-6 impact on the STAT competition (Fig 3F), our model shows that IL-6 also affects the STAT-mediated production of IFN-γ and IL-10 (Fig 3G). In particular, our model predicts that due to the redistribution between STAT3 and STAT5, increased concentrations of IL-6 lead to the reduced level of pro-inflammatory IFN-γ production. This result might be clinically important since it may offer novel strategies for reducing IFN-γ, which is essential in abnormal production of this pro-inflammatory cytokine during inflammation [24].

**Comparative analysis of STAT3-STAT4 versus STAT3-STAT5 machinery**

In this section, we combine the STAT3-STAT5 (Fig 1B) and STAT3-STAT4 (Fig 1C) modules, the first of which was thoroughly investigated in the previous sections, with the aim of understanding the integral properties that emerge from the STAT-STAT circuit pairings. To this end, we first highlight the structural differences between the STAT3-STAT4 and STAT3-STAT5 circuits. The most crucial difference between the two circuits is in the role of IL-2. In the STAT3-STAT5 subsystem, IL-2 activates both STAT3 and STAT5, while in the STAT3-STAT4 subsystem, IL-2 activates only STAT3 but not STAT4. The other cytokines, that activate STAT4, include IL-12 and IL-35 (Fig 1C). In our *in silico* experiment, we assume that the concentrations of IL-12, IL-35 and STAT3-activating IL-6 are maintained constant while IL-2 is varied. This implies that only the amount of phosphorylated STAT3 in the STAT3-STAT4 pairing can be directly changed by varied IL-2 concentration. In order to study the STAT redistribution and cytokines interdependence, we next investigated the STAT3-STAT4 circuit for various IL-2 concentrations. A detailed description of the model for STAT3-STAT4 subsystem can be found in Supplementary Materials.

In absence of experimental data for STAT4 signaling, we assume that the parameters in the STAT3-STAT4 subsystem are similar to the parameters in the STAT3-STAT5 subsystem (Table S4). Fig 4A-Fig 4C represent the model predictions for the STAT3-STAT4 subsystem (Fig 1C). The figures include STAT3, STAT4 monomers (Fig 4A), STAT3:STAT3, STAT4:STAT4 homodimers (Fig 4B) as well as IFN-γ and IL-10 dependences on IL-2 concentration (Fig 4C). Despite the fact that IL-2 does not affect phosphorylation of STAT4 in the STAT3-STAT4 circuit, the switching is nonetheless present in this submodule.

The model suggests that the STAT and subsequent phenotype switching is due to the competition between STAT3 and STAT4 species rather than competition for the source of IL-2.

This suggestion is supported by the structure of our model (Fig 1C), where IL-2 activates STAT3 and not STAT4 in the pairing and therefore there is no competition for the source. Mechanistically, the competition between STAT3 and STAT4 includes redistribution between these species that is implemented by the formation of STAT3:STAT4 heterodimer.

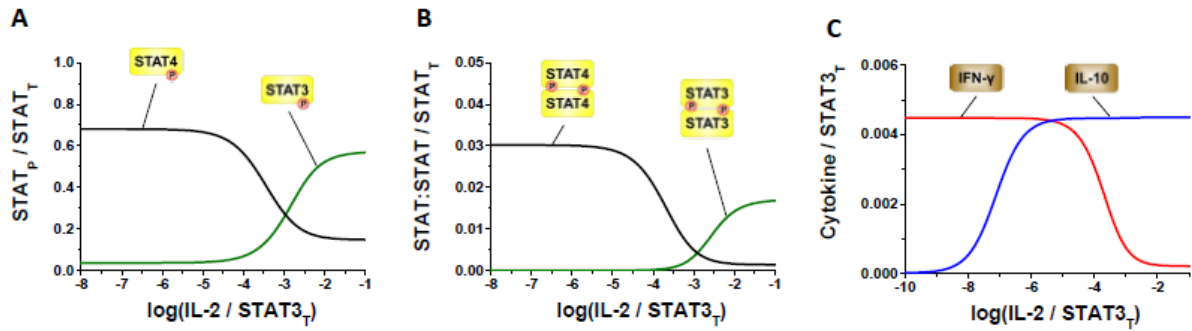

**Fig 4. Model predictions for the STAT3-STAT4 circuit.** Redistribution between STAT3 and STAT4 monomers (A) as well as STAT3:STAT3 and STAT5:STAT5 homodimers (B) leads to the IFN-γ to IL-10 switching for higher IL-2 concentrations shown in (C). Our model predicts that there is a significant basal level of STAT4p and thereby STAT4:STAT4 homodimer for low IL-2. This is due to the fact that in our model IL-12 and IL-35 are maintained constant while varied IL-2 activates only STAT3 and not STAT4 in this pairing. The basal STAT4:STAT4 homodimer level leads to the background level of IFN-γ production and the lack of initial co-expression between produced IFN-γ and IL-10 (C).

Despite the fact that the switching is still present in both the STAT3-STAT5 and STAT3-STAT4 circuits, the model predictions significantly differ between these two circuits for low IL-2 concentrations. The STAT3-STAT5 circuit demonstrates the bell shaped characteristic with the low IFN-γ production for the low amounts of IL-2 (Fig 2A), whereas STAT3-STAT4 circuit reveals the significant background level of IFN-γ (Fig 4C). Fig 4A shows that for low IL-2 concentrations there is also a basal phosphorylation level of STAT4. Our model suggests that these background levels of STAT4p (Fig 4A) and subsequent IFN-γ production (Fig 4C) are due to the STAT4 activation by the maintained concentrations of IL-12 and IL-35 [30].

After establishing the individual properties of the STAT pathways, we next combined the two STAT3-STAT4 and STAT3-STAT5 circuits (Fig 5A). The detailed description of the model for STAT3-STAT4-STAT5 subsystem is shown in Supplementary Materials. The parameters are taken from the corresponding individual modules and are shown in Tables S1 and S2 in Supplementary Materials.

Fig 5B-Fig 5G illustrate the model predictions for the combined STAT3-STAT4-STAT5 circuit. Our model demonstrates that the IFN-γ to IL-10 switching is still present in the combined model (Fig 5D and Fig 5G). One of the major differences between the full circuit and the smaller submodules is that in the full circuit STAT3 competes with both STAT4 and STAT5 at the same time (Fig 5B and Fig 5E) while in the smaller submodules it competes only with either at a time.

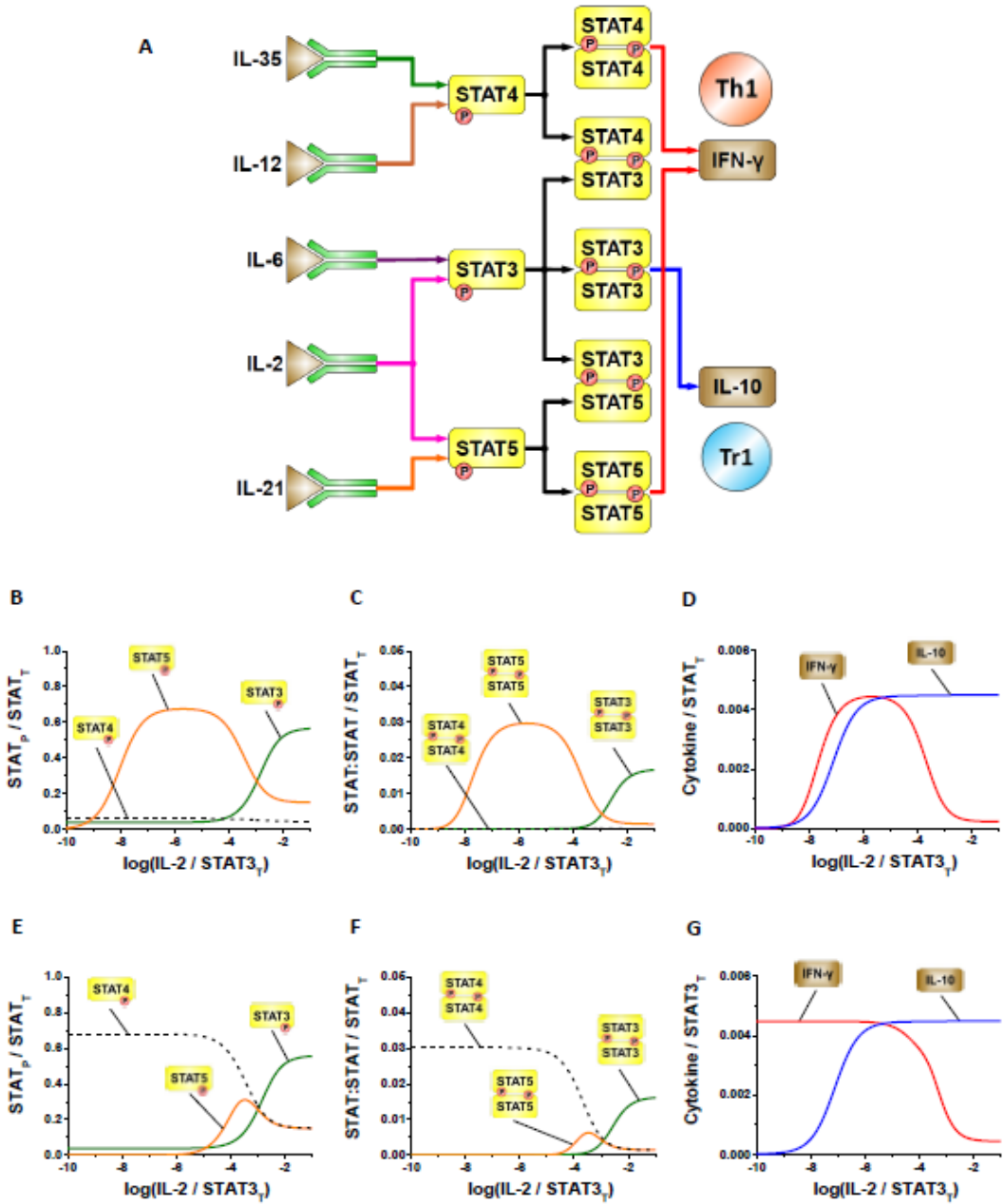

**Fig 5. The model predictions for the combined STAT3-STAT4-STAT5 circuit.** A. Schematic diagram for the combined STAT3-STAT4-STAT5 model. B-D. The dose-response profiles show that for significantly dephosphorylated STAT4, the STAT3-STAT5 like responses prevail and the selective production of IFN-γ arises. E-G. The model predictions for dephosphorylated STAT5 reveal STAT3-STAT4 like responses with no selectivity for IFN-γ production observed.

Previously we showed that the two subsystems, namely STAT3-STAT4 and STAT3-STAT5, have similar dose-response characteristics for medium and high IL-2, whereas the dose-response curves for low IL-2 are different (Fig 2 and Fig 4). In the range of low IL-2 concentrations our model demonstrates two possible scenarios for IFN-γ and IL-10 production depending on which of the two submodules, STAT3-STAT4 or STAT3-STAT5, prevails. Fig 5B-Fig 5D show that the combined model predictions qualitatively coincide with the predictions for the STAT3-STAT5 subsystem (Fig 2) when the amount of phosphorylated STAT4 is significantly reduced in comparison to phosphorylated STAT5. In our model, the reduction of STAT4p is the result of an increase in the total PTP phosphatase concentration. The case shown in Fig 5E-Fig 5G, where STAT5p is strongly dephosphorylated by increased SHP-2, suggests that the response of the combined model is similar to that of STAT3-STAT4 circuit considered previously (Fig 4). For this system we observe a significant (compared with phosphorylated STAT5) basal level of STAT4 phosphorylation (Fig 5E) as well as the background IFN-γ production (Fig 5G). Our model predicts that for the low IL-2 concentrations the phosphorylation level of STAT4 and STAT5 controls the balance between the competing STAT3-STAT4 and STAT3-STAT5 modules.

## Discussion

In this paper, we developed a new integrative modeling approach to study STAT signaling. The approach encompasses the network representation of the interactions between cytokines and JAK-STAT pathways. Based on this approach, we built a mathematical model for STAT signaling in T cells (Fig 1). The proposed model was employed to explain the T cell phenotype plasticity effects using the example of Th1 to Tr1 switching [24]. While it is widely accepted that JAK-STAT pathways regulate the phenotype switching, the underlying mechanisms and possible interdependence effects between the JAK-STAT pathways need further clarification. In order to propose potential underlying mechanisms, we considered the immune response from the systems perspective. The model developed in this paper to the best of our knowledge for the first time takes into account the interdependence effects between the JAK-STAT pathways and predicts the conditions, for which the phenotype switching occurs. The proposed model explains how the same cytokines can activate different STATs and induce the production of other cytokines with opposite immune function. The model predictions are consistent with the experimental data for IFN-γ and IL-10 production (Fig 2A), which demonstrate IL-2 dependent Th1 to Tr1 switching [64].

The molecular mechanisms underlying T cell plasticity have been a subject of extensive systems biology research in recent years. Our model extends the previously published mathematical models that analyze JAK-STAT signaling [51-54]. While the previous models studied cytokine production considering single JAK-STAT activation pathways only, our model introduces cross-regulation effects and STAT-related modulation of signals. The developed model establishes relationship between cytokines, STAT proteins and the overall immune response. Our model predicts that the Th1 to Tr1 phenotype switching is due to the competition between the STAT proteins. The parameter sensitivity analysis (Fig S3) demonstrates that this conclusion is valid with the probabilities of 78% within the 2-fold, 45% within the 3-fold and 28% within the 4-fold parametric changes.

The analysis of our model shows that depending on extracellular cytokine concentrations the competing STAT species can indirectly inhibit each other (Fig 2B). Another finding gives an insight into the role of STAT heterodimer complexes that has been remained unclear [30]. We suggest here that STAT homodimers rather than heterodimers induce the cytokines production, which is consistent with [65], and that the heterodimers can serve as a "buffer" between the various STAT homodimer complexes. The presence of heterodimers allows the redistribution of STAT homodimers and therefore causes the switching of produced cytokines and T cell phenotype. The suggested regulatory role of STAT heterodimers could be studied experimentally by proteomic identification of nuclear STAT heterodimers. This would allow the model predictions regarding the correlation between the STAT homodimers shown in Fig 2C and Fig 4B to be confirmed.

Our model suggests new explanations for JAK-STAT signaling regulation involved in the T cell phenotype switching. A number of external and internal factors can alter the model parameters and biomolecular interactions in JAK-STAT signaling pathways and as a result influence the levels of produced cytokines as well as the T cell phenotype. We investigated how these alterations (possibly caused by genetic mutations) can lead to various pathological states without changing the structure of the model (Fig 3). Our model predicts that inappropriate regulation of the IL-2 receptor system leads to the dysfunctions of the IFN-γ to IL-10 switching (Fig 3C), which may in turn mediate the autoimmune and IBD states [70].

Immunity-related pathologies caused by *Leishmania major* or Epstein-virus are associated with an inappropriate balance between pro-inflammatory and anti-inflammatory cytokines [76, 77]. The proposed model suggests strategies for the regulation of IFN-γ and IL-10 production during disease. We showed that the IFN-γ to IL-10 switching can be controlled biochemically by enhancing or reducing signaling through certain JAK-STAT pathways. Our model predicts that a reduction of uncontrolled inflammation could be achieved by reducing the role of the STAT5 pathway or by enhancement of IL-6-induced STAT3 phosphorylation, which up-regulates anti-inflammatory IL-10 production (Fig 3B and Fig 3G respectively). The predictions of our model might have clinical applications in drug discovery and could be tested experimentally.

The model proposed in this study shows that the alterations in other cytokine signaling, for example IL-6, may also lead to the lack of T cell phenotype switching, which is associated with immunity-related pathologies. Fig 3F- Fig 3G show that an increased IL-6 concentration leads to the lack of switching in phosphorylated STAT proteins, whereas the switching is stronger for reduced IL-6 concentrations. Due to the high dependence of cytokine production on the STAT balance, the lack of STAT switching leads to the lack of switching between IFN-γ and IL-10 production.

A number of pharmacological and clinical studies observed aberrant STATs activation in human tumor diseases. The inhibition of certain STATs and cytokine production might have clinical applications in immunity-related pathologies including cancer [22, 23, 78] and uncontrolled inflammation [79]. It was reported that STAT3 is a promising target for anti-cancer treatment [22, 23, 78]. The number of experimental studies investigating STAT3 inhibition has grown rapidly in recent years [80-83]. However, there is still a limited understanding of the underlying mechanisms of the inhibition. Our model proposes new interdependent strategies for STAT3 and IFN-γ inhibition, which include the activation of competing STAT pathways (Fig 3). The results of the proposed model suggest that the inhibitor selectivity to specific STAT proteins might enhance the anti-cancer effect.

Systems modeling can offer new insights into the interpretation of experimental data. The fact that our model considers multiple JAK-STAT pathways at a time provides an explanation for the conflicting experimental results [25, 47-49, 84]. For example it was shown in [25, 47, 48] that the activated STAT5 leads to the production of IFN-γ while in [49] it was demonstrated that IL-10 production is also enhanced through STAT5 activation. The proposed model can explain this duality in experimental data by introducing the selectivity in JAK-STAT pathways depending on IL-2 concentration (Fig 2). Thus the experimental data cannot be interpreted in the way that certain input cytokine can activate other cytokine production but the concentration of other input extracellular cytokines should also be considered.

In this study, we used a systems biology approach, according to which all sophisticated systems of biomolecular interactions can be divided into subsystems if it is physiologically meaningful. Using this method, we extracted the STAT3-STAT5 (Fig 1B) and STAT3-STAT4 (Fig 1C) subsystems from the scheme shown in Fig 1A, analyzed them separately and then combined them together in Fig 5A and studied the combined module. The analysis revealed that all three subsystems can reproduce the phenotype switching for certain amounts of IL-2, however the model predictions differ for low IL-2 concentrations. The model predictions for the low IL-2 concentrations showed selective activation of IFN-γ production for the STAT3-STAT5 subsystem (Fig 2A), whereas the STAT3-STAT4 subsystem demonstrated no selective activation due to the by basal, relatively significant level of IFN-γ production (Fig 4C). The combined model suggests that extracellular cytokines can switch the overall response of the system to selective or non-selective responses for low concentrations of IL-2 (Fig 5). Thus our model predicts that there is a competition not only between the STATs but also between the STAT subsystems.

T cells can be divided into the phenotypes that produce pro-inflammatory and anti-inflammatory cytokines. One of such examples was investigated in this paper in the context of Th1/Tr1 phenotype switching. We focused on pro-inflammatory IFN-γ and regulatory IL-10 production in the Th1/Tr1 phenotype switching via STAT proteins. However, it should be noted that the considered STATs shown in Fig 1A, can also induce the production of cytokines other than IFN-γ and IL-10. The proposed model can be applied to describe the plasticity effects and the switching not only between Th1 and Tr1, but also other T cell phenotypes. The basic principles of our model might be potentially applied to the Th1/Th2 [85], Treg/Th17 and Th17/Th2 [86] phenotype switching.

## Materials and Methods

In this section, a mathematical description of our model for the STAT3-STAT5 subsystem shown in Fig 1B is provided. We investigated the steady state solutions of the phosphorylated proteins (STATs) and produced cytokines. The full information about all reactions and equations governing STAT3-STAT5, STAT3-STAT4 and the combined STAT3-STAT4-STAT5 subsystems can be found in Supplementary Materials.

**Cytokine-receptor interactions**

Concentrations of phosphorylated Receptor-JAK complexes activated by IL-2, IL-6 and IL-21, respectively, as functions of the corresponding cytokine concentration can be written as follows:

$$[w2] = -\frac{M_2 - r2_t + p2_t \left( \frac{M_1}{n_1 [i2]} + \frac{1}{n_1} + 1 \right)}{2} + \frac{\sqrt{\left( M_2 - r2_t + p2_t \left( \frac{M_1}{n_1 [i2]} + \frac{1}{n_1} + 1 \right) \right)^2 + 4 r2_t M_2}}{2},$$

$$[w6] = -\frac{M_4 - r6_t + p6_t \left( \frac{M_3}{n_2 [i6]} + \frac{1}{n_2} + 1 \right)}{2} + \frac{\sqrt{\left( M_4 - r6_t + p6_t \left( \frac{M_3}{n_2 [i6]} + \frac{1}{n_2} + 1 \right) \right)^2 + 4 r6_t M_4}}{2},$$

$$[w21] = -\frac{M_6 - r21_t + p21_t \left( \frac{M_5}{n_3 [i21]} + \frac{1}{n_3} + 1 \right)}{2} + \frac{\sqrt{\left( M_6 - r21_t + p21_t \left( \frac{M_5}{n_3 [i21]} + \frac{1}{n_3} + 1 \right) \right)^2 + 4 r21_t M_6}}{2},$$

(1)

where $[i2] = \dfrac{[\text{IL-2}]}{STAT3_T}$, $[w2] = \dfrac{[\text{IL2Rp:JAK}]}{STAT3_T}$, $r2_t = \dfrac{IL2R_T}{STAT3_T}$, $p2_t = \dfrac{SHP\text{-}1_T}{STAT3_T}$, $[i6] = \dfrac{[\text{IL-6}]}{STAT3_T}$,

$[w6] = \dfrac{[\text{IL6Rp:JAK}]}{STAT3_T}$, $r6_t = \dfrac{IL6R_T}{STAT3_T}$, $p6_t = \dfrac{SHP\text{-}2_T}{STAT3_T}$, $[i21] = \dfrac{[\text{IL-21}]}{STAT3_T}$,

$[w21] = \dfrac{[\text{IL21Rp:JAK}]}{STAT3_T}$, $r21_t = \dfrac{IL21R_T}{STAT3_T}$, $p21_t = \dfrac{P21_T}{STAT3_T}$ (here subscripts $t$ and $T$ denote the total amount of protein in non-dimensional and dimensional forms, respectively, small $p$ stands for the phosphorylated state), $M_1$, $M_2$, $M_3$, $M_4$, $M_5$ and $M_6$ are the non-dimensional Michaelis constants, $n_1$, $n_2$ and $n_3$ denote the ratio of receptor phosphorylation/dephosphorylation rates. Equations (1) follow from the stationary conditions of the phosphorylation and dephosphorylation of the receptors mentioned previously. The derivation of Equations (1) can be found in Supplementary Materials, Equations (S2)-(S13).

**STAT phosphorylation and dimerization**

STAT proteins are phosphorylated by the activated interleukin-receptor-JAK complexes IL2Rp:JAK, IL6Rp:JAK and IL21Rp:JAK (Fig 1B). The biochemical reactions involved in STAT activation are given by:

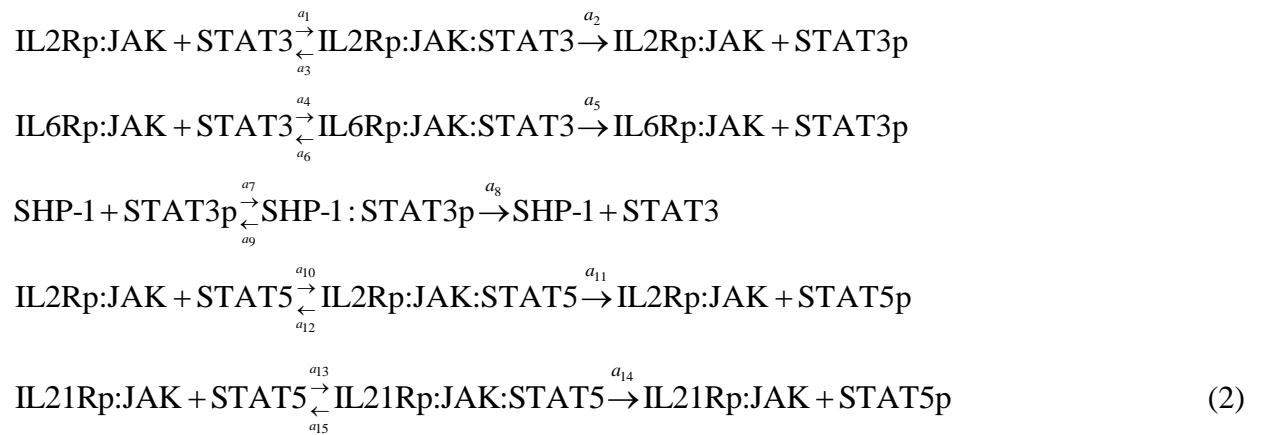

$$\text{IL2Rp:JAK} + \text{STAT3} \underset{a_3}{\overset{a_1}{\rightleftarrows}} \text{IL2Rp:JAK:STAT3} \overset{a_2}{\rightarrow} \text{IL2Rp:JAK} + \text{STAT3p}$$

$$\text{IL6Rp:JAK} + \text{STAT3} \underset{a_6}{\overset{a_4}{\rightleftarrows}} \text{IL6Rp:JAK:STAT3} \overset{a_5}{\rightarrow} \text{IL6Rp:JAK} + \text{STAT3p}$$

$$\text{SHP-1} + \text{STAT3p} \underset{a_9}{\overset{a_7}{\rightleftarrows}} \text{SHP-1:STAT3p} \overset{a_8}{\rightarrow} \text{SHP-1} + \text{STAT3}$$

$$\text{IL2Rp:JAK} + \text{STAT5} \underset{a_{12}}{\overset{a_{10}}{\rightleftarrows}} \text{IL2Rp:JAK:STAT5} \overset{a_{11}}{\rightarrow} \text{IL2Rp:JAK} + \text{STAT5p}$$

$$\text{IL21Rp:JAK} + \text{STAT5} \underset{a_{15}}{\overset{a_{13}}{\rightleftarrows}} \text{IL21Rp:JAK:STAT5} \overset{a_{14}}{\rightarrow} \text{IL21Rp:JAK} + \text{STAT5p} \qquad (2)$$

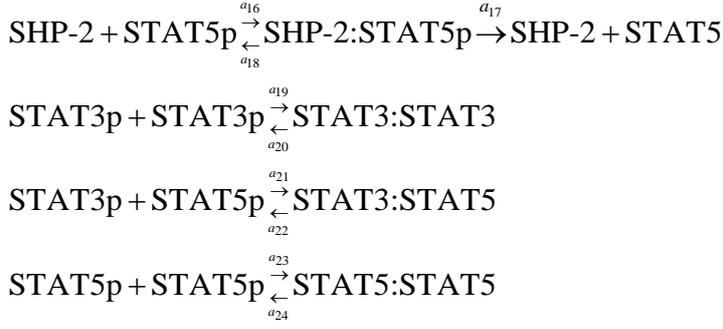

$$\text{SHP-2} + \text{STAT5p} \underset{a_{18}}{\overset{a_{16}}{\rightleftarrows}} \text{SHP-2:STAT5p} \overset{a_{17}}{\rightarrow} \text{SHP-2} + \text{STAT5}$$

$$\text{STAT3p} + \text{STAT3p} \underset{a_{20}}{\overset{a_{19}}{\rightleftarrows}} \text{STAT3:STAT3}$$

$$\text{STAT3p} + \text{STAT5p} \underset{a_{22}}{\overset{a_{21}}{\rightleftarrows}} \text{STAT3:STAT5}$$

$$\text{STAT5p} + \text{STAT5p} \underset{a_{24}}{\overset{a_{23}}{\rightleftarrows}} \text{STAT5:STAT5}$$

The corresponding Ordinary Differential Equations (ODEs) for Reactions (2) can be written as follows:

$$\frac{d}{dt}[IL2Rp{:}JAK{:}STAT3] = a_1[IL2Rp{:}JAK][STAT3] - (a_2 + a_3)[IL2Rp{:}JAK{:}STAT3],$$

$$\frac{d}{dt}[IL6Rp{:}JAK{:}STAT3] = a_4[IL6Rp{:}JAK][STAT3] - (a_5 + a_6)[IL6Rp{:}JAK{:}STAT3],$$

$$\frac{d}{dt}[STAT3p] = a_2[IL2Rp{:}JAK{:}STAT3] + a_5[IL6Rp{:}JAK{:}STAT3] -$$
$$- a_7[SHP\text{-}1][STAT3p] + a_9[SHP\text{-}1{:}STAT3p] - 2a_{19}[STAT3p]^2 +$$
$$+ 2a_{20}[STAT3{:}STAT3] - a_{21}[STAT3p][STAT5p] + a_{22}[STAT3{:}STAT5],$$

$$\frac{d}{dt}[SHP\text{-}1{:}STAT3p] = a_7[SHP\text{-}1][STAT3p] - (a_8 + a_9)[SHP\text{-}1{:}STAT3p],$$

$$\frac{d}{dt}[IL2Rp{:}JAK{:}STAT5] = a_{10}[IL2Rp{:}JAK][STAT5] - (a_{11} + a_{12})[IL2Rp{:}JAK{:}STAT5],$$

$$\frac{d}{dt}[IL21Rp{:}JAK{:}STAT5] = a_{13}[IL21Rp{:}JAK][STAT5] - (a_{14} + a_{15})[IL21Rp{:}JAK{:}STAT5], \qquad (3)$$

$$\frac{d}{dt}[STAT5p] = a_{11}[IL2Rp{:}JAK{:}STAT5] + a_{14}[IL21Rp{:}JAK{:}STAT5] -$$
$$- a_{16}[SHP\text{-}2][STAT5p] + a_{18}[SHP\text{-}2{:}STAT5p] - a_{21}[STAT3p][STAT5p] +$$
$$+ a_{22}[STAT3{:}STAT5] - 2a_{23}[STAT5p]^2 + 2a_{24}[STAT5{:}STAT5],$$

$$\frac{d}{dt}[SHP\text{-}2{:}STAT5p] = a_{16}[SHP\text{-}2][STAT5p] - (a_{17} + a_{18})[SHP\text{-}2{:}STAT5p],$$

$$\frac{d}{dt}[STAT3{:}STAT3] = a_{19}[STAT3p]^2 - a_{20}[STAT3{:}STAT3],$$

$$\frac{d}{dt}[STAT3{:}STAT5] = a_{21}[STAT3p][STAT5p] - a_{22}[STAT3{:}STAT5],$$

$$\frac{d}{dt}[STAT5{:}STAT5] = a_{23}[STAT5p]^2 - a_{24}[STAT5{:}STAT5].$$

Conservation equations of the total proteins concentrations are given by:

$$\begin{aligned}
STAT3_T &= [STAT3]+[STAT3p]+2[STAT3{:}STAT3]+[STAT3{:}STAT5]+\\
&\quad +[IL2Rp{:}JAK{:}STAT3]+[IL6Rp{:}JAK{:}STAT3]+[SHP\text{-}1{:}STAT3p],\\
STAT5_T &= [STAT5]+[STAT5p]+2[STAT5{:}STAT5]+[STAT3{:}STAT5]+\\
&\quad +[IL2Rp{:}JAK{:}STAT5]+[IL21Rp{:}JAK{:}STAT5]+[SHP\text{-}2{:}STAT5p],\\
SHP\text{-}1_T &= [SHP\text{-}1]+[SHP\text{-}1{:}STAT3p],\\
SHP\text{-}2_T &= [SHP\text{-}2]+[SHP\text{-}2{:}STAT5p].
\end{aligned} \quad (4)$$

Conservation Equations (4) can be written in a non-dimensional form:

$$\begin{aligned}
1 &= [s3]+[s3p]+2[s33]+[s35]+[w2s3]+[w6s3]+[p3s3p],\\
s5_t &= [s5]+[s5p]+2[s55]+[s35]+[w2s5]+[w21s5]+[p5s5p],\\
p3_t &= [p3]+[p3s3p],\\
p5_t &= [p5]+[p5s5p],\\
pp_t &= [pp]+[pps3p],
\end{aligned} \quad (5)$$

where we introduced the non-dimensional concentrations of the proteins normalized by

$$STAT3_T: [s3]=\frac{[STAT3]}{STAT3_T},\ [s3p]=\frac{[STAT3p]}{STAT3_T},\ [w2s3]=\frac{[IL2Rp{:}JAK{:}STAT3]}{STAT3_T},$$

$$[w6s3]=\frac{[IL6Rp{:}JAK{:}STAT3]}{STAT3_T},\ [p3s3p]=\frac{[SHP\text{-}1{:}STAT3p]}{STAT3_T},\ s5_t=\frac{STAT5_T}{STAT3_T},\ [s5]=\frac{[STAT5]}{STAT3_T},$$

$$[s5p]=\frac{[STAT5p]}{STAT3_T},\ [w2s5]=\frac{[IL2Rp{:}JAK{:}STAT5]}{STAT3_T},\ [w21s5]=\frac{[IL21Rp{:}JAK{:}STAT5]}{STAT3_T},$$

$$[p5s5p]=\frac{[SHP\text{-}2{:}STAT5p]}{STAT3_T},\ [s33]=\frac{[STAT3{:}STAT3]}{STAT3_T},\ [s35]=\frac{[STAT3{:}STAT5]}{STAT3_T},$$

$$[s55]=\frac{[STAT5{:}STAT5]}{STAT3_T},\ [p3]=\frac{[SHP\text{-}1]}{STAT3_T},\ [p5]=\frac{[SHP\text{-}2]}{STAT3_T},\ p3_t=\frac{SHP\text{-}1_T}{STAT3_T},\ p5_t=\frac{SHP\text{-}2_T}{STAT3_T}.$$

In order to find the steady-state solutions of Equations (3) we need to solve the following system of algebraic equations:

$$\begin{cases} 0=[s3p]+2\dfrac{[s3p]^2}{M_{13}}+\dfrac{[s3p][s5p]}{M_{14}}+\dfrac{p3_t[s3p]}{M_9+[s3p]}\left(1+\dfrac{M_7+[w2]+M_7 Q_6}{n_4[w2]+n_5 M_7 Q_6}\right)-1,\\[2mm] 0=[s5p]+2\dfrac{[s5p]^2}{M_{15}}+\dfrac{[s3p][s5p]}{M_{14}}+\dfrac{p5_t[s5p]}{M_{12}+[s5p]}\left(1+\dfrac{M_{10}+[w2]+M_{10}Q_{21}}{n_6[w2]+n_7 M_{10}Q_{21}}\right)-s5_t, \end{cases} \quad (6)$$

where $M_7 = \dfrac{a_2 + a_3}{a_1 STAT3_T}$, $\quad M_8 = \dfrac{a_5 + a_6}{a_4 STAT3_T}$, $\quad M_9 = \dfrac{a_8 + a_9}{a_7 STAT3_T}$, $\quad M_{10} = \dfrac{a_{11} + a_{12}}{a_{10} STAT3_T}$,

$M_{11} = \dfrac{a_{14} + a_{15}}{a_{13} STAT3_T}$, $\quad M_{12} = \dfrac{a_{17} + a_{18}}{a_{16} STAT3_T}$, $\quad M_{13} = \dfrac{a_{20}}{a_{19} STAT3_T}$, $\quad M_{14} = \dfrac{a_{22}}{a_{21} STAT3_T}$,

$M_{15} = \dfrac{a_{24}}{a_{23} STAT3_T}$ are the non-dimensional Michaelis constants, $Q_6 = \dfrac{[w6]}{M_8}$ and $Q_{21} = \dfrac{[w21]}{M_{11}}$.

System (6) has been solved numerically to obtain steady-state concentrations of STAT proteins as a function of IL-2.

**SP1 activation**

Experimental data on the molecular mechanism of how CD46 enhances IL-10 production are not available at present. However it was established that CD46 can facilitate the secretion of IL-10 through the SPAK-ERK pathway and SP1 transcription factor only in the presence of high concentrations of IL-2 [24]. This dependence is described by hypothetical enzymatic reactions shown in supplementary Equations (S33). Thus, it can be written for the concentration of the active SP1 in non-dimensional form:

$$[sp1a] = sp1_t \dfrac{[i2]}{M_{16} + [i2]} \dfrac{[cd46]}{M_{17} + [cd46]}, \qquad (7)$$

where $[sp1a] = \dfrac{[SP1a]}{STAT3_T}$, $sp1_t = \dfrac{SP1_T}{STAT3_T}$, $[cd46] = \dfrac{[CD46]}{STAT3_T}$, $M_{16}$ and $M_{17}$ are the Michaelis constants.

**Cytokine production**

The production of IFN-γ and IL-10 is induced by STAT dimer interactions with the genes responsible for production of IFN-γ and IL-10 [30]. The produced cytokine can be degraded by a metalloprotease Mp [34]. The concentration of the produced IFN-γ in non-dimensional form can be written as follows:

$$[ig] = \frac{M_{18}}{\frac{mp1_t}{n_8 gg_t \frac{[s55]}{M_{19}+[s55]}} - 1}, \qquad (8)$$

where $[ig] = \frac{[IFN\text{-}\gamma]}{STAT3_T}$, $mp1_t = \frac{Mp1_T}{STAT3_T}$, $gg_t = \frac{Gene_T^{IFN\text{-}\gamma}}{STAT3_T}$, $M_{18}$ and $M_{19}$ are the Michaelis constants, $n_8$ is the ratio of IFN-γ production to degradation rates. In order to achieve the steady-state, IFN-γ production rate should be less than its maximal degradation rate, which implies $n_8 < 1$.

Both STAT3:STAT3 homodimer and CD46 (through SPAK-ERK pathway) can activate the same IL-10 gene but they bind different binding regions, as shown in [87]. IL-10 is produced after the binding of either of the transcription factors to the gene, which corresponds to [88]. Thus it can be written for IL-10 concentration:

$$[i10] = \frac{M_{20}}{\frac{mp2_t}{n_9 g10_t \left(\frac{[s33]}{M_{21}+[s33]} + \frac{[sp1a]}{M_{22}+[sp1a]} - \frac{[s33]}{M_{21}+[s33]}\frac{[sp1a]}{M_{22}+[sp1a]}\right)} - 1}, \qquad (9)$$

where $[i10] = \frac{[IL\text{-}10]}{STAT3_T}$, $mp2_t = \frac{Mp2_T}{STAT3_T}$, $g10_t = \frac{Gene_T^{IL\text{-}10}}{STAT3_T}$, $n_9 = \frac{l_6}{k_{11}}$, $M_{20}$, $M_{21}$ and $M_{22}$ are the Michaelis constants, $n_9$ is the ratio of IL-10 production to degradation rates, $n_9 < 1$. Equations (8) and (9) are used to describe the concentration of produced IFN-γ and IL-10 as a function of IL-2.

Equations (7), (8) and (9) are derived in Supplementary Materials (Equations (S42), (S54) and (S55), respectively).

# Acknowledgements

This work was carried out under Severnside Alliance for Translational Research (SARTRE) grant (GIW) and NSFC grant 61374053 (MZQC). This work was co-funded by the subsidy allocated to the Kazan Federal University for state assignment in the sphere of scientific activities and performed according to the 'Russian Government Program of Competitive Growth' of the Kazan Federal University.

# Supplementary Materials

Here we provide mathematical details of the model for STAT-STAT interactions described in the main text. We derive the equations employed in the STAT3-STAT5 (Fig 1B), STAT3-STAT4 (Fig 1C) and combined STAT3-STAT4-STAT5 (Fig 5A) circuits. Table S1 shows the short names and abbreviations used in our model.

**Table S1. Abbreviations used in the STAT phosphorylation model.**

| Abbreviation | Meaning |
| --- | --- |
| I2 | IL-2 |
| RJ2 | IL-2 Receptor complex with JAK |
| I2RJ2 | IL-2 Receptor:JAK complex with bound IL-2 |
| RpJ2 | Phosphorylated IL-2 Receptor:JAK complex |
| P2 | SHP-1 phosphatase |
| RpJ2P2 | Phosphorylated IL-2 Receptor:JAK complex with SHP-1 |
| I6 | IL-6 |
| RJ6 | IL-6 Receptor:JAK complex |
| I6RJ6 | IL-6 Receptor:JAK complex with bound IL-6 |
| RpJ6 | Phosphorylated IL-6 Receptor:JAK complex |
| P6 | SHP-2 phosphatase |
| RpJ6P6 | Phosphorylated IL-6 Receptor:JAK complex with SHP-2 |
| I12 | IL-12 |
| RJ12 | IL-12 Receptor:JAK complex |
| I12RJ12 | IL-12 Receptor:JAK complex with bound IL-12 |
| RpJ12 | Phosphorylated IL-12 Receptor:JAK complex |
| P12 | JAK phosphatase |
| RpJ12P12 | Phosphorylated IL-12 Receptor:JAK complex with P12 phosphatase |
| I35 | IL-35 |
| RJ35 | IL-35 Receptor:JAK complex |
| I35RJ35 | IL-35 Receptor:JAK complex with bound IL-35 |
| RpJ35 | Phosphorylated IL-35 Receptor:JAK complex |
| P35 | JAK phosphatase |
| RpJ35P35 | Phosphorylated IL-35 Receptor:JAK complex with P35 phosphatase |

| I21 | IL-21 |
| --- | --- |
| RJ21 | IL-21 Receptor:JAK complex |
| I21RJ21 | IL-21 Receptor:JAK complex with bound IL-21 |
| RpJ21 | Phosphorylated IL-21 Receptor:JAK complex |
| P21 | JAK phosphatase |
| RpJ21P21 | Phosphorylated IL-21 Receptor:JAK complex with P21 phosphatase |
| S3 | STAT3 |
| RpJ2S3 | Phosphorylated IL-2 Receptor:JAK complex with STAT3 |
| RpJ6S3 | Phosphorylated IL-6 Receptor:JAK complex with STAT3 |
| S3p | Phosphorylated STAT3 |
| P3 | SHP-1 phosphatase |
| P3S3p | Phosphorylated STAT3 complex with SHP-1 |
| S4 | STAT4 |
| S4p | Phosphorylated STAT4 |
| RpJ12S4 | Phosphorylated IL-12 Receptor:JAK complex with STAT4 |
| RpJ35S4 | Phosphorylated IL-35 Receptor:JAK complex with STAT4 |
| P4 | PTP phosphatase |
| P4S4p | Phosphorylated STAT4 complex with PTP phosphatase |
| S5 | STAT5 |
| S5p | Phosphorylated STAT5 |
| RpJ2S5 | Phosphorylated IL-2 Receptor:JAK complex with STAT5 |
| RpJ21S5 | Phosphorylated IL-21 Receptor:JAK complex with STAT5 |
| P5 | SHP-2 phosphatase |
| P5S5p | Phosphorylated STAT5 complex with SHP-2 phosphatase |
| S33 | STAT3:STAT3 homodimer |
| S34 | STAT3:STAT4 heterodimer |
| S44 | STAT4:STAT4 homodimer |
| S35 | STAT3:STAT5 heterodimer |
| S55 | STAT5:STAT5 homodimer |
| Gg | Gene responsible for IFN-γ production |
| S44Gg | IFN-γ gene complex with STAT4:STAT4 homodimer |
| S55Gg | IFN-γ gene complex with STAT5:STAT5 homodimer |
| S44S55Gg | IFN-γ gene complex with STAT4:STAT4 and STAT5:STAT5 |
| Ig | IFN-γ |
| Mp1 | Metalloprotease that cleaves IFN-γ |
| IgMp1 | Metalloprotease complex with IFN-γ gene |
| Ign | Non-active IFN-γ |
| G10 | IL-10 gene |
| S33G10 | IL-10 gene complex with STAT3:STAT3 homodimer |
| Sp1 | SP1 transcription factor |
| Sp1a | SP1 transcription factor in active form |
| S33Sp1aG10 | IL-10 gene complex with STAT3:STAT3 and Sp1a |
| Sp1aG10 | Complex of Sp1a with IL-10 gene |
| Mp2 | Metalloprotease that cleaves IL-10 |
| I10Mp2 | Metalloprotease complex with IL-10 gene |
| I10n | Non-active IL-10 |

# 1 Model for the STAT3-STAT5 circuit

The biochemical reactions involved in the STAT3-STAT5 circuit (Fig 1B) are given by:

$$I2 + RJ2 \underset{f_3}{\overset{f_1}{\rightleftarrows}} I2RJ2 \overset{f_2}{\rightarrow} RpJ2$$

$$RpJ2 + P2 \underset{f_6}{\overset{f_4}{\rightleftarrows}} RpJ2P2 \overset{f_5}{\rightarrow} RJ2 + P2$$

$$I6 + RJ6 \underset{f_9}{\overset{f_7}{\rightleftarrows}} I6RJ6 \overset{f_8}{\rightarrow} RpJ6$$

$$RpJ6 + P6 \underset{f_{12}}{\overset{f_{10}}{\rightleftarrows}} RpJ6P6 \overset{f_{11}}{\rightarrow} RJ6 + P6$$

$$I21 + RJ21 \underset{f_{15}}{\overset{f_{13}}{\rightleftarrows}} I21RJ21 \overset{f_{14}}{\rightarrow} RpJ21$$

$$RpJ21 + P21 \underset{f_{18}}{\overset{f_{16}}{\rightleftarrows}} RpJ21P21 \overset{f_{17}}{\rightarrow} RJ21 + P21$$

$$RpJ2 + S3 \underset{a_3}{\overset{a_1}{\rightleftarrows}} RpJ2S3 \overset{a_2}{\rightarrow} RpJ2 + S3p$$

$$RpJ6 + S3 \underset{a_6}{\overset{a_4}{\rightleftarrows}} RpJ6S3 \overset{a_5}{\rightarrow} RpJ6 + S3p$$

$$P3 + S3p \underset{a_9}{\overset{a_7}{\rightleftarrows}} P3S3p \overset{a_8}{\rightarrow} P3 + S3$$

$$RpJ2 + S5 \underset{a_{12}}{\overset{a_{10}}{\rightleftarrows}} RpJ2S5 \overset{a_{11}}{\rightarrow} RpJ2 + S5p$$

$$RpJ21 + S5 \underset{a_{15}}{\overset{a_{13}}{\rightleftarrows}} RpJ21S5 \overset{a_{14}}{\rightarrow} RpJ21 + S5p$$

$$P5 + S5p \underset{a_{18}}{\overset{a_{16}}{\rightleftarrows}} P5S5p \overset{a_{17}}{\rightarrow} P5 + S5$$

$$S3p + S3p \underset{a_{20}}{\overset{a_{19}}{\rightleftarrows}} S33$$

$$S3p + S5p \underset{a_{22}}{\overset{a_{21}}{\rightleftarrows}} S35$$

$$S5p + S5p \underset{a_{24}}{\overset{a_{23}}{\rightleftarrows}} S55$$

$$I2 + Sp1 \underset{l_2}{\overset{l_1}{\rightleftarrows}} I2Sp1$$

(S1)

$$I2Sp1 + CD46 \underset{l_4}{\overset{l_3}{\rightleftarrows}} Sp1a$$

$$CD46 + Sp1 \underset{l_4}{\overset{l_3}{\rightleftarrows}} CD46Sp1$$

$$CD46Sp1 + I2 \underset{l_2}{\overset{l_1}{\rightleftarrows}} Sp1a$$

$$S55 + Gg \underset{k_2}{\overset{k_1}{\rightleftarrows}} S55Gg$$

$$S55Gg \overset{l_5}{\rightarrow} Ig$$

$$Ig + Mp1 \underset{k_5}{\overset{k_3}{\rightleftarrows}} IgMp1 \overset{k_4}{\rightarrow} Ign + Mp1$$

$$S33 + G10 \underset{k_7}{\overset{k_6}{\rightleftarrows}} S33G10$$

$$S33G10 + Sp1a \underset{k_9}{\overset{k_8}{\rightleftarrows}} S33Sp1aG10$$

$$Sp1a + G10 \underset{k_9}{\overset{k_8}{\rightleftarrows}} Sp1aG10$$

$$Sp1aG10 + S33 \underset{k_7}{\overset{k_6}{\rightleftarrows}} S33Sp1aG10$$

$$S33G10 \overset{l_6}{\rightarrow} I10$$

$$Sp1aG10 \overset{l_6}{\rightarrow} I10$$

$$S33Sp1aG10 \overset{l_6}{\rightarrow} I10$$

$$I10 + Mp2 \underset{k_{12}}{\overset{k_{10}}{\rightleftarrows}} I10Mp2 \overset{k_{11}}{\rightarrow} I10n + Mp2$$

The system of reactions (S1) can be divided into three major subsystems of interactions: i) Cytokine-receptor interactions, ii) STAT phosphorylation and dimerization, iii) Cytokine production.

## 1.1 Cytokine-receptor interactions

In the most general case the reactions can be written as follows:

$$C + RJ \underset{q_3}{\overset{q_1}{\rightleftarrows}} CRJ \overset{q_2}{\rightarrow} RpJ,$$

$$RpJ + P \underset{q_6}{\overset{q_4}{\rightleftarrows}} RpJP \overset{q_5}{\rightarrow} RJ + P,$$

(S2)

where C is cytokine, RJ is Receptor:JAK complex, P is phosphatase and small p denotes phosphorylated state.

The ODEs for the system (S2):

$$\frac{d}{dt}[CRJ] = q_1[C][RJ] - (q_2 + q_3)[CRJ],$$

$$\frac{d}{dt}[RpJ] = q_2[CRJ] - q_4[RpJ][P] + q_6[RpJP], \quad (S3)$$

$$\frac{d}{dt}[RpJP] = q_4[RpJ][P] - (q_5 + q_6)[RpJP].$$

Corresponding conservation equations:

$$R_T = [RJ] + [CRJ] + [RpJ] + [RpJP],$$
$$P_T = [P] + [RpJP], \quad (S4)$$

where $R_T$ and $P_T$ are the total amounts of receptors and phosphatase, respectively. Here we neglect STAT-receptor interactions since STAT proteins do not have a significant impact on receptor dephosphorylation.

Equations (S4) can be written as follows:

$$R_T = \alpha + \beta + \gamma + \omega,$$
$$P_T = p + \omega, \quad (S5)$$

where $\alpha = [RJ], \beta = [CRJ], \gamma = [RpJ], \omega = [RpJP], p = [P], c = [C]$.

The ODEs (S3) can be rewritten in the following way:

$$\frac{d}{dt}\beta = q_1 c\alpha - (q_2 + q_3)\beta,$$
$$\frac{d}{dt}\gamma = q_2\beta - q_4\gamma p + q_6\omega, \quad (S6)$$
$$\frac{d}{dt}\omega = q_4\gamma p - (q_5 + q_6)\omega.$$

We need to find steady-state solutions of Equations (S6):

$$0 = q_1 c\alpha - (q_2 + q_3)\beta,$$
$$0 = q_2\beta - q_4\gamma p + q_6\omega,$$
$$0 = q_4\gamma p - (q_5 + q_6)\omega, \quad (S7)$$
$$R_T = \alpha + \beta + \gamma + \omega,$$
$$P_T = p + \omega.$$

We found the concentrations of the complexes:

$$\omega = P_T \frac{\gamma}{\frac{q_5+q_6}{q_4}+\gamma} = \frac{P_T\gamma}{Q_2+\gamma},$$

$$\beta = \frac{q_5}{q_2}\omega = \frac{q_5}{q_2}\frac{P_T\gamma}{Q_2+\gamma} = \frac{P_T\gamma}{Q_3(Q_2+\gamma)}, \qquad (S8)$$

$$\alpha = \frac{q_2+q_3}{q_1}\frac{\beta}{c} = \frac{Q_1 P_T \gamma}{cQ_3(Q_2+\gamma)},$$

where $Q_1 = \frac{q_2+q_3}{q_1}$ and $Q_2 = \frac{q_5+q_6}{q_4}$ are the Michaelis constants for phosphorylation and dephosphorylation, respectively, and $Q_3 = \frac{q_2}{q_5}$.

We can write the following equation using the conservation Equation for the receptor (S5):

$$R_T = \gamma + \frac{P_T\gamma}{Q_2+\gamma}\left(\frac{Q_1}{cQ_3}+\frac{1}{Q_3}+1\right). \qquad (S9)$$

As a result, we obtain a quadratic equation:

$$0 = \gamma^2 + \gamma\chi - \delta, \qquad (S10)$$

where $\chi = Q_2 - R_T + P_T\left(\frac{Q_1}{cQ_3}+\frac{1}{Q_3}+1\right)$, $\delta = R_T Q_2$.

The solution of Equation (S10) is:

$$\gamma = -\frac{\chi}{2} + \frac{\sqrt{\chi^2+4\delta}}{2} \qquad (S11)$$

Equation (S11) can be rewritten as follows:

$$\gamma = \frac{1}{2}\left(\sqrt{\left(\lambda+\frac{\rho}{c}\right)^2+4\delta}-\lambda-\frac{\rho}{c}\right), \qquad (S12)$$

where $\lambda = Q_2 - R_T + P_T\left(\frac{1}{Q_3}+1\right)$ and $\rho = \frac{P_T Q_1}{Q_3}$.

Using Equation (S12) we can now write for $[RpJ2], [RpJ6]$ and $[RpJ21]$ in non-dimensional form respectively:

$$[w2] = -\frac{M_2 - r2_t + p2_t\left(\dfrac{M_1}{n_1[i2]} + \dfrac{1}{n_1} + 1\right)}{2} +$$

$$+ \frac{\sqrt{\left(M_2 - r2_t + p2_t\left(\dfrac{M_1}{n_1[i2]} + \dfrac{1}{n_1} + 1\right)\right)^2 + 4r2_t M_2}}{2},$$

$$[w6] = -\frac{M_4 - r6_t + p6_t\left(\dfrac{M_3}{n_2[i6]} + \dfrac{1}{n_2} + 1\right)}{2} +$$

$$+ \frac{\sqrt{\left(M_4 - r6_t + p6_t\left(\dfrac{M_3}{n_2[i6]} + \dfrac{1}{n_2} + 1\right)\right)^2 + 4r6_t M_4}}{2},$$

$$[w21] = -\frac{M_6 - r21_t + p21_t\left(\dfrac{M_5}{n_3[i21]} + \dfrac{1}{n_3} + 1\right)}{2} +$$

$$+ \frac{\sqrt{\left(M_6 - r21_t + p21_t\left(\dfrac{M_5}{n_3[i21]} + \dfrac{1}{n_3} + 1\right)\right)^2 + 4r21_t M_6}}{2},$$

(S13)

where

$$[i2] = \frac{[I2]}{S3_T}, [w2] = \frac{[RpJ2]}{S3_T}, r2_t = \frac{R2_T}{S3_T}, p2_t = \frac{P2_T}{S3_T}, M_1 = \frac{f_2 + f_3}{f_1 S3_T}, M_2 = \frac{f_5 + f_6}{f_4 S3_T}, n_1 = \frac{f_2}{f_5}, [i6] = \frac{[I6]}{S3_T},$$

$$[w6] = \frac{[RpJ6]}{S3_T}, r6_t = \frac{R6_T}{S3_T}, p6_t = \frac{P6_T}{S3_T}, M_3 = \frac{f_8 + f_9}{f_7 S3_T}, M_4 = \frac{f_{11} + f_{12}}{f_{10} S3_T}, n_2 = \frac{f_8}{f_{11}}, [i21] = \frac{[I21]}{S3_T},$$

$$[w21] = \frac{[RpJ21]}{S3_T}, r21_t = \frac{R21_T}{S3_T}, p21_t = \frac{P21_T}{S3_T}, M_5 = \frac{f_{14} + f_{15}}{f_{13} S3_T}, M_6 = \frac{f_{17} + f_{18}}{f_{16} S3_T}, n_3 = \frac{f_{14}}{f_{17}}.$$

## 1.2 STAT phosphorylation and dimerization

The ODEs describing biochemical reactions in the STAT subsystem:

$$\frac{d}{dt}[RpJ2S3] = a_1[RpJ2][S3] - (a_2 + a_3)[RpJ2S3],$$

$$\frac{d}{dt}[RpJ6S3] = a_4[RpJ6][S3] - (a_5 + a_6)[RpJ6S3],$$

$$\frac{d}{dt}[S3p] = a_2[RpJ2S3] + a_5[RpJ6S3] - a_7[P3][S3p] + a_9[P3S3p] -$$

$$-2a_{19}[S3p]^2 + 2a_{20}[S33] - a_{21}[S3p][S5p] + a_{22}[S35],$$

$$\frac{d}{dt}[P3S3p] = a_7[P3][S3p] - (a_8 + a_9)[P3S3p],$$

$$\frac{d}{dt}[RpJ2S5] = a_{10}[RpJ2][S5] - (a_{11} + a_{12})[RpJ2S5],$$

$$\frac{d}{dt}[RpJ21S5] = a_{13}[RpJ21][S5] - (a_{14} + a_{15})[RpJ21S5],$$

$$\frac{d}{dt}[S5p] = a_{11}[RpJ2S5] + a_{14}[RpJ21S5] - a_{16}[P5][S5p] + a_{18}[P5S5p] -$$

$$-a_{21}[S3p][S5p] + a_{22}[S35] - 2a_{23}[S5p]^2 + 2a_{24}[S55],$$

$$\frac{d}{dt}[P5S5p] = a_{16}[P5][S5p] - (a_{17} + a_{18})[P5S5p],$$

$$\frac{d}{dt}[S33] = a_{19}[S3p]^2 - a_{20}[S33], \quad \text{(S14)}$$

$$\frac{d}{dt}[S35] = a_{21}[S3p][S5p] - a_{22}[S35],$$

$$\frac{d}{dt}[S55] = a_{23}[S5p]^2 - a_{24}[S55].$$

Conservation equations are given by:

$$\begin{aligned} S3_T &= [S3] + [S3p] + 2[S33] + [S35] + [RpJ2S3] + [RpJ6S3] + [P3S3p], \\ S5_T &= [S5] + [S5p] + 2[S55] + [S35] + [RpJ2S5] + [RpJ21S5] + [P5S5p], \\ P3_T &= [P3] + [P3S3p], \\ P5_T &= [P5] + [P5S5p]. \end{aligned} \quad \text{(S15)}$$

Then we can normalize Equations (S15) to $S3_T$:

$$\begin{aligned} 1 &= [s3] + [s3p] + 2[s33] + [s35] + [w2s3] + [w6s3] + [p3s3p], \\ s5_t &= [s5] + [s5p] + 2[s55] + [s35] + [w2s5] + [w21s5] + [p5s5p], \\ p3_t &= [p3] + [p3s3p], \\ p5_t &= [p5] + [p5s5p], \end{aligned} \quad \text{(S16)}$$

where

$$[s3]=\frac{[S3]}{S3_T},[s3p]=\frac{[S3p]}{S3_T},[w2s3]=\frac{[RpJ2S3]}{S3_T},[w6s3]=\frac{[RpJ6S3]}{S3_T},[p3s3p]=\frac{[P3S3p]}{S3_T},s5_t=\frac{S5_T}{S3_T},$$

$$[s5]=\frac{[S5]}{S3_T},[s5p]=\frac{[S5p]}{S3_T},[w2s5]=\frac{[RpJ2S5]}{S3_T},[w21s5]=\frac{[RpJ21S5]}{S3_T},[p5s5p]=\frac{[P5S5p]}{S3_T},$$

$$[s33]=\frac{[S33]}{S3_T},[s35]=\frac{[S35]}{S3_T},[s55]=\frac{[S55]}{S3_T},[p3]=\frac{[P3]}{S3_T},[p5]=\frac{[P5]}{S3_T},\ p3_t=\frac{P3_T}{S3_T},p5_t=\frac{P5_T}{S3_T}.$$

The ODEs (S14) can be written in non-dimensional form as follows:

$$\frac{d}{d\tau}[w2s3]=m_1[w2][s3]-[w2s3],$$

$$\frac{d}{d\tau}[w6s3]=m_3[w6][s3]-(m_4+m_5)[w6s3],$$

$$\frac{d}{d\tau}[s3p]=m_2[w2s3]+m_4[w6s3]-m_6[p3][s3p]+m_8[p3s3p]-$$

$$-2m_{18}[s3p]^2+2m_{19}[s33]-m_{20}[s3p][s5p]+m_{21}[s35],$$

$$\frac{d}{d\tau}[p3s3p]=m_6[p3][s3p]-(m_7+m_8)[p3s3p],$$

$$\frac{d}{d\tau}[w2s5]=m_9[w2][s5]-(m_{10}+m_{11})[w2s5],$$

$$\frac{d}{d\tau}[w21s5]=m_{12}[w21][s5]-(m_{13}+m_{14})[w21s5],$$

$$\frac{d}{d\tau}[s5p]=m_{10}[w2s5]+m_{13}[w21s5]-m_{15}[p5][s5p]+m_{17}[p5s5p]-$$

$$-m_{20}[s3p][s5p]+m_{21}[s35]-2m_{22}[s5p]^2+2m_{23}[s55],$$

$$\frac{d}{d\tau}[p5s5p]=m_{15}[p5][s5p]-(m_{16}+m_{17})[p5s5p],$$

$$\frac{d}{d\tau}[s33]=m_{18}[s3p]^2-m_{19}[s33], \quad (S17)$$

$$\frac{d}{d\tau}[s35]=m_{20}[s3p][s5p]-m_{21}[s35],$$

$$\frac{d}{d\tau}[s55]=m_{22}[s5p]^2-m_{23}[s55],$$

where

$$\tau=t(a_2+a_3), m_1=\frac{a_1}{a_2+a_3}S3_T, m_2=\frac{a_2}{a_2+a_3}, m_3=\frac{a_4}{a_2+a_3}S3_T, m_4=\frac{a_5}{a_2+a_3}, m_5=\frac{a_6}{a_2+a_3},$$

$$m_6=\frac{a_7}{a_2+a_3}S3_T, m_7=\frac{a_8}{a_2+a_3}, m_8=\frac{a_9}{a_2+a_3},\ m_9=\frac{a_{10}}{a_2+a_3}S3_T, m_{10}=\frac{a_{11}}{a_2+a_3}, m_{11}=\frac{a_{12}}{a_2+a_3},$$

$$m_{12}=\frac{a_{13}}{a_2+a_3}S3_T, m_{13}=\frac{a_{14}}{a_2+a_3}, m_{14}=\frac{a_{15}}{a_2+a_3},\ m_{15}=\frac{a_{16}}{a_2+a_3}S3_T, m_{16}=\frac{a_{17}}{a_2+a_3}, m_{17}=\frac{a_{18}}{a_2+a_3},$$

$$m_{18}=\frac{a_{19}}{a_2+a_3}S3_T, m_{19}=\frac{a_{20}}{a_2+a_3}, m_{20}=\frac{a_{21}}{a_2+a_3}S3_T, m_{21}=\frac{a_{22}}{a_2+a_3}, m_{22}=\frac{a_{23}}{a_2+a_3}S3_T, m_{23}=\frac{a_{24}}{a_2+a_3}.$$

We need to find steady-state solutions of Equations (S17):

$$0 = m_1[w2][s3] - [w2s3],$$
$$0 = m_3[w6][s3] - (m_4 + m_5)[w6s3],$$
$$\frac{d}{d\tau}[s3p] = m_2[w2s3] + m_4[w6s3] - m_6[p3][s3p] + m_8[p3s3p] -$$
$$- 2m_{18}[s3p]^2 + 2m_{19}[s33] - m_{20}[s3p][s5p] + m_{21}[s35],$$
$$0 = m_6[p3][s3p] - (m_7 + m_8)[p3s3p],$$
$$0 = m_9[w2][s5] - (m_{10} + m_{11})[w2s5],$$
$$0 = m_{12}[w21][s5] - (m_{13} + m_{14})[w21s5],$$
$$\frac{d}{d\tau}[s5p] = m_{10}[w2s5] + m_{13}[w21s5] - m_{15}[p5][s5p] + m_{17}[p5s5p] -$$
$$- m_{20}[s3p][s5p] + m_{21}[s35] - 2m_{22}[s5p]^2 + 2m_{23}[s55],$$
$$0 = m_{15}[p5][s5p] - (m_{16} + m_{17})[p5s5p],$$
$$0 = m_{18}[s3p]^2 - m_{19}[s33],$$
$$0 = m_{20}[s3p][s5p] - m_{21}[s35],$$
$$0 = m_{22}[s5p]^2 - m_{23}[s55].$$
(S18)

Equations (S18) can be simplified as follows:

$$0 = m_1[w2][s3] - [w2s3],$$
$$0 = m_3[w6][s3] - (m_4 + m_5)[w6s3],$$
$$\frac{d}{d\tau}[s3p] = m_2[w2s3] + m_4[w6s3] - m_7[p3s3p],$$
$$0 = m_6[p3][s3p] - (m_7 + m_8)[p3s3p],$$
$$0 = m_9[w2][s5] - (m_{10} + m_{11})[w2s5],$$
$$0 = m_{12}[w21][s5] - (m_{13} + m_{14})[w21s5],$$
$$\frac{d}{d\tau}[s5p] = m_{10}[w2s5] + m_{13}[w21s5] - m_{16}[p5s5p],$$
$$0 = m_{15}[p5][s5p] - (m_{16} + m_{17})[p5s5p],$$
$$0 = m_{18}[s3p]^2 - m_{19}[s33],$$
$$0 = m_{20}[s3p][s5p] - m_{21}[s35],$$
$$0 = m_{22}[s5p]^2 - m_{23}[s55].$$
(S19)

Then using conservation Equations (S16) and System (S19) we obtain:

$$\frac{d}{d\tau}[s3p] = m_2[w2s3] + m_4[w6s3] - m_7[p3s3p],$$
$$\frac{d}{d\tau}[s5p] = m_{10}[w2s5] + m_{13}[w21s5] - m_{16}[p5s5p],$$
(S20)

where

$$[p3s3p] = p3_t \frac{[s3p]}{\frac{m_7 + m_8}{m_6} + [s3p]},$$

$$[p5s5p] = p5_t \frac{[s5p]}{\frac{m_{16} + m_{17}}{m_{15}} + [s5p]},$$

$$[s33] = \frac{m_{18}}{m_{19}}[s3p]^2,$$

$$[s35] = \frac{m_{20}}{m_{21}}[s3p][s5p],$$

$$[s55] = \frac{m_{22}}{m_{23}}[s5p]^2,$$

$$[w2s3] = m_1[w2][s3],$$

$$[w6s3] = \frac{m_3}{m_4 + m_5}[w6][s3],$$

$$[w2s5] = \frac{m_9}{m_{10} + m_{11}}[w2][s5],$$

$$[w21s5] = \frac{m_{12}}{m_{13} + m_{14}}[w21][s5],$$

$$[s3] = \frac{1 - [s3p] - 2[s33] - [s35] - [p3s3p]}{1 + m_1[w2] + \frac{m_3}{m_4 + m_5}[w6]},$$

$$[s5] = \frac{s5_t - [s5p] - 2[s55] - [s35] - [p5s5p]}{1 + \frac{m_9}{m_{10} + m_{11}}[w2] + \frac{m_{12}}{m_{13} + m_{14}}[w21]}.$$

(S21)

We can rewrite Equations (S21):

$$[p3s3p] = p3_t \frac{[s3p]}{M_9 + [s3p]},$$

$$[p5s5p] = p5_t \frac{[s5p]}{M_{12} + [s5p]},$$

$$[s33] = \frac{[s3p]^2}{M_{13}},$$

$$[s35] = \frac{[s3p][s5p]}{M_{14}},$$

$$[s55] = \frac{[s5p]^2}{M_{15}},$$

$$[w2s3] = \frac{[w2][s3]}{M_7},$$

$$[w6s3] = \frac{[w6][s3]}{M_8},$$

$$[w2s5] = \frac{[w2][s5]}{M_{10}},$$

$$[w21s5] = \frac{[w21][s5]}{M_{11}},$$

$$[s3] = \frac{1 - [s3p] - 2[s33] - [s35] - [p3s3p]}{1 + \frac{[w2]}{M_7} + \frac{[w6]}{M_8}},$$

$$[s5] = \frac{s5_t - [s5p] - 2[s55] - [s35] - [p5s5p]}{1 + \frac{[w2]}{M_{10}} + \frac{[w21]}{M_{11}}},$$

(S22)

where we denote the Michaelis constants

$$M_7 = \frac{a_2 + a_3}{a_1 S3_T}, M_8 = \frac{a_5 + a_6}{a_4 S3_T}, M_9 = \frac{a_8 + a_9}{a_7 S3_T}, M_{10} = \frac{a_{11} + a_{12}}{a_{10} S3_T}, M_{11} = \frac{a_{14} + a_{15}}{a_{13} S3_T}, M_{12} = \frac{a_{17} + a_{18}}{a_{16} S3_T},$$

$$M_{13} = \frac{a_{20}}{a_{19} S3_T}, M_{14} = \frac{a_{22}}{a_{21} S3_T}, M_{15} = \frac{a_{24}}{a_{23} S3_T}.$$

We look for steady-state solutions of System (S20):

$$0 = m_2[w2s3] + m_4[w6s3] - m_7[p3s3p],$$
$$0 = m_{10}[w2s5] + m_{13}[w21s5] - m_{16}[p5s5p].$$

(S23)

We can rewrite Equations (S23) as follows:

$$0 = n_4[w2s3] + n_5[w6s3] - [p3s3p],$$
$$0 = n_6[w2s5] + n_7[w21s5] - [p5s5p],$$

(S24)

where $n_4 = \dfrac{a_2}{a_8}, n_5 = \dfrac{a_5}{a_8}, n_6 = \dfrac{a_{11}}{a_{17}}, n_7 = \dfrac{a_{14}}{a_{17}}$.

System (S24) can be rewritten after substituting solutions from Equation (S22):

$$
\begin{aligned}
0 &= [s3p] + 2\dfrac{[s3p]^2}{M_{13}} + \dfrac{[s3p][s5p]}{M_{14}} + \\
&\quad + p3_t \dfrac{[s3p]}{M_9 + [s3p]}\left(1 + \dfrac{M_7 M_8 + M_8[w2] + M_7[w6]}{n_4 M_8[w2] + n_5 M_7[w6]}\right) - 1, \\
0 &= [s5p] + 2\dfrac{[s5p]^2}{M_{15}} + \dfrac{[s3p][s5p]}{M_{14}} + \\
&\quad + p5_t \dfrac{[s5p]}{M_{12} + [s5p]}\left(1 + \dfrac{M_{10}M_{11} + M_{11}[w2] + M_{10}[w21]}{n_6 M_{11}[w2] + n_7 M_{10}[w21]}\right) - s5_t,
\end{aligned}
\tag{S25}
$$

We solved System (S25) for $[s3p]$ and $[s5p]$ numerically.

## 1.3 Cytokine production

In general case, transcription factor T can activate gene G by forming a complex with the gene TG:

$$\text{T} + \text{G} \underset{h_2}{\overset{h_1}{\rightleftarrows}} \text{TG} \tag{S26}$$

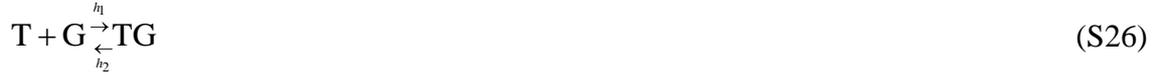

The ODEs for Equation (S26):

$$
\begin{aligned}
\dfrac{d}{dt}[G] &= -h_1[T][G] + h_2[TG], \\
\dfrac{d}{dt}[TG] &= h_1[T][G] - h_2[TG].
\end{aligned}
\tag{S27}
$$

Conservation equation that follows from Equations (S27):

$$G_T = [G] + [TG], \tag{S28}$$

where $G_T$ is the total concentration of the gene.

Equation (S28) can be written as follows:

$$G_T = \lambda + \theta, \tag{S29}$$

where $\lambda = [G], \theta = [TG]$.

The ODEs (S27) can be rewritten in the following way:

$$\frac{d}{dt}\lambda = -h_1 T \lambda + h_2 \theta,$$
$$\frac{d}{dt}\theta = h_1 T \lambda - h_2 \theta,$$
(S30)

where $T = [T]$.

We need to find steady-state solutions of Equation (S30):

$$0 = h_1 T \lambda - h_2 \theta,$$
$$G_T = \lambda + \theta.$$
(S31)

We can find $\theta$ from Equations (S31):

$$\theta = G_T \frac{T}{Qh + T},$$
(S32)

where $Qh = \dfrac{h_2}{h_1}$ is the Michaelis constant.

In the most general case the reactions of the activation of a gene G by two transcription factors T1 and T2 are:

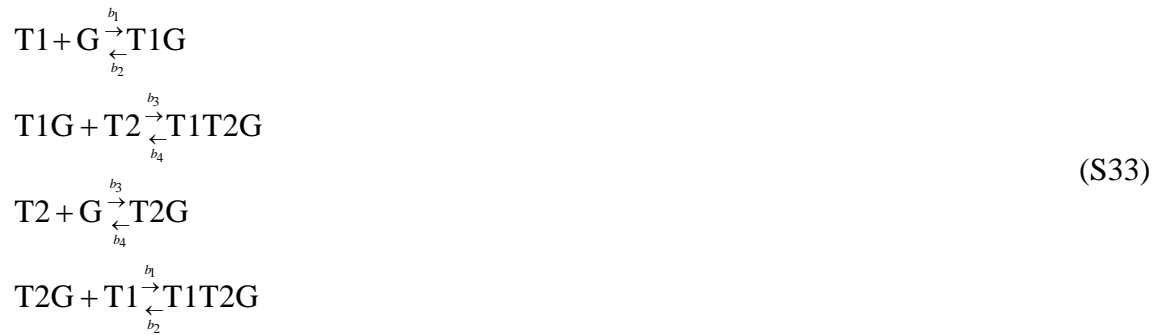

$$\begin{aligned}
&\text{T1} + \text{G} \underset{b_2}{\overset{b_1}{\rightleftarrows}} \text{T1G} \\
&\text{T1G} + \text{T2} \underset{b_4}{\overset{b_3}{\rightleftarrows}} \text{T1T2G} \\
&\text{T2} + \text{G} \underset{b_4}{\overset{b_3}{\rightleftarrows}} \text{T2G} \\
&\text{T2G} + \text{T1} \underset{b_2}{\overset{b_1}{\rightleftarrows}} \text{T1T2G}
\end{aligned}$$
(S33)

The ODEs for System (S33):

$$\frac{d}{dt}[G] = -b_1[T1][G] + b_2[T1G] - b_3[T2][G] + b_4[T2G],$$

$$\frac{d}{dt}[T1G] = b_1[T1][G] - b_2[T1G] - b_3[T1G][T2] + b_4[T1T2G],$$

$$\frac{d}{dt}[T2G] = b_3[T2][G] - b_4[T2G] - b_1[T2G][T1] + b_2[T1T2G],$$

$$\frac{d}{dt}[T1T2G] = b_3[T1G][T2] - b_4[T1T2G] + b_1[T2G][T1] - b_2[T1T2G].$$

(S34)

Conservation equation that follows from Equations (S34):

$$G_T = [G] + [T1G] + [T2G] + [T1T2G],$$

(S35)

where $G_T$ is the total amount of the gene.

Equation (S35) can be written as follows:

$$G_T = \psi + \xi + \varphi + \nu,$$

(S36)

where $\psi = [G], \xi = [T1G], \varphi = [T2G], \nu = [T1T2G]$.

The ODEs (S34) can be rewritten in the following way:

$$\frac{d}{dt}\psi = -b_1 T1 \psi + b_2 \xi - b_3 T2 \psi + b_4 \varphi,$$

$$\frac{d}{dt}\xi = b_1 T1 \psi - b_2 \xi - b_3 \xi T2 + b_4 \nu,$$

$$\frac{d}{dt}\varphi = b_3 T2 \psi - b_4 \varphi - b_1 \varphi T1 + b_2 \nu,$$

$$\frac{d}{dt}\nu = b_3 \xi T2 - b_4 \nu + b_1 \varphi T1 - b_2 \nu,$$

(S37)

where $T1 = [T1], T2 = [T2]$.

We find steady-state solutions of Equations (S37):

$$0 = b_1 T1 \psi - b_2 \xi - b_3 \xi T2 + b_4 \nu,$$
$$0 = b_3 T2 \psi - b_4 \varphi - b_1 \varphi T1 + b_2 \nu,$$
$$0 = b_3 \xi T2 - b_4 \nu + b_1 \varphi T1 - b_2 \nu,$$
$$G_T = \psi + \xi + \varphi + \nu.$$

(S38)

Next, we find concentrations of the complexes:

$$\xi = G_T \frac{\frac{b_4}{b_3}T1}{\left(\frac{b_2}{b_1}+T1\right)\left(\frac{b_4}{b_3}+T2\right)} = G_T \frac{Qb_2 T1}{(Qb_1+T1)(Qb_2+T2)},$$

$$\varphi = G_T \frac{\frac{b_2}{b_1}T2}{\left(\frac{b_2}{b_1}+T1\right)\left(\frac{b_4}{b_3}+T2\right)} = G_T \frac{Qb_1 T2}{(Qb_1+T1)(Qb_2+T2)},\quad\quad\text{(S39)}$$

$$\nu = G_T \frac{T1T2}{\left(\frac{b_2}{b_1}+T1\right)\left(\frac{b_4}{b_3}+T2\right)} = G_T \frac{T1T2}{(Qb_1+T1)(Qb_2+T2)},$$

where $Qb_1 = \frac{b_2}{b_1}, Qb_2 = \frac{b_4}{b_3}$ are the Michaelis constants.

If a protein is activated by the first and the second transcription factors at the same time, then its concentration is proportional to the concentration $\nu$ only:

$$\nu = G_T \frac{T1}{Qb_1+T1} \cdot \frac{T2}{Qb_2+T2},\quad\quad\text{(S40)}$$

which is a probability of the two transcription factors to be bound to the same gene.

If a protein is activated by the first or the second transcription factors, then it is proportional to the sum of concentrations $\xi + \varphi + \nu$:

$$\xi+\varphi+\nu = G_T \frac{Qb_2 T1 + Qb_1 T2 + T1T2}{(Qb_1+T1)(Qb_2+T2)},$$

$$\xi+\varphi+\nu = G_T \frac{Qb_2 T1 + Qb_1 T2 + T1T2 + T1T2 - T1T2}{(Qb_1+T1)(Qb_2+T2)},$$

$$\xi+\varphi+\nu = G_T \frac{T1(Qb_2+T2) + T2(Qb_1+T1) - T1T2}{(Qb_1+T1)(Qb_2+T2)},\quad\quad\text{(S41)}$$

$$\xi+\varphi+\nu = G_T \left(\frac{T1}{Qb_1+T1} + \frac{T2}{Qb_2+T2} - \frac{T1T2}{(Qb_1+T1)(Qb_2+T2)}\right),$$

$$\xi+\varphi+\nu = G_T \left(\frac{T1}{Qb_1+T1} + \frac{T2}{Qb_2+T2} - \frac{T1}{Qb_1+T1}\frac{T2}{Qb_2+T2}\right),$$

which is a probability of either of the two transcription factors to be bound to the same gene.

## CD46

The full mechanism of how CD46 enhances IL-10 production is still not clear. We assume here that the mechanism of reactions is similar to the one described in Equation (S33). Thus it can be written for the concentration of SP1 in non-dimensional form according to Equation (S40):

$$[sp1a] = sp1_t \frac{[i2]}{M_{16} + [i2]} \cdot \frac{[cd46]}{M_{17} + [cd46]}, \tag{S42}$$

where $[sp1a] = \frac{[SP1a]}{S3_T}$, $sp1_t = \frac{SP1_T}{S3_T}$, $[cd46] = \frac{[CD46]}{S3_T}$, $M_{16} = \frac{l_2}{l_1 S3_T}$, and $M_{17} = \frac{l_4}{l_3 S3_T}$.

We assume here that the gene interaction with the transcription factor and its subsequent expression lead to the mRNA translation and certain cytokine secretion. The produced cytokine then can be degraded by a metalloprotease. In this case the biochemical reactions can be written as follows:

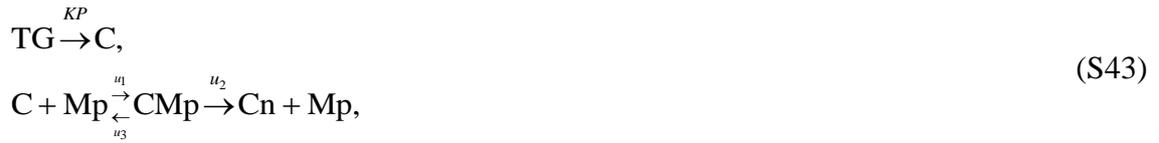

$$\begin{aligned}&TG \xrightarrow{KP} C, \\ &C + Mp \underset{u_3}{\overset{u_1}{\rightleftarrows}} CMp \xrightarrow{u_2} Cn + Mp,\end{aligned} \tag{S43}$$

where TG is the transcription factor complex with gene, C is the active cytokine, Mp is the metalloprotease, CMp is cytokine-metalloprotease complex and Cn is a non-active cytokine.

The ODEs for the reactions in System (S43):

$$\begin{aligned}\frac{d}{dt}[C] &= KP[TG] - u_1[C][Mp] + u_3[CMp], \\ \frac{d}{dt}[CMp] &= u_1[C][Mp] - (u_2 + u_3)[CMp], \\ \frac{d}{dt}[Mp] &= -u_1[C][Mp] + (u_2 + u_3)[CMp].\end{aligned} \tag{S44}$$

Conservation equation that follows from Equations (S44):

$$Mp_T = [Mp] + [CMp] \tag{S45}$$

We find steady-state solutions of System (S44):

$$\begin{aligned}0 &= KP[TG] - u_1[C][Mp] + u_3[CMp], \\ 0 &= u_1[C][Mp] - (u_2 + u_3)[CMp].\end{aligned} \tag{S46}$$

Thus we obtain a system of equations:

$$0 = KP\vartheta - u_2\sigma,$$
$$0 = u_1 C\zeta - (u_2 + u_3)\sigma, \qquad (S47)$$
$$Mp_T = \zeta + \sigma,$$

where $\vartheta = [TG], C = [C], \zeta = [Mp], \sigma = [CMp]$.

We can find $\sigma$ from this system of Equations (S47):

$$\sigma = Mp_T \frac{C}{Qu + C}, \qquad (S48)$$

where $Qu = \dfrac{u_2 + u_3}{u_1}$ is the Michaelis constant.

We next substitute $\sigma$ from equation (S48) to the first equation in System (S47):

$$0 = KP\vartheta - u_2 Mp_T \frac{C}{Qu + C},$$
$$C = \frac{KP\vartheta Qu}{u_2 Mp_T - KP\vartheta}, \qquad (S49)$$
$$C = \frac{Qu}{\dfrac{u_2 Mp_T}{KP\vartheta} - 1}.$$

Since $C$ should be positive and the maximum value for $\dfrac{Mp_T}{\vartheta}$ is 1, $KP < u_2$, which implies that the rate of the cytokine production should be less than its maximum rate of the degradation by metalloprotease.

Equation (S49) can be written as follows:

$$C = \frac{Qu}{\dfrac{Mp_T}{Qp\vartheta} - 1}, \qquad (S50)$$

where $Qp = \dfrac{KP}{u_2}$, $Qp < 1$.

When the cytokine production is up-regulated by one transcription factor only, $\vartheta = \theta$ from Equation (S32) and thus it can be written:

$$C = \frac{Qu}{\dfrac{Mp_T}{QpG_T \dfrac{T}{Qh+T}} - 1}. \tag{S51}$$

If the cytokine production is up-regulated by two transcription factors at the same time, $\vartheta = \nu$ as shown in Equation (S40), it can be written:

$$C = \frac{Qu}{\dfrac{Mp_T}{QpG_T \dfrac{T1}{Qb_1+T1} \dfrac{T2}{Qb_2+T2}} - 1}. \tag{S52}$$

If the cytokine production is up-regulated by either of the two transcription factors, $\vartheta = \xi + \varphi + \nu$ as shown in Equation (S41), and thus it can be written:

$$C = \frac{Qu}{\dfrac{Mp_T}{QpG_T \left(\dfrac{T1}{Qb_1+T1} + \dfrac{T2}{Qb_2+T2} - \dfrac{T1}{Qb_1+T1}\dfrac{T2}{Qb_2+T2}\right)} - 1}. \tag{S53}$$

**IFN-γ and IL-10 production**

Since IFN-γ is activated by STAT55 only (Fig 1B) we can write using Equation (S51):

$$[ig] = \frac{M_{18}}{\dfrac{mp1_t}{n_8 gg_t \dfrac{[s55]}{M_{19}+[s55]}} - 1}, \tag{S54}$$

where $[ig] = \dfrac{[Ig]}{S3_T}, mp1_t = \dfrac{Mp1_T}{S3_T}, gg_t = \dfrac{Gg_T}{S3_T}, M_{18} = \dfrac{k_4+k_5}{k_3 S3_T}, M_{19} = \dfrac{k_2}{k_1 S3_T}, n_8 = \dfrac{l_5}{k_4}, n_8 < 1$.

IL-10 gene can be activated by either STAT33 or CD46. Thus it can be written according to Equation (S53):

$$[i10] = \cfrac{M_{20}}{n_9 g10_t \left( \cfrac{[s33]}{M_{21}+[s33]} + \cfrac{[sp1a]}{M_{22}+[sp1a]} - \cfrac{[s33]}{M_{21}+[s33]}\cfrac{[sp1a]}{M_{22}+[sp1a]} \right) - 1}, \qquad (S55)$$

where $mp2_t = \dfrac{Mp2_t}{S3_T}, g10_t = \dfrac{G10_t}{S3_T}, M_{20} = \dfrac{k_{11}+k_{12}}{k_{10}S3_T}, M_{21} = \dfrac{k_7}{k_6 S3_T}, M_{22} = \dfrac{k_9}{k_8 S3_T}, n_9 = \dfrac{l_6}{k_{11}}, n_9 < 1$.

## 2 Model for the STAT3-STAT4 circuit

The biochemical reactions involved in the STAT3-STAT4 circuit (Fig 1C) are as follows:

$$I2 + RJ2 \underset{f_3}{\overset{f_1}{\rightleftarrows}} I2RJ2 \overset{f_2}{\rightarrow} RpJ2$$

$$RpJ2 + P2 \underset{f_6}{\overset{f_4}{\rightleftarrows}} RpJ2P2 \overset{f_5}{\rightarrow} RJ2 + P2$$

$$I6 + RJ6 \underset{f_9}{\overset{f_7}{\rightleftarrows}} I6RJ6 \overset{f_8}{\rightarrow} RpJ6$$

$$RpJ6 + P6 \underset{f_{12}}{\overset{f_{10}}{\rightleftarrows}} RpJ6P6 \overset{f_{11}}{\rightarrow} RJ6 + P6$$

$$I12 + RJ12 \underset{f_{15}}{\overset{f_{13}}{\rightleftarrows}} I12RJ12 \overset{f_{14}}{\rightarrow} RpJ12$$

$$RpJ12 + P12 \underset{f_{18}}{\overset{f_{16}}{\rightleftarrows}} RpJ12P12 \overset{f_{17}}{\rightarrow} RJ12 + P12$$

$$I35 + RJ35 \underset{f_{21}}{\overset{f_{19}}{\rightleftarrows}} I35RJ35 \overset{f_{20}}{\rightarrow} RpJ35$$

$$\text{RpJ35} + \text{P35} \underset{f_{24}}{\overset{f_{22}}{\rightleftarrows}} \text{RpJ35P35} \overset{f_{23}}{\rightarrow} \text{RJ35} + \text{P35}$$

$$\text{RpJ2} + \text{S3} \underset{a_3}{\overset{a_1}{\rightleftarrows}} \text{RpJ2S3} \overset{a_2}{\rightarrow} \text{RpJ2} + \text{S3p}$$

$$\text{RpJ6} + \text{S3} \underset{a_6}{\overset{a_4}{\rightleftarrows}} \text{RpJ6S3} \overset{a_5}{\rightarrow} \text{RpJ6} + \text{S3p}$$

$$\text{P3} + \text{S3p} \underset{a_9}{\overset{a_7}{\rightleftarrows}} \text{P3S3p} \overset{a_8}{\rightarrow} \text{P3} + \text{S3}$$

$$\text{RpJ12} + \text{S4} \underset{a_{12}}{\overset{a_{10}}{\rightleftarrows}} \text{RpJ12S4} \overset{a_{11}}{\rightarrow} \text{RpJ12} + \text{S4p}$$

$$\text{RpJ35} + \text{S4} \underset{a_{15}}{\overset{a_{13}}{\rightleftarrows}} \text{RpJ35S4} \overset{a_{14}}{\rightarrow} \text{RpJ35} + \text{S4p}$$

$$\text{P4} + \text{S4p} \underset{a_{18}}{\overset{a_{16}}{\rightleftarrows}} \text{P4S4p} \overset{a_{17}}{\rightarrow} \text{P4} + \text{S4}$$

$$\text{S3p} + \text{S3p} \underset{a_{20}}{\overset{a_{19}}{\rightleftarrows}} \text{S33}$$

$$\text{S3p} + \text{S4p} \underset{a_{22}}{\overset{a_{21}}{\rightleftarrows}} \text{S34}$$

$$\text{S4p} + \text{S4p} \underset{a_{24}}{\overset{a_{23}}{\rightleftarrows}} \text{S44}$$

$$\text{I2} + \text{Sp1} \underset{l_2}{\overset{l_1}{\rightleftarrows}} \text{I2Sp1}$$

$$\text{I2Sp1} + \text{CD46} \underset{l_4}{\overset{l_3}{\rightleftarrows}} \text{Sp1a}$$

$$\text{CD46} + \text{Sp1} \underset{l_4}{\overset{l_3}{\rightleftarrows}} \text{CD46Sp1}$$

$$\text{CD46Sp1} + \text{I2} \underset{l_2}{\overset{l_1}{\rightleftarrows}} \text{Sp1a}$$

$$\text{S44} + \text{Gg} \underset{k_2}{\overset{k_1}{\rightleftarrows}} \text{S44Gg}$$

$$\text{S44Gg} \overset{l_5}{\rightarrow} \text{Ig}$$

$$\text{Ig} + \text{Mp1} \underset{k_5}{\overset{k_3}{\rightleftarrows}} \text{IgMp1} \overset{k_4}{\rightarrow} \text{Ign} + \text{Mp1}$$

$$\text{S33} + \text{G10} \underset{k_7}{\overset{k_6}{\rightleftarrows}} \text{S33G10}$$

(S56)

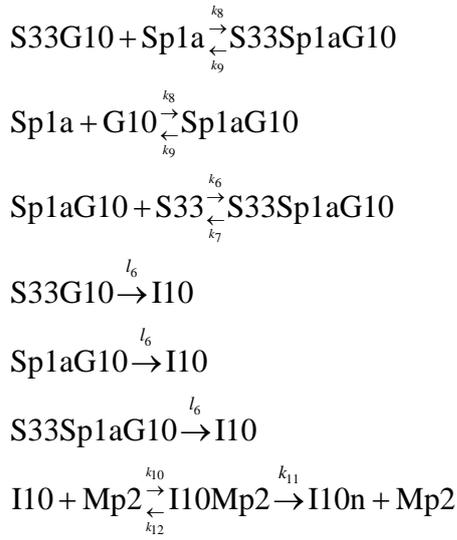

$$S33G10 + Sp1a \underset{k_9}{\overset{k_8}{\rightleftarrows}} S33Sp1aG10$$

$$Sp1a + G10 \underset{k_9}{\overset{k_8}{\rightleftarrows}} Sp1aG10$$

$$Sp1aG10 + S33 \underset{k_7}{\overset{k_6}{\rightleftarrows}} S33Sp1aG10$$

$$S33G10 \overset{l_6}{\rightarrow} I10$$

$$Sp1aG10 \overset{l_6}{\rightarrow} I10$$

$$S33Sp1aG10 \overset{l_6}{\rightarrow} I10$$

$$I10 + Mp2 \underset{k_{12}}{\overset{k_{10}}{\rightleftarrows}} I10Mp2 \overset{k_{11}}{\rightarrow} I10n + Mp2$$

## 2.1 Cytokine-receptor interactions

According to Equations (1.12) we can write for $[RpJ2]$, $[RpJ6]$, $[RpJ12]$ and $[RpJ35]$ in non-dimensional form respectively:

$$[w2] = -\frac{M_2 - r2_t + p2_t\left(\frac{M_1}{n_1[i2]} + \frac{1}{n_1} + 1\right)}{2} + \frac{\sqrt{\left(M_2 - r2_t + p2_t\left(\frac{M_1}{n_1[i2]} + \frac{1}{n_1} + 1\right)\right)^2 + 4r2_t M_2}}{2},$$

$$[w6] = -\frac{M_4 - r6_t + p6_t\left(\frac{M_3}{n_2[i6]} + \frac{1}{n_2} + 1\right)}{2} + \frac{\sqrt{\left(M_4 - r6_t + p6_t\left(\frac{M_3}{n_2[i6]} + \frac{1}{n_2} + 1\right)\right)^2 + 4r6_t M_4}}{2},$$

$$[w12] = -\frac{M_6 - r12_t + p12_t\left(\dfrac{M_5}{n_3[i12]} + \dfrac{1}{n_3} + 1\right)}{2} +$$

$$+\frac{\sqrt{\left(M_6 - r12_t + p12_t\left(\dfrac{M_5}{n_3[i12]} + \dfrac{1}{n_3} + 1\right)\right)^2 + 4r12_t M_6}}{2},$$

$$[w35] = -\frac{M_8 - r35_t + p35_t\left(\dfrac{M_7}{n_4[i35]} + \dfrac{1}{n_4} + 1\right)}{2} +$$

$$+\frac{\sqrt{\left(M_8 - r35_t + p35_t\left(\dfrac{M_7}{n_4[i35]} + \dfrac{1}{n_4} + 1\right)\right)^2 + 4r35_t M_8}}{2},$$

(S57)

where

$$[i2] = \frac{[I2]}{S3_T}, [w2] = \frac{[RpJ2]}{S3_T}, r2_t = \frac{R2_T}{S3_T}, p2_t = \frac{P2_T}{S3_T}, M_1 = \frac{f_2 + f_3}{f_1 S3_T}, M_2 = \frac{f_5 + f_6}{f_4 S3_T}, n_1 = \frac{f_2}{f_5}, [i6] = \frac{[I6]}{S3_T},$$

$$[w6] = \frac{[RpJ6]}{S3_T}, r6_t = \frac{R6_T}{S3_T}, p6_t = \frac{P6_T}{S3_T}, M_3 = \frac{f_8 + f_9}{f_7 S3_T}, M_4 = \frac{f_{11} + f_{12}}{f_{10} S3_T}, n_2 = \frac{f_8}{f_{11}}, [i12] = \frac{[I12]}{S3_T},$$

$$[w12] = \frac{[RpJ12]}{S3_T}, r12_t = \frac{R12_T}{S3_T}, p12_t = \frac{P12_T}{S3_T}, M_5 = \frac{f_{14} + f_{15}}{f_{13} S3_T}, M_6 = \frac{f_{17} + f_{18}}{f_{16} S3_T}, n_3 = \frac{f_{14}}{f_{17}},$$

$$[i35] = \frac{[I35]}{S3_T}, [w35] = \frac{[RpJ35]}{S3_T}, r35_t = \frac{R35_T}{S3_T}, p35_t = \frac{P35_T}{S3_T}, M_7 = \frac{f_{20} + f_{21}}{f_{19} S3_T}, M_8 = \frac{f_{23} + f_{24}}{f_{22} S3_T}, n_4 = \frac{f_{20}}{f_{23}}.$$

## 2.2 STAT phosphorylation and subsequent dimerization

ODEs for the STAT phosphorylation and dimerization module are given by:

$$\frac{d}{dt}[RpJ2S3] = a_1[RpJ2][S3] - (a_2 + a_3)[RpJ2S3],$$

$$\frac{d}{dt}[RpJ6S3] = a_4[RpJ6][S3] - (a_5 + a_6)[RpJ6S3],$$

$$\frac{d}{dt}[S3p] = a_2[RpJ2S3] + a_5[RpJ6S3] - a_7[P3][S3p] + a_9[P3S3p] -$$

$$-2a_{19}[S3p]^2 + 2a_{20}[S33] - a_{21}[S3p][S5p] + a_{22}[S35],$$

$$\frac{d}{dt}[P3S3p] = a_7[P3][S3p] - (a_8 + a_9)[P3S3p],$$

$$\frac{d}{dt}[RpJ12S4] = a_{10}[RpJ12][S4] - (a_{11} + a_{12})[RpJ12S4],$$

$$\frac{d}{dt}[RpJ35S4] = a_{13}[RpJ35][S4] - (a_{14} + a_{15})[RpJ35S4],$$

$$\frac{d}{dt}[S4p] = a_{11}[RpJ12S4] + a_{14}[RpJ35S4] - a_{16}[P4][S4p] + a_{18}[P4S4p] -$$

$$-a_{21}[S3p][S4p] + a_{22}[S34] - 2a_{23}[S4p]^2 + 2a_{24}[S44],$$

$$\frac{d}{dt}[P4S4p] = a_{16}[P4][S4p] - (a_{17} + a_{18})[P4S4p],$$

$$\frac{d}{dt}[S33] = a_{19}[S3p]^2 - a_{20}[S33],$$

$$\frac{d}{dt}[S34] = a_{21}[S3p][S4p] - a_{22}[S34],$$

$$\frac{d}{dt}[S44] = a_{23}[S4p]^2 - a_{24}[S44].$$

(S58)

Conservation equations:

$$S3_T = [S3] + [S3p] + 2[S33] + [S34] + [RpJ2S3] + [RpJ6S3] + [P3S3p],$$
$$S4_T = [S4] + [S4p] + 2[S44] + [S34] + [RpJ12S4] + [RpJ35S4] + [P4S4p],$$
$$P3_T = [P3] + [P3S3p],$$
$$P4_T = [P4] + [P4S4p].$$

(S59)

Conservation Equations (S59) in non-dimensional form:

$$1 = [s3] + [s3p] + 2[s33] + [s34] + [w2s3] + [w6s3] + [p3s3p],$$
$$s4_t = [s4] + [s4p] + 2[s44] + [s34] + [w12s4] + [w35s4] + [p4s4p],$$
$$p3_t = [p3] + [p3s3p],$$
$$p4_t = [p4] + [p4s4p],$$

(S60)

where

$$[s3] = \frac{[S3]}{S3_T}, [s3p] = \frac{[S3p]}{S3_T}, [w2s3] = \frac{[RpJ2S3]}{S3_T}, [w6s3] = \frac{[RpJ6S3]}{S3_T}, [p3s3p] = \frac{[P3S3p]}{S3_T}, s4_t = \frac{S4_T}{S3_T},$$

$$[s4] = \frac{[S4]}{S3_T}, [s4p] = \frac{[S4p]}{S3_T}, [w12s4] = \frac{[RpJ12S4]}{S3_T}, [w35s4] = \frac{[RpJ35S4]}{S3_T}, [p4s4p] = \frac{[P4S4p]}{S3_T},$$

$$[s33] = \frac{[S33]}{S3_T}, [s34] = \frac{[S34]}{S3_T}, [s44] = \frac{[S44]}{S3_T}, [p3] = \frac{[P3]}{S3_T}, [p4] = \frac{[P4]}{S3_T}, p3_t = \frac{P3_T}{S3_T}, p4_t = \frac{P4_T}{S3_T}.$$

ODEs (S58) in non-dimensional form can be written as follows:

$$\frac{d}{d\tau}[w2s3] = m_1[w2][s3] - [w2s3]$$

$$\frac{d}{d\tau}[w6s3] = m_3[w6][s3] - (m_4 + m_5)[w6s3]$$

$$\frac{d}{d\tau}[s3p] = m_2[w2s3] + m_4[w6s3] - m_6[p3][s3p] + m_8[p3s3p] -$$
$$-2m_{18}[s3p]^2 + 2m_{19}[s33] - m_{20}[s3p][s4p] + m_{21}[s34]$$

$$\frac{d}{d\tau}[p3s3p] = m_6[p3][s3p] - (m_7 + m_8)[p3s3p]$$

$$\frac{d}{d\tau}[w12s4] = m_9[w12][s4] - (m_{10} + m_{11})[w12s4]$$

$$\frac{d}{d\tau}[w35s4] = m_{12}[w35][s4] - (m_{13} + m_{14})[w35s4]$$

$$\frac{d}{d\tau}[s4p] = m_{10}[w12s4] + m_{13}[w35s4] - m_{15}[p4][s4p] + m_{17}[p4s4p] -$$
$$-m_{20}[s3p][s4p] + m_{21}[s34] - 2m_{22}[s4p]^2 + 2m_{23}[s44]$$

$$\frac{d}{d\tau}[p4s4p] = m_{15}[p4][s4p] - (m_{16} + m_{17})[p4s4p]$$

$$\frac{d}{d\tau}[s33] = m_{18}[s3p]^2 - m_{19}[s33] \qquad (S61)$$

$$\frac{d}{d\tau}[s34] = m_{20}[s3p][s4p] - m_{21}[s34]$$

$$\frac{d}{d\tau}[s44] = m_{22}[s4p]^2 - m_{23}[s44]$$

where

$$\tau = t(a_2 + a_3), m_1 = \frac{a_1}{a_2 + a_3}S3_T, m_2 = \frac{a_2}{a_2 + a_3}, m_3 = \frac{a_4}{a_2 + a_3}S3_T, m_4 = \frac{a_5}{a_2 + a_3}, m_5 = \frac{a_6}{a_2 + a_3},$$

$$m_6 = \frac{a_7}{a_2 + a_3}S3_T, m_7 = \frac{a_8}{a_2 + a_3}, m_8 = \frac{a_9}{a_2 + a_3}, m_9 = \frac{a_{10}}{a_2 + a_3}S3_T, m_{10} = \frac{a_{11}}{a_2 + a_3}, m_{11} = \frac{a_{12}}{a_2 + a_3},$$

$$m_{12} = \frac{a_{13}}{a_2 + a_3}S3_T, m_{13} = \frac{a_{14}}{a_2 + a_3}, m_{14} = \frac{a_{15}}{a_2 + a_3}, m_{15} = \frac{a_{16}}{a_2 + a_3}S3_T, m_{16} = \frac{a_{17}}{a_2 + a_3}, m_{17} = \frac{a_{18}}{a_2 + a_3},$$

$$m_{18} = \frac{a_{19}}{a_2 + a_3}S3_T, m_{19} = \frac{a_{20}}{a_2 + a_3}, m_{20} = \frac{a_{21}}{a_2 + a_3}S3_T, m_{21} = \frac{a_{22}}{a_2 + a_3}, m_{22} = \frac{a_{23}}{a_2 + a_3}S3_T, m_{23} = \frac{a_{24}}{a_2 + a_3}.$$

We need to find steady-state solution of System (S61):

$$0 = m_1[w2][s3] - [w2s3]$$
$$0 = m_3[w6][s3] - (m_4 + m_5)[w6s3]$$
$$\frac{d}{d\tau}[s3p] = m_2[w2s3] + m_4[w6s3] - m_6[p3][s3p] + m_8[p3s3p] -$$
$$-2m_{18}[s3p]^2 + 2m_{19}[s33] - m_{20}[s3p][s4p] + m_{21}[s34]$$
$$0 = m_6[p3][s3p] - (m_7 + m_8)[p3s3p]$$
$$0 = m_9[w12][s4] - (m_{10} + m_{11})[w12s4]$$
$$0 = m_{12}[w35][s4] - (m_{13} + m_{14})[w35s4]$$
$$\frac{d}{d\tau}[s4p] = m_{10}[w12s4] + m_{13}[w35s4] - m_{15}[p4][s4p] + m_{17}[p4s4p] -$$
$$-m_{20}[s3p][s4p] + m_{21}[s34] - 2m_{22}[s4p]^2 + 2m_{23}[s44]$$
$$0 = m_{15}[p4][s4p] - (m_{16} + m_{17})[p4s4p]$$
$$0 = m_{18}[s3p]^2 - m_{19}[s33]$$
$$0 = m_{20}[s3p][s4p] - m_{21}[s34]$$
$$0 = m_{22}[s4p]^2 - m_{23}[s44]$$
(S62)

We can simplify System (S62):

$$0 = m_1[w2][s3] - [w2s3]$$
$$0 = m_3[w6][s3] - (m_4 + m_5)[w6s3]$$
$$\frac{d}{d\tau}[s3p] = m_2[w2s3] + m_4[w6s3] - m_7[p3s3p]$$
$$0 = m_6[p3][s3p] - (m_7 + m_8)[p3s3p]$$
$$0 = m_9[w12][s4] - (m_{10} + m_{11})[w12s4]$$
$$0 = m_{12}[w35][s4] - (m_{13} + m_{14})[w35s4]$$
$$\frac{d}{d\tau}[s4p] = m_{10}[w12s4] + m_{13}[w35s4] - m_{16}[p4s4p]$$
$$0 = m_{15}[p4][s4p] - (m_{16} + m_{17})[p4s4p]$$
$$0 = m_{18}[s3p]^2 - m_{19}[s33]$$
$$0 = m_{20}[s3p][s4p] - m_{21}[s34]$$
$$0 = m_{22}[s4p]^2 - m_{23}[s44]$$
(S63)

Then using conservation Equations (S60) and System (S63) we obtain:

$$\frac{d}{d\tau}[s3p] = m_2[w2s3] + m_4[w6s3] - m_7[p3s3p],$$
$$\frac{d}{d\tau}[s4p] = m_{10}[w12s4] + m_{13}[w35s4] - m_{16}[p4s4p],$$
(S64)

where

$$[p3s3p] = p3_t \frac{[s3p]}{\frac{m_7 + m_8}{m_6} + [s3p]}$$

$$[p4s4p] = p4_t \frac{[s4p]}{\frac{m_{16} + m_{17}}{m_{15}} + [s4p]}$$

$$[s33] = \frac{m_{18}}{m_{19}}[s3p]^2$$

$$[s34] = \frac{m_{20}}{m_{21}}[s3p][s4p]$$

$$[s44] = \frac{m_{22}}{m_{23}}[s4p]^2$$

$$[w2s3] = m_1[w2][s3]$$

$$[w6s3] = \frac{m_3}{m_4 + m_5}[w6][s3]$$

$$[w12s4] = \frac{m_9}{m_{10} + m_{11}}[w12][s4]$$

$$[w35s4] = \frac{m_{12}}{m_{13} + m_{14}}[w35][s4]$$

$$[s3] = \frac{1 - [s3p] - 2[s33] - [s34] - [p3s3p]}{1 + m_1[w2] + \frac{m_3}{m_4 + m_5}[w6]}$$

$$[s4] = \frac{s4_t - [s4p] - 2[s44] - [s34] - [p4s4p]}{1 + \frac{m_9}{m_{10} + m_{11}}[w12] + \frac{m_{12}}{m_{13} + m_{14}}[w35]}$$

Or we can rewrite it as follows:

$$[p3s3p] = p3_t \frac{[s3p]}{M_{11} + [s3p]}$$

$$[p4s4p] = p4_t \frac{[s4p]}{M_{14} + [s4p]}$$

$$[s33] = \frac{[s3p]^2}{M_{15}}$$

$$[s34] = \frac{[s3p][s4p]}{M_{16}}$$

$$[s44] = \frac{[s4p]^2}{M_{17}}$$

$$[w2s3] = \frac{[w2][s3]}{M_9}$$

$$[w6s3] = \frac{[w6][s3]}{M_{10}}$$

$$[w12s4] = \frac{[w12][s4]}{M_{12}}$$

$$[w35s4] = \frac{[w35][s4]}{M_{13}}$$

$$[s3] = \frac{1 - [s3p] - 2[s33] - [s34] - [p3s3p]}{1 + \frac{[w2]}{M_9} + \frac{[w6]}{M_{10}}}$$

$$[s4] = \frac{s4_t - [s4p] - 2[s44] - [s34] - [p4s4p]}{1 + \frac{[w12]}{M_{12}} + \frac{[w35]}{M_{13}}}$$

(S65)

where we denote the Michaelis constants

$$M_9 = \frac{a_2 + a_3}{a_1 S3_T}, M_{10} = \frac{a_5 + a_6}{a_4 S3_T}, M_{11} = \frac{a_8 + a_9}{a_7 S3_T}, M_{12} = \frac{a_{11} + a_{12}}{a_{10} S3_T}, M_{13} = \frac{a_{14} + a_{15}}{a_{13} S3_T},$$

$$M_{14} = \frac{a_{17} + a_{18}}{a_{16} S3_T}, M_{15} = \frac{a_{20}}{a_{19} S3_T}, M_{16} = \frac{a_{22}}{a_{21} S3_T}, M_{17} = \frac{a_{24}}{a_{23} S3_T}.$$

When considering steady-state solutions of System (S64) we can write:

$$0 = m_2[w2s3] + m_4[w6s3] - m_7[p3s3p],$$
$$0 = m_{10}[w12s4] + m_{13}[w35s4] - m_{16}[p4s4p].$$

(S66)

Or we can rewrite Equations (S66) as follows:

$$0 = n_5[w2s3] + n_6[w6s3] - [p3s3p],$$
$$0 = n_7[w12s4] + n_8[w35s4] - [p4s4p],$$

(S67)

where $n_5 = \dfrac{a_2}{a_8}, n_6 = \dfrac{a_5}{a_8}, n_7 = \dfrac{a_{11}}{a_{17}}, n_8 = \dfrac{a_{14}}{a_{17}}$.

We can rewrite System (S67) substituting Equations (S65):

$$0 = [s3p] + 2\dfrac{[s3p]^2}{M_{15}} + \dfrac{[s3p][s4p]}{M_{16}} + $$
$$+ p3_t \dfrac{[s3p]}{M_{11} + [s3p]}\left(1 + \dfrac{M_9 M_{10} + M_{10}[w2] + M_9[w6]}{n_5 M_{10}[w2] + n_6 M_9[w6]}\right) - 1,$$
$$0 = [s4p] + 2\dfrac{[s4p]^2}{M_{17}} + \dfrac{[s3p][s4p]}{M_{16}} + $$
$$+ p4_t \dfrac{[s4p]}{M_{14} + [s4p]}\left(1 + \dfrac{M_{12} M_{13} + M_{13}[w12] + M_{12}[w35]}{n_7 M_{13}[w12] + n_8 M_{12}[w35]}\right) - s4_t,$$

(S68)

We find $[s3p]$ and $[s4p]$ in System (S68) numerically.

**CD46**

It can be written for SP1 in non-dimensional form according to Equation (S40):

$$[sp1a] = sp1_t \dfrac{[i2]}{M_{18} + [i2]} \dfrac{[cd46]}{M_{19} + [cd46]},$$
(S69)

where $[sp1a] = \dfrac{[SP1a]}{S3_T}$, $sp1_t = \dfrac{SP1_T}{S3_T}$, $[cd46] = \dfrac{[CD46]}{S3_T}$, $M_{18} = \dfrac{l_2}{l_1 S3_T}$, and $M_{19} = \dfrac{l_4}{l_3 S3_t}$.

## 2.3 IFN-γ and IL-10 production

Since IFN-γ is activated in this module by STAT44 only (Fig 1C) we can write using Equation (S51):

$$[ig] = \dfrac{M_{20}}{\dfrac{mp1_t}{n_9 gg_t \dfrac{[s44]}{M_{21} + [s44]}} - 1},$$
(S70)

where $[ig] = \dfrac{[Ig]}{S3_T}, mp1_t = \dfrac{Mp1_T}{S3_T}, gg_t = \dfrac{Gg_T}{S3_T}, M_{20} = \dfrac{k_4 + k_5}{k_3 S3_T}, M_{21} = \dfrac{k_2}{k_1 S3_T}, n_9 = \dfrac{l_5}{k_4}, n_9 < 1$.

According to Equation (S53) we can write:

$$[i10] = \frac{M_{22}}{\dfrac{mp2_t}{n_{10}g10_t\left(\dfrac{[s33]}{M_{23}+[s33]}+\dfrac{[sp1a]}{M_{24}+[sp1a]}-\dfrac{[s33]}{M_{23}+[s33]}\dfrac{[sp1a]}{M_{24}+[sp1a]}\right)}-1}, \qquad (S71)$$

where

$$mp2_t = \frac{Mp2_T}{S3_T}, g10_t = \frac{G10_T}{S3_T}, M_{22} = \frac{k_{11}+k_{12}}{k_{10}S3_T}, M_{23} = \frac{k_7}{k_6 S3_T}, M_{24} = \frac{k_9}{k_8 S3_T}, n_{10} = \frac{l_6}{k_{11}}, n_{10} < 1.$$

## 3 Combined STAT3-STAT4-STAT5 model

The reactions involved in STAT3-STAT4-STAT5 circuit (Fig 5A):

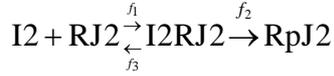

$$I2 + RJ2 \underset{f_3}{\overset{f_1}{\rightleftarrows}} I2RJ2 \overset{f_2}{\rightarrow} RpJ2$$

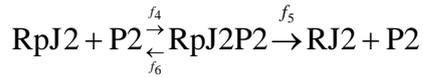

$$RpJ2 + P2 \underset{f_6}{\overset{f_4}{\rightleftarrows}} RpJ2P2 \overset{f_5}{\rightarrow} RJ2 + P2$$

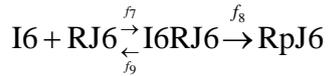

$$I6 + RJ6 \underset{f_9}{\overset{f_7}{\rightleftarrows}} I6RJ6 \overset{f_8}{\rightarrow} RpJ6$$

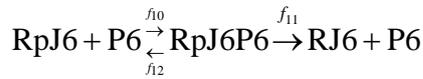

$$RpJ6 + P6 \underset{f_{12}}{\overset{f_{10}}{\rightleftarrows}} RpJ6P6 \overset{f_{11}}{\rightarrow} RJ6 + P6$$

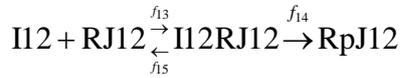

$$I12 + RJ12 \underset{f_{15}}{\overset{f_{13}}{\rightleftarrows}} I12RJ12 \overset{f_{14}}{\rightarrow} RpJ12$$

$$\text{RpJ12} + \text{P12} \underset{f_{18}}{\overset{f_{16}}{\rightleftarrows}} \text{RpJ12P12} \overset{f_{17}}{\rightarrow} \text{RJ12} + \text{P12}$$

$$\text{I35} + \text{RJ35} \underset{f_{21}}{\overset{f_{19}}{\rightleftarrows}} \text{I35RJ35} \overset{f_{20}}{\rightarrow} \text{RpJ35}$$

$$\text{RpJ35} + \text{P35} \underset{f_{24}}{\overset{f_{22}}{\rightleftarrows}} \text{RpJ35P35} \overset{f_{23}}{\rightarrow} \text{RJ35} + \text{P35}$$

$$\text{I21} + \text{RJ21} \underset{f_{27}}{\overset{f_{25}}{\rightleftarrows}} \text{I21RJ21} \overset{f_{26}}{\rightarrow} \text{RpJ21}$$

$$\text{RpJ21} + \text{P21} \underset{f_{30}}{\overset{f_{28}}{\rightleftarrows}} \text{RpJ21P21} \overset{f_{29}}{\rightarrow} \text{RJ21} + \text{P21}$$

$$\text{RpJ2} + \text{S3} \underset{a_{3}}{\overset{a_{1}}{\rightleftarrows}} \text{RpJ2S3} \overset{a_{2}}{\rightarrow} \text{RpJ2} + \text{S3p}$$

$$\text{RpJ6} + \text{S3} \underset{a_{6}}{\overset{a_{4}}{\rightleftarrows}} \text{RpJ6S3} \overset{a_{5}}{\rightarrow} \text{RpJ6} + \text{S3p}$$

$$\text{P3} + \text{S3p} \underset{a_{9}}{\overset{a_{7}}{\rightleftarrows}} \text{P3S3p} \overset{a_{8}}{\rightarrow} \text{P3} + \text{S3}$$

$$\text{RpJ12} + \text{S4} \underset{a_{12}}{\overset{a_{10}}{\rightleftarrows}} \text{RpJ12S4} \overset{a_{11}}{\rightarrow} \text{RpJ12} + \text{S4p}$$

$$\text{RpJ35} + \text{S4} \underset{a_{15}}{\overset{a_{13}}{\rightleftarrows}} \text{RpJ35S4} \overset{a_{14}}{\rightarrow} \text{RpJ35} + \text{S4p}$$

$$\text{P4} + \text{S4p} \underset{a_{18}}{\overset{a_{16}}{\rightleftarrows}} \text{P4S4p} \overset{a_{17}}{\rightarrow} \text{P4} + \text{S4}$$

$$\text{RpJ2} + \text{S5} \underset{a_{21}}{\overset{a_{19}}{\rightleftarrows}} \text{RpJ2S5} \overset{a_{20}}{\rightarrow} \text{RpJ2} + \text{S5p}$$

$$\text{RpJ21} + \text{S5} \underset{a_{24}}{\overset{a_{22}}{\rightleftarrows}} \text{RpJ21S5} \overset{a_{23}}{\rightarrow} \text{RpJ21} + \text{S5p}$$

$$\text{P5} + \text{S5p} \underset{a_{27}}{\overset{a_{25}}{\rightleftarrows}} \text{P5S5p} \overset{a_{26}}{\rightarrow} \text{P5} + \text{S5}$$

$$\text{S3p} + \text{S3p} \underset{a_{29}}{\overset{a_{28}}{\rightleftarrows}} \text{S33}$$

$$\text{S3p} + \text{S4p} \underset{a_{31}}{\overset{a_{30}}{\rightleftarrows}} \text{S34}$$

(S72)

$$S4p + S4p \underset{a_{33}}{\overset{a_{32}}{\rightleftarrows}} S44$$

$$S3p + S5p \underset{a_{35}}{\overset{a_{34}}{\rightleftarrows}} S35$$

$$S5p + S5p \underset{a_{37}}{\overset{a_{36}}{\rightleftarrows}} S55$$

$$S44 + Gg \underset{k_{2}}{\overset{k_{1}}{\rightleftarrows}} S44Gg$$

$$S44Gg + S55 \underset{k_{4}}{\overset{k_{3}}{\rightleftarrows}} S44S55Gg$$

$$S55 + Gg \underset{k_{4}}{\overset{k_{3}}{\rightleftarrows}} S55Gg$$

$$S55Gg + S44 \underset{k_{2}}{\overset{k_{1}}{\rightleftarrows}} S44S55Gg$$

$$S44Gg \xrightarrow{l_{1}} Ig$$

$$S55Gg \xrightarrow{l_{1}} Ig$$

$$S44S55Gg \xrightarrow{l_{1}} Ig$$

$$Ig + Mp1 \underset{k_{7}}{\overset{k_{5}}{\rightleftarrows}} IgMp1 \xrightarrow{k_{6}} Ign + Mp1$$

$$I2 + Sp1 \underset{l_{3}}{\overset{l_{2}}{\rightleftarrows}} I2Sp1$$

$$I2Sp1 + CD46 \underset{l_{5}}{\overset{l_{4}}{\rightleftarrows}} Sp1a$$

$$CD46 + Sp1 \underset{l_{5}}{\overset{l_{4}}{\rightleftarrows}} CD46Sp1$$

$$CD46Sp1 + I2 \underset{l_{3}}{\overset{l_{2}}{\rightleftarrows}} Sp1a$$

$$S33 + G10 \underset{k_{9}}{\overset{k_{8}}{\rightleftarrows}} S33G10$$

$$S33G10 + Sp1a \underset{k_{11}}{\overset{k_{10}}{\rightleftarrows}} S33Sp1aG10$$

$$Sp1a + G10 \underset{k_{11}}{\overset{k_{10}}{\rightleftarrows}} Sp1aG10$$

$$Sp1aG10 + S33 \underset{k_{9}}{\overset{k_{8}}{\rightleftarrows}} S33Sp1aG10$$

$$S33G10 \xrightarrow{l_{6}} I10$$

$$Sp1aG10 \xrightarrow{l_{6}} I10$$

$$S33Sp1aG10 \xrightarrow{l_{6}} I10$$

$$I10 + Mp2 \underset{k_{14}}{\overset{k_{12}}{\rightleftarrows}} I10Mp2 \xrightarrow{k_{13}} I10n + Mp2$$

## 3.1 Cytokine-receptor interactions

According to Equation (S12) we can write for $[RpJ2]$, $[RpJ6]$, $[RpJ12]$, $[RpJ35]$ and $[RpJ21]$ in non-dimensional form respectively:

$$[w2] = -\frac{M_2 - r2_t + p2_t\left(\frac{M_1}{n_1[i2]} + \frac{1}{n_1} + 1\right)}{2} + \frac{\sqrt{\left(M_2 - r2_t + p2_t\left(\frac{M_1}{n_1[i2]} + \frac{1}{n_1} + 1\right)\right)^2 + 4r2_t M_2}}{2},$$

$$[w6] = -\frac{M_4 - r6_t + p6_t\left(\frac{M_3}{n_2[i6]} + \frac{1}{n_2} + 1\right)}{2} + \frac{\sqrt{\left(M_4 - r6_t + p6_t\left(\frac{M_3}{n_2[i6]} + \frac{1}{n_2} + 1\right)\right)^2 + 4r6_t M_4}}{2},$$

$$[w12] = -\frac{M_6 - r12_t + p12_t\left(\frac{M_5}{n_3[i12]} + \frac{1}{n_3} + 1\right)}{2} + \frac{\sqrt{\left(M_6 - r12_t + p12_t\left(\frac{M_5}{n_3[i12]} + \frac{1}{n_3} + 1\right)\right)^2 + 4r12_t M_6}}{2},$$

$$[w35] = -\frac{M_8 - r35_t + p35_t\left(\frac{M_7}{n_4[i35]} + \frac{1}{n_4} + 1\right)}{2} + \frac{\sqrt{\left(M_8 - r35_t + p35_t\left(\frac{M_7}{n_4[i35]} + \frac{1}{n_4} + 1\right)\right)^2 + 4r35_t M_8}}{2},$$

$$[w21] = -\frac{M_{10} - r21_t + p21_t\left(\frac{M_9}{n_5[i21]} + \frac{1}{n_5} + 1\right)}{2} + \frac{\sqrt{\left(M_{10} - r21_t + p21_t\left(\frac{M_9}{n_5[i21]} + \frac{1}{n_5} + 1\right)\right)^2 + 4r21_t M_{10}}}{2}, \quad (S73)$$

where
$$[i2] = \frac{[I2]}{S3_T}, [w2] = \frac{[RpJ2]}{S3_T}, r2_t = \frac{R2_T}{S3_T}, p2_t = \frac{P2_T}{S3_T}, M_1 = \frac{f_2 + f_3}{f_1 S3_T}, M_2 = \frac{f_5 + f_6}{f_4 S3_T}, n_1 = \frac{f_2}{f_5}, [i6] = \frac{[I6]}{S3_T},$$
$$[w6] = \frac{[RpJ6]}{S3_T}, r6_t = \frac{R6_T}{S3_T}, p6_t = \frac{P6_T}{S3_T}, M_3 = \frac{f_8 + f_9}{f_7 S3_T}, M_4 = \frac{f_{11} + f_{12}}{f_{10} S3_T}, n_2 = \frac{f_8}{f_{11}}, [i12] = \frac{[I12]}{S3_T},$$
$$[w12] = \frac{[RpJ12]}{S3_T}, r12_t = \frac{R12_T}{S3_T}, p12_t = \frac{P12_T}{S3_T}, M_5 = \frac{f_{14} + f_{15}}{f_{13} S3_T}, M_6 = \frac{f_{17} + f_{18}}{f_{16} S3_T}, n_3 = \frac{f_{14}}{f_{17}}, [i35] = \frac{[I35]}{S3_T},$$
$$[w35] = \frac{[RpJ35]}{S3_T}, r35_t = \frac{R35_T}{S3_T}, p35_t = \frac{P35_T}{S3_T}, M_7 = \frac{f_{20} + f_{21}}{f_{19} S3_T}, M_8 = \frac{f_{23} + f_{24}}{f_{22} S3_T}, n_4 = \frac{f_{20}}{f_{23}},$$
$$[i21] = \frac{[I21]}{S3_T}, [w21] = \frac{[RpJ21]}{S3_T}, r21_t = \frac{R21_T}{S3_T}, p21_t = \frac{P21_T}{S3_T}, M_9 = \frac{f_{26} + f_{27}}{f_{25} S3_T}, M_{10} = \frac{f_{29} + f_{30}}{f_{28} S3_T}, n_5 = \frac{f_{26}}{f_{29}}.$$

## 3.2 STAT phosphorylation and dimerization

ODEs for the STAT module are given by:

$$\frac{d}{dt}[RpJ2S3] = a_1[RpJ2][S3] - (a_2 + a_3)[RpJ2S3],$$

$$\frac{d}{dt}[RpJ6S3] = a_4[RpJ6][S3] - (a_5 + a_6)[RpJ6S3],$$

$$\frac{d}{dt}[S3p] = a_2[RpJ2S3] + a_5[RpJ6S3] - a_7[P3][S3p] + a_9[P3S3p] -$$
$$- 2a_{28}[S3p]^2 + 2a_{29}[S33] - a_{30}[S3p][S4p] + a_{31}[S34] -$$
$$- a_{34}[S3p][S5p] + a_{35}[S35],$$

$$\frac{d}{dt}[P3S3p] = a_7[P3][S3p] - (a_8 + a_9)[P3S3p],$$

$$\frac{d}{dt}[RpJ12S4] = a_{10}[RpJ12][S4] - (a_{11} + a_{12})[RpJ12S4],$$

$$\frac{d}{dt}[RpJ35S4] = a_{13}[RpJ35][S4] - (a_{14} + a_{15})[RpJ35S4],$$

$$\frac{d}{dt}[S4p] = a_{11}[RpJ12S4] + a_{14}[RpJ35S4] - a_{16}[P4][S4p] + a_{18}[P4S4p] -$$
$$- a_{30}[S3p][S4p] + a_{31}[S34] - 2a_{32}[S4p]^2 + 2a_{33}[S44],$$

$$\frac{d}{dt}[P4S4p] = a_{16}[P4][S4p] - (a_{17} + a_{18})[P4S4p],$$

$$\frac{d}{dt}[RpJ2S5] = a_{19}[RpJ2][S5] - (a_{20} + a_{21})[RpJ2S5],$$

$$\frac{d}{dt}[RpJ21S5] = a_{22}[RpJ21][S5] - (a_{23} + a_{24})[RpJ21S5],$$

$$\frac{d}{dt}[S5p] = a_{20}[RpJ2S5] + a_{23}[RpJ21S5] - a_{25}[P5][S5p] + a_{27}[P5S5p] -$$
$$- a_{34}[S3p][S5p] + a_{35}[S35] - 2a_{36}[S5p]^2 + 2a_{37}[S55],$$

$$\frac{d}{dt}[P5S5p] = a_{25}[P5][S5p] - (a_{26} + a_{27})[P5S5p],$$

$$\frac{d}{dt}[S33] = a_{28}[S3p]^2 - a_{29}[S33],$$

$$\frac{d}{dt}[S34] = a_{30}[S3p][S4p] - a_{31}[S34], \tag{S74}$$

$$\frac{d}{dt}[S44] = a_{32}[S4p]^2 - a_{33}[S44],$$

$$\frac{d}{dt}[S35] = a_{34}[S3p][S5p] - a_{35}[S35],$$

$$\frac{d}{dt}[S55] = a_{36}[S5p]^2 - a_{37}[S55].$$

Conservation equations:

$$\begin{aligned}
S3_T &= [S3] + [S3p] + 2[S33] + [S34] + [S35] + [RpJ2S3] + [RpJ6S3] + [P3S3p], \\
S4_T &= [S4] + [S4p] + 2[S44] + [S34] + [RpJ12S4] + [RpJ35S4] + [P4S4p], \\
S5_T &= [S5] + [S5p] + 2[S55] + [S35] + [RpJ2S5] + [RpJ21S5] + [P5S5p], \\
P3_T &= [P3] + [P3S3p], \\
P4_T &= [P4] + [P4S4p], \\
P5_T &= [P5] + [P5S5p].
\end{aligned} \tag{S75}$$

Conservation Equations (S75) in non-dimensional form:

$$1 = [s3] + [s3p] + 2[s33] + [s34] + [s35] + [w2s3] + [w6s3] + [p3s3p],$$
$$s4_t = [s4] + [s4p] + 2[s44] + [s34] + [w12s4] + [w35s4] + [p4s4p],$$
$$s5_t = [s5] + [s5p] + 2[s55] + [s35] + [w2s5] + [w21s5] + [p5s5p],$$
$$p3_t = [p3] + [p3s3p],$$
$$p4_t = [p4] + [p4s4p],$$
$$p5_t = [p5] + [p5s5p],$$
(S76)

where

$$[s3] = \frac{[S3]}{S3_T}, [s3p] = \frac{[S3p]}{S3_T}, [w2s3] = \frac{[RpJ2S3]}{S3_T}, [w6s3] = \frac{[RpJ6S3]}{S3_T}, [p3s3p] = \frac{[P3S3p]}{S3_T}, s4_t = \frac{S4_t}{S3_T},$$

$$[s4] = \frac{[S4]}{S3_T}, [s4p] = \frac{[S4p]}{S3_T}, [w12s4] = \frac{[RpJ12S4]}{S3_T}, [w35s4] = \frac{[RpJ35S4]}{S3_T}, [p4s4p] = \frac{[P4S4p]}{S3_T},$$

$$s5_t = \frac{S5_T}{S3_T}, [s5] = \frac{[S5]}{S3_T}, [s5p] = \frac{[S5p]}{S3_T}, [w2s5] = \frac{[RpJ2S5]}{S3_T}, [w21s5] = \frac{[RpJ21S5]}{S3_T}, [p5s5p] = \frac{[P5S5p]}{S3_T},$$

$$[s33] = \frac{[S33]}{S3_T}, [s34] = \frac{[S34]}{S3_T}, [s44] = \frac{[S44]}{S3_T}, [s35] = \frac{[S35]}{S3_T}, [s55] = \frac{[S55]}{S3_T}, [p3] = \frac{[P3]}{S3_T}, [p4] = \frac{[P4]}{S3_T},$$

$$[p5] = \frac{[P5]}{S3_T}, p3_t = \frac{P3_T}{S3_T}, p4_t = \frac{P4_T}{S3_T}, p5_t = \frac{P5_T}{S3_T}.$$

ODEs (S74) in non-dimensional form are given by:

$$\frac{d}{d\tau}[w2s3] = m_1[w2][s3] - [w2s3]$$

$$\frac{d}{d\tau}[w6s3] = m_3[w6][s3] - (m_4 + m_5)[w6s3]$$

$$\frac{d}{d\tau}[s3p] = m_2[w2s3] + m_4[w6s3] - m_6[p3][s3p] + m_8[p3s3p] -$$
$$-2m_{27}[s3p]^2 + 2m_{28}[s33] - m_{29}[s3p][s4p] + m_{30}[s34] -$$
$$-m_{33}[s3p][s5p] + m_{34}[s35]$$

$$\frac{d}{d\tau}[p3s3p] = m_6[p3][s3p] - (m_7 + m_8)[p3s3p]$$

$$\frac{d}{d\tau}[w12s4] = m_9[w12][s4] - (m_{10} + m_{11})[w12s4]$$

$$\frac{d}{d\tau}[w35s4] = m_{12}[w35][s4] - (m_{13} + m_{14})[w35s4]$$

$$\frac{d}{d\tau}[s4p] = m_{10}[w12s4] + m_{13}[w35s4] - m_{15}[p4][s4p] +$$
$$+ m_{17}[p4s4p] - m_{29}[s3p][s4p] +$$
$$+ m_{30}[s34] - 2m_{31}[s4p]^2 + 2m_{32}[s44]$$

$$\frac{d}{d\tau}[p4s4p] = m_{15}[p4][s4p] - (m_{16} + m_{17})[p4s4p]$$

$$\frac{d}{d\tau}[w2s5] = m_{18}[w2][s5] - (m_{19} + m_{20})[w2s5]$$

$$\frac{d}{d\tau}[w21s5] = m_{21}[w21][s5] - (m_{22} + m_{23})[w21s5]$$

$$\frac{d}{d\tau}[s5p] = m_{19}[w2s5] + m_{22}[w21s5] - m_{24}[p5][s5p] + m_{26}[p5s5p] -$$
$$-m_{33}[s3p][s5p] + m_{34}[s35] - 2m_{35}[s5p]^2 + 2m_{36}[s55]$$

$$\frac{d}{d\tau}[p5s5p] = m_{24}[p5][s5p] - (m_{25} + m_{26})[p5s5p]$$

$$\frac{d}{d\tau}[s33] = m_{27}[s3p]^2 - m_{28}[s33]$$

$$\frac{d}{d\tau}[s34] = m_{29}[s3p][s4p] - m_{30}[s34]$$

$$\frac{d}{d\tau}[s44] = m_{31}[s4p]^2 - m_{32}[s44] \quad \text{(S77)}$$

$$\frac{d}{d\tau}[s35] = m_{33}[s3p][s5p] - m_{34}[s35]$$

$$\frac{d}{d\tau}[s55] = m_{35}[s5p]^2 - m_{36}[s55]$$

where

$$\tau = t(a_2 + a_3), m_1 = \frac{a_1}{a_2 + a_3} S3_t, m_2 = \frac{a_2}{a_2 + a_3}, m_3 = \frac{a_4}{a_2 + a_3} S3_t, m_4 = \frac{a_5}{a_2 + a_3}, m_5 = \frac{a_6}{a_2 + a_3}, m_6 = \frac{a_7}{a_2 + a_3} S3_t,$$

$$m_7 = \frac{a_8}{a_2 + a_3}, m_8 = \frac{a_9}{a_2 + a_3}, m_9 = \frac{a_{10}}{a_2 + a_3} S3_t, m_{10} = \frac{a_{11}}{a_2 + a_3}, m_{11} = \frac{a_{12}}{a_2 + a_3}, m_{12} = \frac{a_{13}}{a_2 + a_3} S3_t, m_{13} = \frac{a_{14}}{a_2 + a_3},$$

$$m_{14} = \frac{a_{15}}{a_2 + a_3}, m_{15} = \frac{a_{16}}{a_2 + a_3} S3_t, m_{16} = \frac{a_{17}}{a_2 + a_3}, m_{17} = \frac{a_{18}}{a_2 + a_3}, m_{18} = \frac{a_{19}}{a_2 + a_3} S3_t, m_{19} = \frac{a_{20}}{a_2 + a_3}, m_{20} = \frac{a_{21}}{a_2 + a_3},$$

$$m_{21} = \frac{a_{22}}{a_2 + a_3} S3_t, m_{22} = \frac{a_{23}}{a_2 + a_3}, m_{23} = \frac{a_{24}}{a_2 + a_3}, m_{24} = \frac{a_{25}}{a_2 + a_3} S3_t, m_{25} = \frac{a_{26}}{a_2 + a_3}, m_{26} = \frac{a_{27}}{a_2 + a_3},$$

$$m_{27} = \frac{a_{28}}{a_2 + a_3} S3_t, m_{28} = \frac{a_{29}}{a_2 + a_3}, m_{29} = \frac{a_{30}}{a_2 + a_3} S3_t, m_{30} = \frac{a_{31}}{a_2 + a_3}, m_{31} = \frac{a_{32}}{a_2 + a_3} S3_t, m_{32} = \frac{a_{33}}{a_2 + a_3},$$

$$m_{33} = \frac{a_{34}}{a_2 + a_3} S3_t, m_{34} = \frac{a_{35}}{a_2 + a_3}, m_{35} = \frac{a_{36}}{a_2 + a_3} S3_t, m_{36} = \frac{a_{37}}{a_2 + a_3}.$$

We need to find a steady-state solution of System (S77):

$$0 = m_1[w2][s3] - [w2s3]$$

$$0 = m_3[w6][s3] - (m_4 + m_5)[w6s3]$$

$$\frac{d}{d\tau}[s3p] = m_2[w2s3] + m_4[w6s3] - m_6[p3][s3p] + m_8[p3s3p] - 2m_{27}[s3p]^2 +$$
$$+ 2m_{28}[s33] - m_{29}[s3p][s4p] + m_{30}[s34] -$$
$$- m_{33}[s3p][s5p] + m_{34}[s35]$$

$$0 = m_6[p3][s3p] - (m_7 + m_8)[p3s3p]$$

$$0 = m_9[w12][s4] - (m_{10} + m_{11})[w12s4]$$

$$0 = m_{12}[w35][s4] - (m_{13} + m_{14})[w35s4]$$

$$\frac{d}{d\tau}[s4p] = m_{10}[w12s4] + m_{13}[w35s4] - m_{15}[p4][s4p] + m_{17}[p4s4p] - m_{29}[s3p][s4p] +$$
$$+ m_{30}[s34] - 2m_{31}[s4p]^2 + 2m_{32}[s44]$$

$$0 = m_{15}[p4][s4p] - (m_{16} + m_{17})[p4s4p]$$

$$0 = m_{18}[w2][s5] - (m_{19} + m_{20})[w2s5]$$

$$0 = m_{21}[w21][s5] - (m_{22} + m_{23})[w21s5]$$

$$\frac{d}{d\tau}[s5p] = m_{19}[w2s5] + m_{22}[w21s5] - m_{24}[p5][s5p] + m_{26}[p5s5p] -$$
$$- m_{33}[s3p][s5p] + m_{34}[s35] - 2m_{35}[s5p]^2 + 2m_{36}[s55]$$

$$0 = m_{24}[p5][s5p] - (m_{25} + m_{26})[p5s5p]$$

$$0 = m_{27}[s3p]^2 - m_{28}[s33]$$

$$0 = m_{29}[s3p][s4p] - m_{30}[s34]$$

$$0 = m_{31}[s4p]^2 - m_{32}[s44]$$

$$0 = m_{33}[s3p][s5p] - m_{34}[s35]$$

$$0 = m_{35}[s5p]^2 - m_{36}[s55] \qquad \text{(S78)}$$

We can simplify System (S78):

$$0 = m_1[w2][s3] - [w2s3]$$

$$0 = m_3[w6][s3] - (m_4 + m_5)[w6s3]$$

$$\frac{d}{d\tau}[s3p] = m_2[w2s3] + m_4[w6s3] - m_7[p3s3p]$$

$$0 = m_6[p3][s3p] - (m_7 + m_8)[p3s3p]$$

$$0 = m_9[w12][s4] - (m_{10} + m_{11})[w12s4]$$

$$0 = m_{12}[w35][s4] - (m_{13} + m_{14})[w35s4]$$

$$\frac{d}{d\tau}[s4p] = m_{10}[w12s4] + m_{13}[w35s4] - m_{16}[p4s4p]$$

$$0 = m_{15}[p4][s4p] - (m_{16} + m_{17})[p4s4p]$$

$$0 = m_{18}[w2][s5] - (m_{19} + m_{20})[w2s5]$$

$$0 = m_{21}[w21][s5] - (m_{22} + m_{23})[w21s5]$$

$$\frac{d}{d\tau}[s5p] = m_{19}[w2s5] + m_{22}[w21s5] - m_{25}[p5s5p]$$

$$0 = m_{24}[p5][s5p] - (m_{25} + m_{26})[p5s5p]$$

$$0 = m_{27}[s3p]^2 - m_{28}[s33]$$

$$0 = m_{29}[s3p][s4p] - m_{30}[s34]$$

$$0 = m_{31}[s4p]^2 - m_{32}[s44]$$

$$0 = m_{33}[s3p][s5p] - m_{34}[s35]$$

$$0 = m_{35}[s5p]^2 - m_{36}[s55]$$

(S79)

Then using conservation Equations (S76) and System (S79) we obtain:

$$\frac{d}{d\tau}[s3p] = m_2[w2s3] + m_4[w6s3] - m_7[p3s3p]$$

$$\frac{d}{d\tau}[s4p] = m_{10}[w12s4] + m_{13}[w35s4] - m_{16}[p4s4p]$$

$$\frac{d}{d\tau}[s5p] = m_{19}[w2s5] + m_{22}[w21s5] - m_{25}[p5s5p]$$

(S80)

where

$$[p3s3p] = p3_t \frac{[s3p]}{\frac{m_7 + m_8}{m_6} + [s3p]}$$

$$[p4s4p] = p4_t \frac{[s4p]}{\frac{m_{16} + m_{17}}{m_{15}} + [s4p]}$$

$$[p5s5p] = p5_t \frac{[s5p]}{\frac{m_{25} + m_{26}}{m_{24}} + [s5p]}$$

$$[s33] = \frac{m_{27}}{m_{28}} [s3p]^2$$

$$[s34] = \frac{m_{29}}{m_{30}} [s3p][s4p]$$

$$[s44] = \frac{m_{31}}{m_{32}} [s4p]^2$$

$$[s35] = \frac{m_{33}}{m_{34}} [s3p][s5p]$$

$$[s55] = \frac{m_{35}}{m_{36}} [s5p]^2$$

$$[w2s3] = m_1 [w2][s3]$$

$$[w6s3] = \frac{m_3}{m_4 + m_5} [w6][s3]$$

$$[w12s4] = \frac{m_9}{m_{10} + m_{11}} [w12][s4]$$

$$[w35s4] = \frac{m_{12}}{m_{13} + m_{14}} [w35][s4]$$

$$[w2s5] = \frac{m_{18}}{m_{19} + m_{20}} [w2][s5]$$

$$[w21s5] = \frac{m_{21}}{m_{22} + m_{23}} [w21][s5]$$

$$[s3] = \frac{1 - [s3p] - 2[s33] - [s34] - [s35] - [p3s3p]}{1 + m_1[w2] + \frac{m_3}{m_4 + m_5}[w6]}$$

$$[s4] = \frac{s4_t - [s4p] - 2[s44] - [s34] - [p4s4p]}{1 + \frac{m_9}{m_{10} + m_{11}}[w12] + \frac{m_{12}}{m_{13} + m_{14}}[w35]}$$

$$[s5] = \frac{s5_t - [s5p] - 2[s55] - [s35] - [p5s5p]}{1 + \frac{m_{18}}{m_{19} + m_{20}}[w2] + \frac{m_{21}}{m_{22} + m_{23}}[w21]}$$

Or we can rewrite it in the following way:

$$[p3s3p] = p3_t \frac{[s3p]}{M_{13} + [s3p]}$$

$$[p4s4p] = p4_t \frac{[s4p]}{M_{16} + [s4p]}$$

$$[p5s5p] = p5_t \frac{[s5p]}{M_{19} + [s5p]}$$

$$[s33] = \frac{[s3p]^2}{M_{20}}$$

$$[s34] = \frac{[s3p][s4p]}{M_{21}}$$

$$[s44] = \frac{[s4p]^2}{M_{22}}$$

$$[s35] = \frac{[s3p][s5p]}{M_{23}}$$

$$[s55] = \frac{[s5p]^2}{M_{24}}$$

$$[w2s3] = \frac{[w2][s3]}{M_{11}}$$

$$[w6s3] = \frac{[w6][s3]}{M_{12}}$$

$$[w12s4] = \frac{[w12][s4]}{M_{14}}$$

$$[w35s4] = \frac{[w35][s4]}{M_{15}}$$

$$[w2s5] = \frac{[w2][s5]}{M_{17}}$$

$$[w21s5] = \frac{[w21][s5]}{M_{18}}$$

$$[s3] = \frac{1 - [s3p] - 2[s33] - [s34] - [s35] - [p3s3p]}{1 + \frac{[w2]}{M_{11}} + \frac{[w6]}{M_{12}}}$$

$$[s4] = \frac{s4_t - [s4p] - 2[s44] - [s34] - [p4s4p]}{1 + \frac{[w12]}{M_{14}} + \frac{[w35]}{M_{15}}}$$

$$[s5] = \frac{s5_t - [s5p] - 2[s55] - [s35] - [p5s5p]}{1 + \frac{[w2]}{M_{17}} + \frac{[w21]}{M_{18}}}$$

(S81)

where we denote the Michaelis constants

$$M_{11} = \frac{a_2 + a_3}{a_1 S3_T}, M_{12} = \frac{a_5 + a_6}{a_4 S3_T}, M_{13} = \frac{a_8 + a_9}{a_7 S3_T}, M_{14} = \frac{a_{11} + a_{12}}{a_{10} S3_T}, M_{15} = \frac{a_{14} + a_{15}}{a_{13} S3_T}, M_{16} = \frac{a_{17} + a_{18}}{a_{16} S3_T},$$

$$M_{17} = \frac{a_{20} + a_{21}}{a_{19} S3_T}, M_{18} = \frac{a_{23} + a_{24}}{a_{22} S3_T}, M_{19} = \frac{a_{26} + a_{27}}{a_{25} S3_T}, M_{20} = \frac{a_{29}}{a_{28} S3_T}, M_{21} = \frac{a_{31}}{a_{30} S3_T}, M_{22} = \frac{a_{33}}{a_{32} S3_T},$$

$$M_{23} = \frac{a_{35}}{a_{34} S3_T}, M_{24} = \frac{a_{37}}{a_{36} S3_T}.$$

We look for steady-state solutions of System (S80):

$$\begin{aligned} 0 &= m_2 [w2s3] + m_4 [w6s3] - m_7 [p3s3p], \\ 0 &= m_{10} [w12s4] + m_{13} [w35s4] - m_{16} [p4s4p], \\ 0 &= m_{19} [w2s5] + m_{22} [w21s5] - m_{25} [p5s5p]. \end{aligned} \quad (S82)$$

We can rewrite Equations (S82) as follows:

$$\begin{aligned} 0 &= n_6 [w2s3] + n_7 [w6s3] - [p3s3p], \\ 0 &= n_8 [w12s4] + n_9 [w35s4] - [p4s4p], \\ 0 &= n_{10} [w2s5] + n_{11} [w21s5] - [p5s5p]. \end{aligned} \quad (S83)$$

where $n_6 = \frac{a_2}{a_8}, n_7 = \frac{a_5}{a_8}, n_8 = \frac{a_{11}}{a_{17}}, n_9 = \frac{a_{14}}{a_{17}}, n_{10} = \frac{a_{20}}{a_{26}}, n_{11} = \frac{a_{23}}{a_{26}}$.

We can rewrite System (S83) substituting Equations (S81):

$$\begin{cases} 0 = [s3p] + 2\frac{[s3p]^2}{M_{20}} + \frac{[s3p][s4p]}{M_{21}} + \frac{[s3p][s5p]}{M_{23}} + \\ \quad + p3_t \frac{[s3p]}{M_{13} + [s3p]} \left( 1 + \frac{M_{11} M_{12} + M_{12}[w2] + M_{11}[w6]}{n_6 M_{12}[w2] + n_7 M_{11}[w6]} \right) - 1, \\ 0 = [s4p] + 2\frac{[s4p]^2}{M_{22}} + \frac{[s3p][s4p]}{M_{21}} + \\ \quad + p4_t \frac{[s4p]}{M_{16} + [s4p]} \left( 1 + \frac{M_{14} M_{15} + M_{15}[w12] + M_{14}[w35]}{n_8 M_{15}[w12] + n_9 M_{14}[w35]} \right) - s4_t, \\ 0 = [s5p] + 2\frac{[s5p]^2}{M_{24}} + \frac{[s3p][s5p]}{M_{23}} + \\ \quad + p5_t \frac{[s5p]}{M_{19} + [s5p]} \left( 1 + \frac{M_{17} M_{18} + M_{18}[w2] + M_{17}[w21]}{n_{10} M_{18}[w2] + n_{11} M_{17}[w21]} \right) - s5_t, \end{cases} \quad (S84)$$

We find $[s3p]$, $[s4p]$ and $[s5p]$ in System (S84) numerically.

**CD46**

It can be written for SP1 in non-dimensional form according to Equation (S40):

$$[sp1a] = sp1_t \frac{[i2]}{M_{25}+[i2]} \frac{[cd46]}{M_{26}+[cd46]}, \quad (S85)$$

where $[sp1a] = \frac{[SP1a]}{S3_T}$, $sp1_t = \frac{SP1_T}{S3_T}$, $[cd46] = \frac{[CD46]}{S3_T}$, $M_{25} = \frac{l_3}{l_2 S3_T}$, and $M_{26} = \frac{l_5}{l_4 S3_T}$.

## 3.3 IFN-γ and IL-10 production

Since IFN-γ gene is activated by either STAT44 or STAT55, it can be written according to Equation (S53):

$$[ig] = \frac{M_{27}}{\frac{mp1_t}{n_{12} gg_t \left( \frac{[s44]}{M_{28}+[s44]} + \frac{[s55]}{M_{29}+[s55]} - \frac{[s44]}{M_{28}+[s44]} \frac{[s55]}{M_{29}+[s55]} \right)} - 1}, \quad (S86)$$

where

$[ig] = \frac{[Ig]}{S3_T}, mp1_t = \frac{Mp1_T}{S3_T}, gg_t = \frac{Gg_T}{S3_T}, M_{27} = \frac{k_6 + k_7}{k_5 S3_T}, M_{28} = \frac{k_2}{k_1 S3_T}, KM_{29} = \frac{k_4}{k_3 S3_T}, n_{12} = \frac{l_1}{k_6}$, and $n_{12} < 1$.

The production of IL-10 can be activated by either STAT33 or SP1 through CD46. Thus, according to Equation (S53), it can be written:

$$[i10] = \frac{M_{30}}{\frac{mp2_t}{n_{13} g10_t \left( \frac{[s33]}{M_{31}+[s33]} + \frac{[sp1a]}{M_{32}+[sp1a]} - \frac{[s33]}{M_{31}+[s33]} \frac{[sp1a]}{M_{32}+[sp1a]} \right)} - 1}, \quad (S87)$$

where $mp2_t = \frac{Mp2_T}{S3_T}, g10_t = \frac{G10_T}{S3_T}, M_{30} = \frac{k_{13}+k_{14}}{k_{12} S3_T}, M_{31} = \frac{k_9}{k_8 S3_T}, M_{32} = \frac{k_{11}}{k_{10} S3_T}, n_{13} = \frac{l_6}{k_{13}}$,

where $n_{13} < 1$.

## Model Parameters

For the parameter fitting in the STAT3-STAT5 model, we used the Genetic Algorithm (GA) tool integrated to MATLAB. As the criterion for fitting we chose the squared error, which can be described as $SM = \sum_{i=1}^{N}(E_i - M_i)^2$, where $E_i$ is experimental data for cytokine concentration (IFN-γ or IL-10) corresponding to the $i$-th value of IL-2 concentration, $M$ is the model predictions for cytokine concentration corresponding to the same IL-2 concentration, $i$ is the number of experimental data point, $N = 4$ is the total number of experimental data points. The GA tool allows minimizing the squared error using the integrated algorithms for the optima search.

We selected a "nominal" set of parameters "by hand", which qualitatively demonstrates the switching between IFN-γ and IL-10. This set of parameters is represented as "Nom" in Table S2. The sets of the optimized parameters and corresponding squared errors are also shown in Table S2. We performed 15 optimization tests setting the allowable ranges for the parameters ten-fold either side of the nominal values of parameters and chose the best fitting (set "O3" in Table S2) with the smallest squared error $SM = 7.34 \cdot 10^{-7}$. Fig S1 shows the distribution of five parameter sets with the closest minimum squared errors $SM$, namely sets "O2", "O3", "O9", "O10" and "O12".

**Table S2. Nominal, optimized parameters and squared error SM.**

| Par | Nom | O1 | O2 | O3 | O4 | O5 | O6 | O7 | O8 | O9 | O10 | O11 | O12 | O13 | O14 | O15 |
|---|---|---|---|---|---|---|---|---|---|---|---|---|---|---|---|---|
| $r2_t$ | 0.003 | 0.003509 | 0.002832 | 0.0027 | 0.000642 | 0.019794 | 0.00045 | 0.019193 | 0.007935 | 0.001365 | 0.0004 | 0.027443 | 0.017109 | 0.004116 | 0.001955 | 0.000483 |
| $p2_t$ | 0.003 | 0.027889 | 0.005466 | 0.0027 | 0.003567 | 0.003013 | 0.00236 | 0.000391 | 0.001873 | 0.003381 | 0.000954 | 0.018632 | 0.006114 | 0.002781 | 0.001408 | 0.00091 |
| $M_1$ | 0.1 | 0.025534 | 0.47271 | 0.1333 | 0.040663 | 0.3277 | 0.026149 | 0.40489 | 0.20611 | 0.033704 | 0.015943 | 0.72946 | 0.056363 | 0.13378 | 0.021674 | 0.019028 |
| $M_2$ | 4.00E-05 | 0.000204 | 0.000161 | 4.25E-05 | 0.000197 | 7.32E-05 | 0.000283 | 1.62E-05 | 0.000343 | 3.83E-05 | 0.000223 | 7.22E-05 | 5.12E-05 | 6.21E-05 | 3.20E-05 | 9.20E-05 |
| $n_1$ | 120 | 134.51 | 644.3 | 118.42 | 72.739 | 103.68 | 560.59 | 820.37 | 200.61 | 862.9 | 283.75 | 105.73 | 274.74 | 34.881 | 230.86 | 14.268 |
| $s5_t$ | 0.025 | 0.024774 | 0.1077 | 0.0247 | 0.002822 | 0.023708 | 0.013273 | 0.018098 | 0.11675 | 0.00651 | 0.004803 | 0.23769 | 0.12053 | 0.023753 | 0.015575 | 0.010874 |
| $p3_t$ | 2.6 | 0.89728 | 1.2949 | 2.5924 | 24.163 | 22.171 | 7.4742 | 10.399 | 5.6766 | 10.873 | 4.9583 | 21.681 | 19.95 | 20.225 | 16.902 | 7.6764 |
| $p5_t$ | 0.001 | 0.001858 | 0.000861 | 0.0012 | 0.000221 | 0.006268 | 0.000882 | 0.00265 | 0.009407 | 0.001797 | 0.000474 | 0.000497 | 0.009068 | 0.000475 | 0.001185 | 0.006386 |
| $M_7$ | 4.00E-04 | 0.001728 | 4.57E-05 | 3.63E-04 | 0.001127 | 0.000878 | 4.59E-05 | 0.001467 | 0.001375 | 0.000268 | 0.001358 | 0.000201 | 0.002265 | 0.000388 | 8.25E-05 | 0.000542 |
| $M_9$ | 48 | 9.6863 | 44.177 | 47.714 | 6.8966 | 69.877 | 16.866 | 6.4433 | 13.955 | 20.904 | 5.4463 | 8.2154 | 13.764 | 10.142 | 8.9364 | 6.9836 |
| $M_{13}$ | 19 | 43.297 | 12.157 | 19.154 | 26.49 | 64.47 | 11.366 | 3.0619 | 6.861 | 29.894 | 4.2357 | 58.53 | 2.3561 | 28.926 | 2.0914 | 54.439 |
| $n_4$ | 0.2 | 0.2107 | 1.2957 | 0.1987 | 0.83912 | 1.0389 | 0.27776 | 0.65197 | 0.13579 | 0.31198 | 1.6554 | 0.99071 | 1.015 | 0.46062 | 0.3713 | 0.94161 |
| $n_5$ | 1.6 | 9.207 | 7.6231 | 1.5589 | 11.252 | 0.79439 | 0.60074 | 0.68472 | 1.4609 | 0.34328 | 0.43991 | 3.5845 | 10.667 | 1.1144 | 0.41878 | 0.22359 |
| $M_{14}$ | 0.1 | 0.052263 | 0.94864 | 0.1002 | 0.14521 | 0.42794 | 0.74908 | 0.68789 | 0.045914 | 0.23346 | 0.64129 | 0.026867 | 0.19001 | 0.71674 | 0.87757 | 0.22681 |
| $M_{10}$ | 0.005 | 0.007492 | 0.001309 | 0.0047 | 0.040772 | 0.005439 | 0.006246 | 0.007474 | 0.001363 | 0.002274 | 0.019538 | 0.002244 | 0.00649 | 0.000814 | 0.01832 | 0.001643 |
| $M_{12}$ | 2.00E+03 | 3354 | 558.06 | 1.96E+03 | 2941.2 | 7507.6 | 643.49 | 542.99 | 712.59 | 1185.3 | 1028.3 | 1391.6 | 248.85 | 257.14 | 17278 | 4567.5 |
| $M_{15}$ | 0.5 | 1.3545 | 1.3625 | 0.3787 | 0.073671 | 0.62091 | 1.3246 | 0.04003 | 0.1918 | 2.5022 | 0.36109 | 1.2976 | 0.29186 | 0.50924 | 0.51241 | 1.027 |
| $n_6$ | 5.5 | 2.7485 | 1.1754 | 5.5056 | 33.297 | 2.4937 | 1.0916 | 1.3194 | 20.942 | 1.999 | 1.5219 | 55 | 17.825 | 13.946 | 0.63868 | 25.109 |
| $n_7$ | 0.03 | 0.14378 | 0.17183 | 0.0322 | 0.14479 | 0.060788 | 0.007304 | 0.022957 | 0.19581 | 0.031909 | 0.033886 | 0.025282 | 0.13774 | 0.006371 | 0.004357 | 0.019089 |
| $Q_6$ | 0.001 | 0.000604 | 0.000987 | 0.0014 | 0.001767 | 0.000368 | 0.002322 | 0.001199 | 0.009353 | 0.000293 | 0.001671 | 0.001688 | 0.002172 | 0.001201 | 0.00045 | 0.000147 |
| $Q_{21}$ | 3.40E-11 | 1.82E-11 | 3.29E-11 | 3.42E-11 | 8.30E-11 | 6.91E-12 | 9.33E-11 | 2.18E-11 | 2.57E-11 | 2.05E-11 | 1.36E-10 | 9.48E-12 | 6.35E-11 | 9.55E-12 | 4.26E-11 | 1.97E-11 |
| $gg_t$ | 0.9 | 0.15956 | 0.28281 | 0.8949 | 5.7607 | 6.1621 | 5.1155 | 0.76901 | 0.12177 | 6.8237 | 5.0325 | 5.1714 | 0.71957 | 0.11343 | 0.5014 | 8.0729 |
| $mp1_t$ | 0.003 | 0.000362 | 0.011831 | 0.0034 | 0.002017 | 0.00276 | 0.000491 | 0.010081 | 0.000543 | 0.001536 | 0.023098 | 0.002676 | 0.01239 | 0.003733 | 0.00066 | 0.019531 |
| $M_{18}$ | 10 | 49.794 | 12.051 | 9.913 | 17.075 | 12.601 | 2.0105 | 16.569 | 1.2898 | 3.2815 | 57.245 | 1.3748 | 4.9301 | 27.867 | 11.032 | 46.883 |
| $M_{19}$ | 5 | 1.0175 | 1.6903 | 4.354 | 7.0324 | 29.257 | 1.8676 | 24.928 | 4.1342 | 3.3328 | 11.128 | 23.95 | 1.5233 | 1.3154 | 22.729 | 19.486 |
| $n_8$ | 0.01 | 0.001766 | 0.007209 | 0.0101 | 0.00607 | 0.005921 | 0.003473 | 0.024054 | 0.018385 | 0.053508 | 0.057119 | 0.001551 | 0.001237 | 0.006047 | 0.023173 | 0.031357 |
| $g10_t$ | 7 | 6.9815 | 8.8946 | 7.2184 | 63.392 | 14.123 | 42.777 | 11.788 | 1.5184 | 2.3453 | 1.1908 | 10.022 | 0.89799 | 25.713 | 13.946 | 13.284 |
| $mp2_t$ | 0.6 | 1.2895 | 0.26955 | 0.5933 | 4.5548 | 3.0464 | 0.78732 | 0.29224 | 0.57425 | 0.060594 | 0.20987 | 0.26053 | 0.36536 | 1.293 | 0.67941 | 4.3577 |
| $M_{20}$ | 0.015 | 0.018369 | 0.019073 | 0.0149 | 0.020308 | 0.030563 | 0.003852 | 0.007404 | 0.020735 | 0.019924 | 0.020195 | 0.014823 | 0.055386 | 0.054662 | 0.02151 | 0.074155 |
| $M_{21}$ | 0.01 | 0.024239 | 0.032193 | 0.0138 | 0.024077 | 0.022656 | 0.00449 | 0.00449 | 0.002048 | 0.001389 | 0.095423 | 0.03017 | 0.001637 | 0.001683 | 0.019085 |
| $M_{22}$ | 0.2 | 0.21423 | 1.2631 | 0.1885 | 0.044434 | 1.0261 | 0.035059 | 0.13186 | 0.10364 | 0.051725 | 0.3348 | 0.1968 | 0.5264 | 0.055339 | 0.18746 | 0.20254 |
| $n_9$ | 0.02 | 0.035162 | 0.006445 | 0.0191 | 0.014896 | 0.031773 | 0.00877 | 0.009492 | 0.06544 | 0.005009 | 0.033464 | 0.00655 | 0.030446 | 0.003868 | 0.007794 | 0.021937 |
| cd46 | 0.7 | 0.073077 | 0.1249 | 0.6826 | 0.17389 | 0.23045 | 0.1164 | 1.9959 | 0.98331 | 2.6895 | 0.60932 | 0.14905 | 0.5349 | 2.0231 | 3.279 | 0.12424 |
| $sp1_t$ | 33 | 89.836 | 37.294 | 33.142 | 9.0941 | 18.309 | 249.25 | 22.863 | 16.009 | 5.3526 | 14.461 | 131.19 | 204.26 | 16.788 | 115.79 | 14.167 |
| $M_{16}$ | 9.00E-06 | 6.77E-05 | 8.85E-06 | 8.96E-06 | 5.10E-06 | 4.47E-06 | 7.37E-06 | 1.21E-05 | 7.99E-06 | 1.28E-05 | 6.40E-06 | 2.10E-05 | 1.55E-05 | 2.26E-05 | 6.83E-05 | 9.60E-07 |
| $M_{17}$ | 0.1 | 0.017299 | 0.013423 | 0.1071 | 0.93739 | 0.059451 | 0.16776 | 0.11441 | 0.014937 | 0.070503 | 0.027861 | 0.076081 | 0.37638 | 0.29741 | 0.031363 | 0.9858 |
| SM | 1.19E-06 | 1.15E-06 | 9.18E-07 | 7.34E-07 | 1.81E-06 | 1.56E-06 | 2.13E-06 | 1.21E-06 | 2.11E-06 | 7.96E-07 | 7.92E-07 | 2.22E-06 | 8.70E-07 | 1.13E-06 | 1.54E-06 | 1.31E-06 |

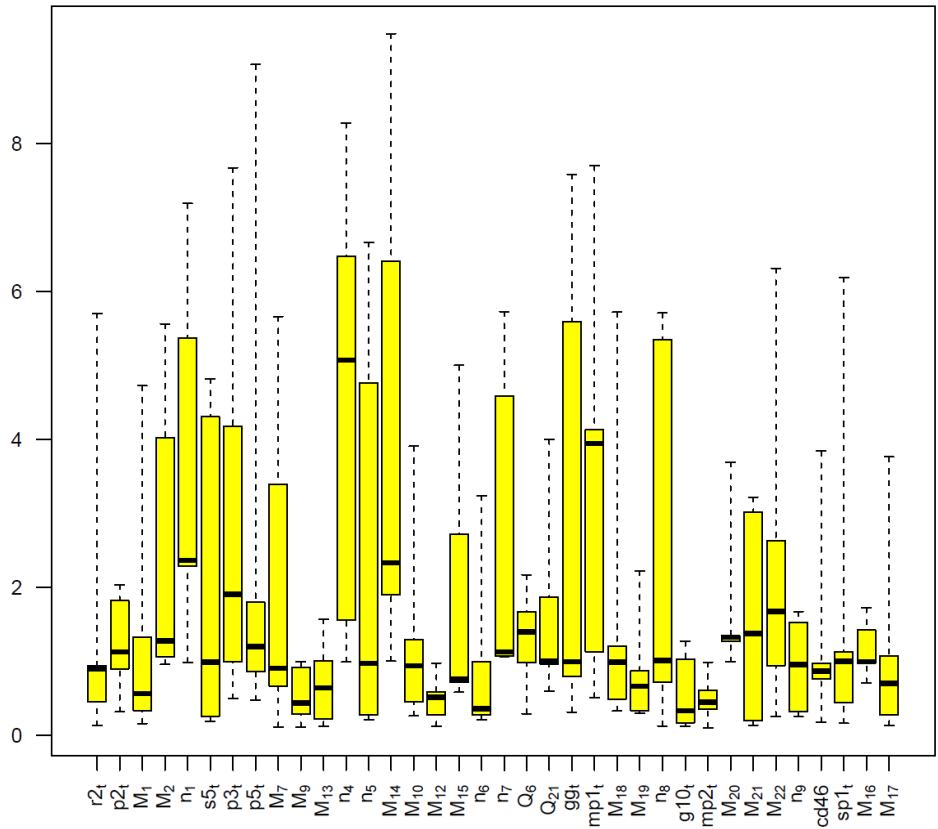

**Fig S1. The distribution of the optimized parameters.** The parameter sets with the closest minimum squared errors $SM$, namely "O2", "O3", "O9", "O10" and "O12".

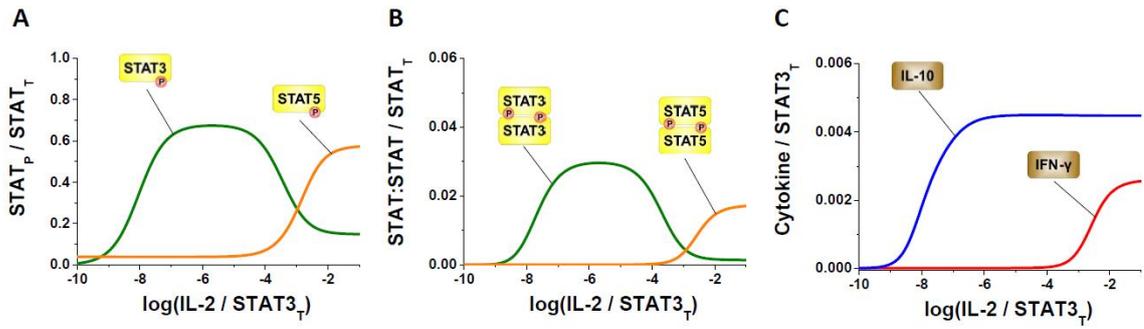

**Fig S2. Model predictions for the swapped parameters.** A. STAT monomers. B. STAT homodimers. C. Produced cytokines.

**Table S3. The effects of the parametric changes on the concentration of produced IFN-γ and IL-10 shown in Fig 3.**

| Fig | Parameters | Perturbation ↓ | thin line (optimized) | ↑ |
|---|---|---|---|---|
| 3A | $n_9$ | $163 \cdot 10^{-4}$ | $191 \cdot 10^{-4}$ | $220 \cdot 10^{-4}$ |
|  | $M_{20}$ | $119 \cdot 10^{-4}$ | $149 \cdot 10^{-4}$ | $179 \cdot 10^{-4}$ |
|  | $g10_t$ | 6.136 | 7.218 | 8.301 |
| 3B | $s5_t$ | $198 \cdot 10^{-4}$ | $247 \cdot 10^{-4}$ | $297 \cdot 10^{-4}$ |
|  | $M_{18}$ | 7.93 | 9.913 | 11.896 |
|  | $n_8$ | $81 \cdot 10^{-4}$ | $101 \cdot 10^{-4}$ | $121 \cdot 10^{-4}$ |
|  | $M_{14}$ | 0.068 | 0.1 | 0.145 |
|  | $M_9$ | 69.185 | 47.714 | 32.445 |
| 3C | $r2_t$ | $2.7 \cdot 10^{-4}$ | $27 \cdot 10^{-4}$ | $268 \cdot 10^{-4}$ |
|  | $n_1$ | 11.842 | 118.42 | 1184.2 |
| 3D | $M_7$ | $3.63 \cdot 10^{-2}$ | $3.63 \cdot 10^{-4}$ | $3.63 \cdot 10^{-6}$ |
| 3E | $M_{10}$ | 470 | $47 \cdot 10^{-4}$ | $47 \cdot 10^{-9}$ |
| 3F, G | $Q_6$ | $1 \cdot 10^{-4}$ | $14 \cdot 10^{-4}$ | $209 \cdot 10^{-4}$ |

**Table S4. Parameters in the STAT3-STAT4 subsystem and their correspondence to the parameters in the STAT3-STAT5 subsystem.**

| Parameters in STAT3-STAT5 | Parameters in STAT3-STAT4 | Values |
|---|---|---|
| $r2_t$ | $r2_t$ | 0.0027 |
| $p2_t$ | $p2_t$ | 0.0027 |
| $M_1$ | $M_1$ | 0.1333 |
| $M_2$ | $M_2$ | 4.25E-05 |
| $n_1$ | $n_1$ | 118.42 |
| $s5_t$ | $s4_t$ | 0.0247 |
| $p3_t$ | $p3_t$ | 2.5924 |
| $p5_t$ | $p4_t$ | 0.0012 |
| $M_7$ | $M_9$ | 3.63E-04 |
| $M_9$ | $M_{11}$ | 47.714 |
| $M_{13}$ | $M_{15}$ | 19.154 |
| $n_4$ | $n_5$ | 0.1987 |
| $n_5$ | $n_6$ | 1.5589 |
| $M_{14}$ | $M_{16}$ | 0.1002 |
| $M_{12}$ | $M_{14}$ | 1.96E+03 |
| $M_{15}$ | $M_{17}$ | 0.3787 |
| $n_6$ | $n_7$ | 5.5056 |
| $n_7$ | $n_8$ | 0.0322 |
| $Q_6$ | $Q_6, Q_{12}$ | 0.0014 |
| $Q_{21}$ | $Q_{35}$ | 3.42E-11 |
| $gg_t$ | $gg_t$ | 0.8949 |
| $mp1_t$ | $mp1_t$ | 0.0034 |
| $M_{18}$ | $M_{20}$ | 9.913 |
| $M_{19}$ | $M_{21}$ | 4.354 |
| $n_8$ | $n_9$ | 0.0101 |
| $g10_t$ | $g10_t$ | 7.2184 |
| $mp2_t$ | $mp2_t$ | 0.5933 |
| $M_{20}$ | $M_{22}$ | 0.0149 |
| $M_{21}$ | $M_{23}$ | 0.0138 |
| $M_{22}$ | $M_{24}$ | 0.1885 |
| $n_9$ | $n_{10}$ | 0.0191 |
| cd46 | cd46 | 0.6826 |
| $sp1_t$ | $sp1_t$ | 33.142 |
| $M_{16}$ | $M_{18}$ | 8.96E-06 |
| $M_{17}$ | $M_{19}$ | 0.1071 |

**Parameter Sensitivity Analysis**

In the main text, we showed that our model predicts that IFN-γ to IL-10 switching (Fig 2) is due to the competition between STAT3 and STAT5 proteins. These model predictions are not obvious from the model structure due to the following. The structure of our model presented in Fig 1 is symmetrical in relation to the varied IL-2, i.e. IL-2 activates both STAT3 and STAT5. The model predictions for the phosphorylated states of STAT3, STAT5 and their homodimers could be swapped in Fig 2 if the parameters for STAT3 and STAT5 are swapped (Fig S2). However, in this case there would be no cytokine switching (Fig S2C). Thus the model predictions depend on the assumed structure of our model as well as on the chosen set of parameters. In this section, we perform parameter sensitivity analysis (SA) [1, 2] to identify the parameter conditions for the conclusions to hold.

First, we test if the IFN-γ to IL-10 and STAT5 to STAT3 switchings are still present if we vary the parameters compared to their optimal values (set "O3" in Table S3). Quantitatively, we define the switching of either IFN-γ to IL-10 or STAT5 to STAT3 as follows. We assume that the switching occurs when the 2-fold changes take place, which is typically considered as significant change in Biology [3, 4]. Thus the changes should include at least 2-fold increase in IL-10 and STAT3 as well as decrease of the peak of IFN-γ and STAT5 concentrations within the range of the tested IL-2concentrations.

We vary the parameters up to 10-fold either side of their values in the optimized set. We use the Latin Hypercube Sampling (LHS), which is considered as one of the most effective strategies for sampling the parameters [2, 5]. We perform the LHS sampling and check the assumed condition for switching for 1000 samples of the optimized parameters. Fig S3 illustrates the probability of the cases, in percent, where the switching of both cytokine and STAT (C+S+), cytokine but not STAT (C+S-), not cytokine but STAT (C-S+), neither cytokine nor STAT (C-S-) occurs for 1-10 fold change of the optimized parameters. The obvious result that follows from Fig S3 is that when the parameters are not perturbed, which corresponds to the 1-fold change, the probability of the presence of both cytokine and STAT switching is 100%. However, with an increase of the fold change up to ten, the probability for switching of both cytokine and STAT decreases down to 2.1% (data shown in Table S5). The probability of either of the cases (cytokine or STAT switching) is almost equal with an increase of the fold change as it can be seen from Fig S3. Thus, we can conclude that the model with the optimized parameters demonstrates both IFN-γ to IL-10 and STAT5 to STAT3 switching with higher probability (more than 50%) when the parameters are perturbed within 2-fold, with modest probability (between 10% and 50%) when the parameters are perturbed within 3-6-fold and with low probability (less than 10%) when the parameters are perturbed within 7-fold and higher.

Next we identify the most sensitive parameters that have the greatest effect on the steady-state concentrations of IFN-γ, IL-10, STAT5 and STAT3 for the three concentrations of IL-2: $10^{-10}$, $10^{-6}$ and $10^{-1}$, which correspond to the three assumed T cell phenotypes shown in Fig 1A: Th1, Th1/Tr1 and Tr1 respectively. We perform the SA using the eFast method [6], because it was reported as one of the most reliable methods of parameter sensitivity analysis [5]. As a tool for the eFast sensitivity analysis, we use the SBToolbox software [7]. We perform the SA over one order of magnitude of perturbation for 10000 simulations.

Fig S4-Fig S6 illustrate the results of the sensitivity analysis for IFN-γ, IL-10, STAT5 and STAT3 by the SBToolbox for the non-dimensional IL-2 concentrations $[i2] = 10^{-10}$ (Fig S4), $[i2] = 10^{-6}$ (Fig S5) and $[i2] = 10^{-1}$ (Fig S6). The bars indicate the sensitivity indices for each of the parameters of our model. Here we classify the parameters as sensitive if the corresponding sensitivity index is more than $0.5$.

It can be seen from Fig S4A, Fig S5A and Fig S6A that IFN-γ production is the most sensitive to the following parameters: $M_{18}$, $gg_t$, $n_8$, $M_{19}$, $mp1_t$ and $s5_t$, which demonstrate high sensitivity indices (more than $0.5$) for all the three IL-2 concentrations. Parameters $n_1$, $r2_t$, $n_6$ and $M_2$ are sensitive for $[i2] = 10^{-10}$ (Fig S4A) and $[i2] = 10^{-6}$ (Fig S5A). Parameter $M_{14}$ is sensitive only for $[i2] = 10^{-6}$ (Fig S5A) and $[i2] = 10^{-1}$ (Fig S6A). Another group of IFN-γ-sensitive parameters that includes $s5_t$ and $n_6$ is involved in the STAT5 pathway activation, which leads to the production of IFN-γ as shown in Fig 1B and Equations (6). There is also the third group of IFN-γ-sensitive parameters consisting of $n_1$, $r2_t$ and $M_2$ that are involved in the upstream activation of IL-2 receptor described by Equation (1). Finally, IFN-γ is sensitive to the Michaelis constant of heterodimerization $M_{14}$.

The parameter sensitivity analysis reveals that the concentration of IL-10 is the most sensitive to $g10_t$ and $n_9$ for $[i2]=10^{-10}$ (Fig S4B), $[i2]=10^{-6}$ (Fig S5B) and $[i2]=10^{-1}$ (Fig S6B). Parameter $M_9$ is the most sensitive for the phosphorylated STAT3 for all the three IL-2 concentrations as it is shown in Fig S4C, Fig S5C and Fig S6C. Two parameters, namely, $n_5$ and $Q_6$, show high sensitivity only for $[i2]=10^{-10}$ (Fig S4C) and $[i2]=10^{-6}$ (Fig S5C).

The concentration of phosphorylated STAT5 is the most sensitive to the total amount of STAT5, modeled by parameter $s5_t$, for all the three tested IL-2 concentrations as it is shown in Fig S4D, Fig S5D and Fig S6D. Parameter $n_6$ demonstrates high sensitivity indices for STAT5p for $[i2]=10^{-10}$ (Fig S4D) and $[i2]=10^{-6}$ (Fig S5D). This parameter is involved in the activation of STAT5 pathway as shown in Fig 1B and Equations (6). The concentration of STAT5p is sensitive to the Michaelis constant of heterodimerization $M_{14}$ for $[i2]=10^{-6}$ (Fig S5D) and $[i2]=10^{-1}$ (Fig S6D). Three parameters, namely $n_1$, $r2_t$ and $M_2$, are sensitive for $[i2]=10^{-10}$ (Fig S4D) and $[i2]=10^{-6}$ (Fig S5D) involved in the activation of the IL-2 receptor and described by Equation (1).

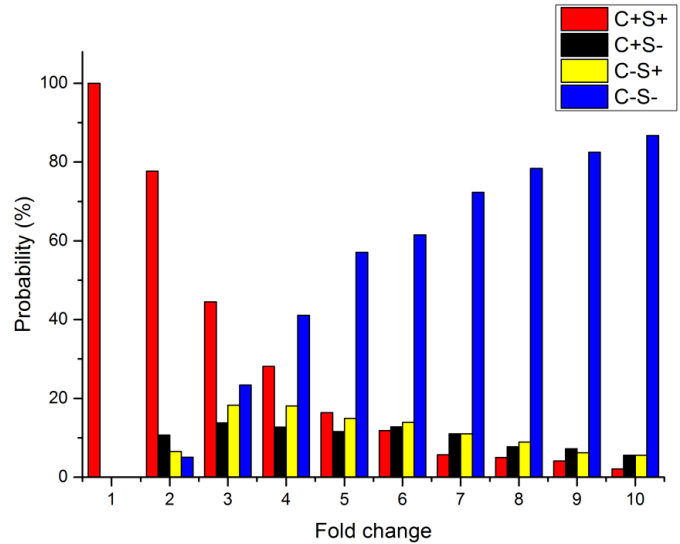

**Fig S3. The probability of cases when the switching occurs.** The percentage of cases when the switching of both cytokine and STAT (C+S+), cytokine but not STAT (C+S-), not cytokine but STAT (C-S+), neither cytokine nor STAT (C-S-) occurs for 1-10 fold change of the optimized parameters.

**Table S5. The percentage of cases when the switching of both cytokine and STAT (C+S+), cytokine but not STAT (C+S-), not cytokine but STAT (C-S+), neither cytokine nor STAT (C-S-) occurs for 1-10 fold change of the optimized parameters.**

| FC | C+S+ | C+S- | C-S+ | C-S- |
|---|---|---|---|---|
| 1 | 100 | 0 | 0 | 0 |
| 2 | 77.7 | 10.7 | 6.5 | 5.1 |
| 3 | 44.5 | 13.8 | 18.3 | 23.4 |
| 4 | 28.1 | 12.7 | 18.1 | 41.1 |
| 5 | 16.4 | 11.6 | 14.9 | 57.1 |
| 6 | 11.8 | 12.8 | 13.9 | 61.5 |
| 7 | 5.7 | 11 | 11 | 72.3 |
| 8 | 5 | 7.7 | 8.9 | 78.4 |
| 9 | 4.1 | 7.2 | 6.2 | 82.5 |
| 10 | 2.1 | 5.6 | 5.6 | 86.7 |

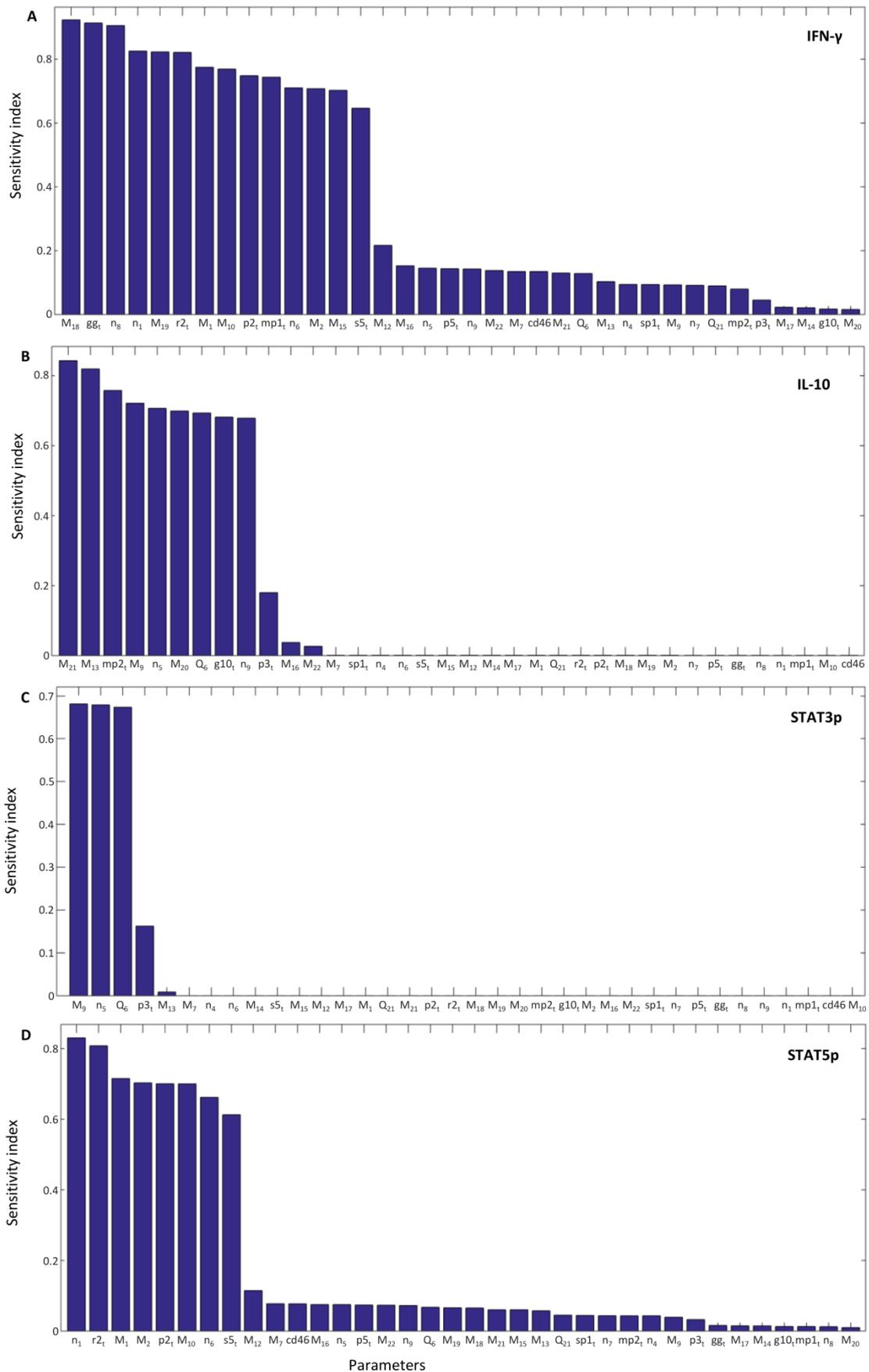

**Fig S4. Parameter sensitivity analysis performed by eFAST for low concentrations of IL-2.** Sensitivity indicators for the developed model for IFN-γ (A), IL-10 (B), STAT3 (C) and STAT5 (D) for non-dimensional IL-2 concentration $[i2] = 10^{-10}$.

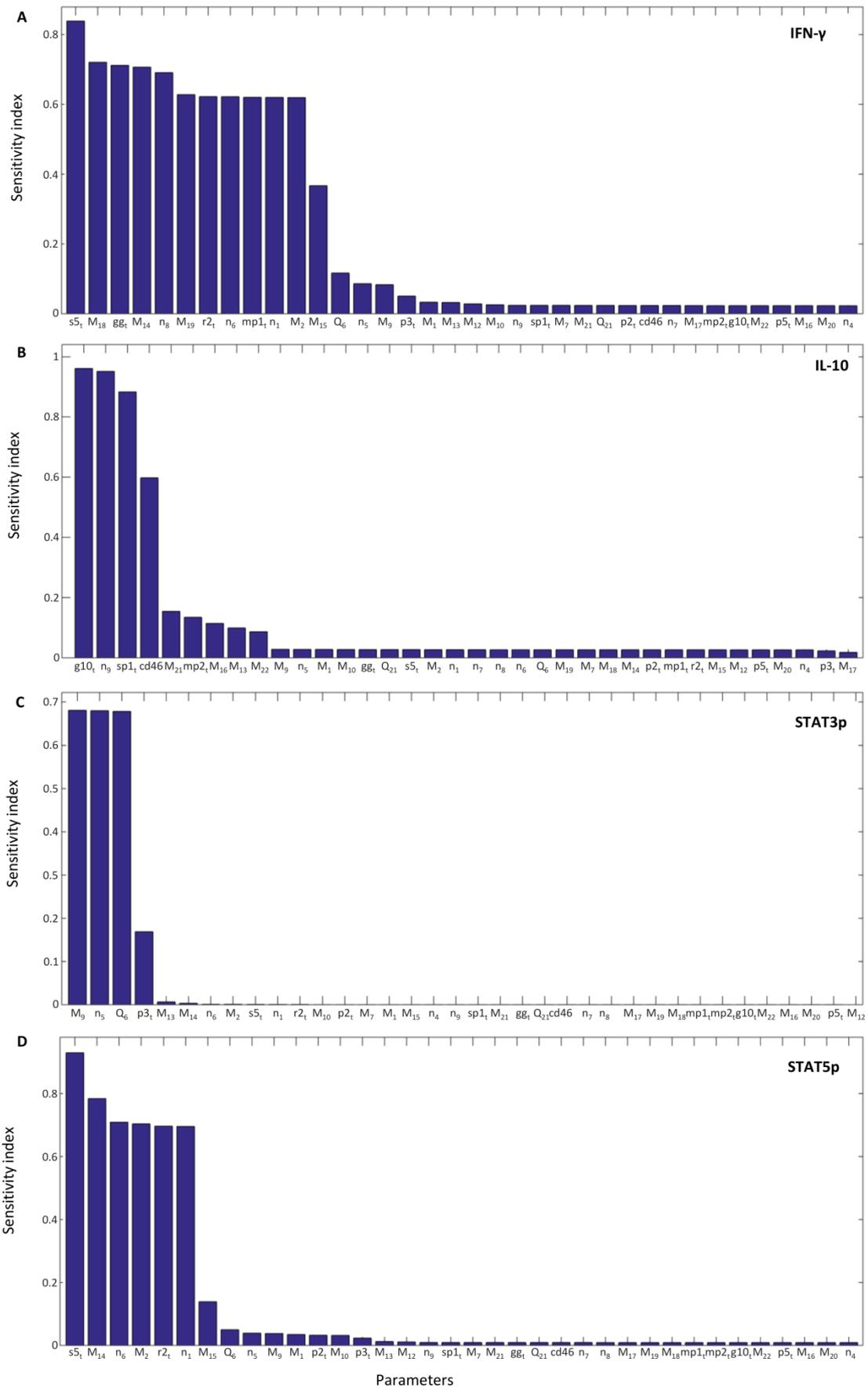

**Fig S5. Parameter sensitivity analysis performed by eFAST for medium concentrations of IL-2.** Sensitivity indicators for the developed model for IFN-γ (A), IL-10 (B), STAT3 (C) and STAT5 (D) for non-dimensional IL-2 concentration $[i2] = 10^{-6}$.

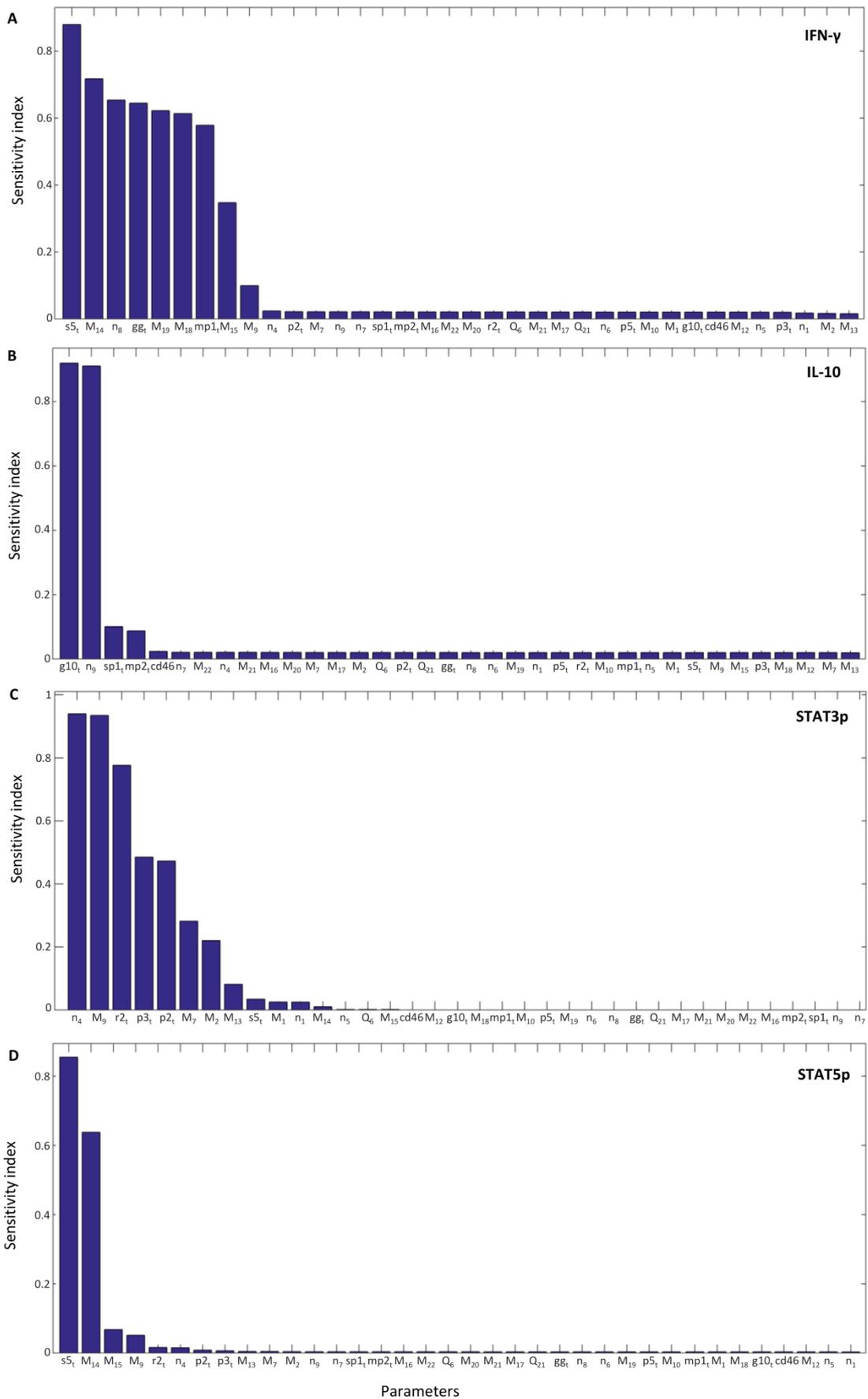

**Fig S6. Parameter sensitivity analysis performed by eFAST for high concentrations of IL-2.**
Sensitivity indicators for the developed model for IFN-γ (A), IL-10 (B), STAT3 (C) and STAT5 (D) for non-dimensional IL-2 concentration $[i2] = 10^{-1}$.